\documentclass[11pt,a4paper]{article}
\pdfoutput=1

\usepackage{jheppub}

\usepackage{amssymb}
\usepackage{amsmath}
\usepackage{braket}
\usepackage{graphicx}
\usepackage{slashed}

\def\cB{\mathcal{B}}

\def\cI{\mathcal{I}}

\def\cL{\mathcal{L}}

\def\cO{\mathcal{O}}
\def\cS{\mathcal{S}}
\def\bS{\mathbb{S}}

\DeclareRobustCommand{\eq}[1]{eq.~\eqref{eq:#1}}
\DeclareRobustCommand{\eqs}[2]{eqs.~\eqref{eq:#1} and \eqref{eq:#2}}
\DeclareRobustCommand{\fig}[1]{fig.~\ref{fig:#1}}

\DeclareRobustCommand{\sec}[1]{sec.~\ref{sec:#1}}
\DeclareRobustCommand{\secs}[2]{secs.~\ref{sec:#1} and \ref{sec:#2}}
\DeclareRobustCommand{\mycite}[1]{ref.~\cite{#1}}
\DeclareRobustCommand{\mycites}[1]{refs.~\cite{#1}}

\newcommand{\kcut}{k_\mathrm{cut}}

\newcommand{\pt}{\vec{p}_T}
\newcommand{\qt}{\vec{q}_T}
\newcommand{\kt}{\vec{k}_T}
\newcommand{\bt}{\vec{b}_T}

\newcommand{\ki}[1]{{\vec k}_{#1}}
\newcommand{\ka}{\ki{a}}
\newcommand{\kb}{\ki{b}}
\newcommand{\ks}{\ki{s}}

\let\oldvec\vec		
\renewcommand*\vec[1]{\oldvec{\kern0pt #1}}
\DeclareMathOperator{\dis}{\!|_+}		
\newcommand{\nn}{\nonumber}
\newcommand{\FT}{\mathrm{FT}}
\newcommand{\del}{\mathrm{d}}

\newcommand{\PlusLog}[1]{\mathcal{L}_{#1}}	
\newcommand{\PlusPol}[1]{\mathcal{L}^{#1}}	

\newcommand{\GeV}{\text{GeV}}
\newcommand{\TeV}{\text{TeV}}
\newcommand{\dR}[2][2]{R_{#1}^{(#2)}}		
\newcommand{\dRc}[2][2]{\tilde R_{#1}^{(#2)}}	
\newcommand{\LQCD}{\Lambda_\text{QCD}}

\newcommand{\GammaC}{\Gamma_\text{cusp}}
\newcommand{\gMuH}{\gamma_{H}}			         
\newcommand{\gMuB}{\gamma_B}			         
\newcommand{\gMuS}{\gamma_S}			         
\newcommand{\gMuSconst}[1]{\gamma_{S\,#1}}	
\newcommand{\gMuSCNC}[1]{\gMuSconst{#1}}           
\newcommand{\gNu}{\gamma_\nu}				            
\newcommand{\tgNu}{\tilde\gamma_\nu}
\newcommand{\gNuNP}{\gamma_\nu^{\text{(np)}}}                      
\newcommand{\gNuC}[1]{\gamma_{\nu}^{(#1)}}		   
\newcommand{\gNuConst}[1]{\gamma_{\nu\,#1}}         
\newcommand{\gNuConstB}[1]{\tilde\gamma_{\nu\,#1}}         
\newcommand{\gNuCNC}[1]{\gNuConst{#1}} 		      

\newcommand{\as}{\alpha_s}
\newcommand{\AS}{\Bigl(\frac{\as}{4\pi}\Bigr)}		
\newcommand{\ASmu}{\biggl[\frac{\as(\mu)}{4\pi}\biggr]}	

\newcommand{\FO}{\mathrm{FO}}

\title{Resummation of Transverse Momentum Distributions in Distribution Space}

\author{Markus A.~Ebert}
\author{and Frank J.~Tackmann}

\affiliation{Theory Group, Deutsches Elektronen-Synchrotron (DESY),\\ D-22607 Hamburg, Germany}

\emailAdd{markus.ebert@desy.de}
\emailAdd{frank.tackmann@desy.de}

\abstract{
Differential spectra in observables that resolve additional soft or collinear QCD emissions
exhibit Sudakov double logarithms in the form of logarithmic plus distributions.
Important examples are the total transverse momentum $q_T$ in color-singlet production, $N$-jettiness (with thrust or beam thrust as special cases), but also jet mass and more complicated jet substructure observables.
The all-order logarithmic structure of such distributions is often fully encoded
in differential equations, so-called (renormalization group) evolution equations.
We introduce a well-defined technique of distributional scale setting, which allows one to treat logarithmic plus distributions like ordinary logarithms when solving these differential equations. In particular, this allows one (through canonical scale choices) to minimize logarithmic contributions in the boundary terms of the solution, and to obtain the full distributional logarithmic structure from the solution's evolution kernel directly in distribution space.
We apply this technique to the $q_T$ distribution, where the two-dimensional nature of
convolutions leads to additional difficulties (compared to one-dimensional cases like thrust),
and for which the resummation in distribution (or momentum) space has been a long-standing open question.
For the first time, we show how to perform the RG evolution fully in momentum space,
thereby directly resumming the logarithms $[\ln^n(q_T^2/Q^2)/q_T^2]_+$
appearing in the physical $q_T$ distribution.
The resummation accuracy is then solely determined by the perturbative expansion of the associated anomalous dimensions.
}

\preprint{\vbox{
\hbox{DESY 16-215}
\hbox{November 25, 2016}
}}

\begin{document}
\maketitle

\section{Introduction}
An important class of differential observables at colliders are those that resolve additional soft or collinear QCD emissions on top of the underlying hard Born process.
They typically exhibit Sudakov double logarithms of the form $\as^n \ln^m(Q/k)$ with $m\le2n$, where $Q$ is the relevant hard scale of the problem and $k$ the differential observable.
The bulk of the cross section is usually contained in the regime $k \ll Q$,
where the double logarithms can become large and eventually spoil the convergence of the fixed-order
perturbative expansion.
The resummation of the Sudakov logarithms to all orders in $\as$ becomes necessary to obtain a stable and reliable prediction in this regime.

Resummation can be carried out either using Monte Carlo techniques such as parton showers,
or analytically based on factorization theorems which can be derived diagrammatically or using effective field theories.
All ingredients of the factorized cross section obey (renormalization group) evolution equations of the form
\begin{equation} \label{eq:intro_RGE}
 \mu \frac{\del F(k,\mu)}{\del \mu} = \gamma_F(\mu) F(k,\mu)
\,,\end{equation}
where $F$ represents an ingredient of the factorized cross section depending on the (momentum space) variable $k$
and the (unphysical) scale $\mu$,
and $\gamma_F$ is its anomalous dimension.
\eq{intro_RGE} encodes the logarithmic structure of $F$ to all orders.
Solving it allows to exponentiate all large logarithms $\ln(k/\mu)$,
\begin{equation} \label{eq:intro_RGevolved}
 F(k,\mu) = F(k,\mu_0) \exp\biggl[\int_{\mu_0}^\mu \frac{\del\mu'}{\mu'} \gamma_F(\mu') \biggr]
\,,\end{equation}
provided that $\mu_0$ is chosen of the order of $k$, $\mu_0 \sim k$.
In this case, $F(k,\mu_0)$ is free of large logarithms and can be reliably calculated in fixed-order perturbation theory,
whereas all logarithms $\ln(\mu/k)$ are explicitly exponentiated.

While this is straightforward for ordinary functions $F$ such as hard functions describing virtual corrections,
it becomes more involved when $F$ contains (plus) distributions.
These naturally arise for many observables in order to properly cancel infrared divergences between the different functions.
Well-known examples are the total transverse momentum $q_T$ in color-singlet production, $N$-jettiness (with thrust or beam thrust as special cases), but also jet mass and more complicated jet substructure observables.
The simplest example to illustrate the complications in such cases is a pure $\delta$-distribution,
which typically arises as the LO boundary term.
In this case, \eq{intro_RGevolved} reads
\begin{equation}
 F(k,\mu)
 = \delta(k) \exp\biggl[\int_{\mu_0}^\mu \frac{\del\mu'}{\mu'} \gamma_F(\mu') \biggr]
 \approx \delta(k) \exp\biggl[ \ln\frac{\mu}{\mu_0} \gamma_F(\mu) \biggr]
\,,\end{equation}
such that it is obviously not possible to choose $\mu_0 = k$ to fully resum all logarithms $\ln(k/\mu)$.
Hence one needs a technique that correctly turns this expression into a sum of plus distributions $\bigl[\ln^n(k/\mu) / k\bigr]_+^\mu$ that are known to arise at higher orders.

In this paper, we derive a well-defined technique to solve RGEs in distribution space by introducing a distributional scale setting $\mu_0 = k\dis$.
It allows to treat evolution equations such as \eq{intro_RGE} like ordinary differential equations, or equivalently treats distributional logarithms like ordinary logarithms.
Hence this technique allows to minimize logarithmic terms in the boundary condition $F(k,\mu_0)$ through distributional canonical scale setting, thereby fully resumming all logarithms directly in distribution space.
We also show that this is fundamentally different from performing the resummation in conjugate space,
which also turns distributions into ordinary functions, but induces subleading terms to all orders.

We then apply this technique to the resummation of the transverse momentum ($\qt$) distribution in color-singlet production. The transverse-momentum spectrum is a key observable for many processes at hadron colliders. In Higgs production, it is one of the primary variables describing the production kinematics \cite{Aad:2014lwa, Aad:2014tca, Aad:2015lha, Aad:2016lvc, ATLAS:2016nke, Khachatryan:2015yvw, Khachatryan:2015rxa, Khachatryan:2016vnn}. In Drell-Yan production it is an important benchmark observable, which has been measured to very high precision~\cite{Aad:2014xaa, Aad:2015auj, Chatrchyan:2011wt, Khachatryan:2015oaa, Khachatryan:2016nbe}.

It has been a long-standing open question whether a direct resummation of the $\qt$ distribution in momentum (or distribution) space is possible at all.
Instead, the resummation is usually carried out in the conjugate Fourier ($\bt)$ space, which however fundamentally resums logarithms $\ln(b_T Q)$ rather than the logarithms $\ln(Q/q_T)$ appearing in the physical $q_T$ spectrum (with $Q$ being the hard scale of the process).
Previous attempts so far have either encountered spurious divergences in the resummed $q_T$ spectrum, see e.g.~\mycites{Frixione:1998dw, Chiu:2012ir}, or the evolution is still partially performed in Fourier space, see e.g.~\mycites{Becher:2010tm, Becher:2011xn, Becher:2012yn, D'Alesio:2014vja, Echevarria:2015uaa},
while \mycites{Kulesza:1999gm,Kulesza:1999sg,Kulesza:2001jc} tried to obtain a closed form of the Fourier-resummed spectrum in momentum space.
Recently, \mycite{Monni:2016ktx} obtained the resummation in momentum space based on the coherent branching formalism \cite{Banfi:2004yd, Banfi:2014sua} rather than solving the evolution equations associated with a factorization theorem.
They showed that the spurious divergences are avoided by expanding around the transverse momentum of the hardest emission instead of $q_T$. We briefly compare our findings to those of \mycite{Monni:2016ktx} in \sec{comparison}.

In this paper, we derive the solution to perform the RG evolution entirely in distribution (momentum) space, thus allowing for the explicit resummation of all logarithmic contributions $[\ln^n(q_T^2/Q^2)/q_T^2]_+$ appearing in the physical $q_T$ distribution.
We show that it intrinsically requires distributional scale setting due to the two-dimensional convolutions appearing for $q_T$, which is not the case for one-dimensional observables such as thrust.
In particular, we find that this is the origin of the divergences observed in previous attempts of momentum-space resummation.

An advantage of performing the resummation via the solution of the $q_T$ evolution equations is that the solution automatically applies to all orders in resummed perturbation theory, and the resummation accuracy is solely defined through the perturbative accuracy of the anomalous dimensions (as well as matching conditions). In particular, this allows one to completely avoid any discussions of how to consistently count explicit logarithms in the cross section (which has been part of the difficulties in previous attempts).
We indeed find that apparent subleading terms in the spectrum become important to obtain a well-defined prediction.
The resummation through RG evolution also provides a natural and convenient way to smoothly turn off the resummation and match to the fixed-order region at large $q_T$, as well as to estimate perturbative uncertainties through the variation of all the appearing resummation scales. Importantly, by performing the evolution in momentum space, these steps can be carried out directly at the level of the physical $q_T$ spectrum.
In this paper, our primary purpose will be to derive the all-order solution for the momentum-space evolution, while we leave its numerical implementation to future work.

The remainder of the paper is organized as follows.
In \sec{distr scale setting}, we introduce the technique of distributional scale setting
and discuss how it is used to solve distributional differential (RG) equations in distribution space.
We also discuss why this is fundamentally different from performing the resummation in conjugate space.
For simplicity, most of this general discussion is for one-dimensional observables, and then
generalized to two-dimensional cases at the end.
The remaining sections are then devoted to the $\qt$ resummation.
In \sec{TMD}, we briefly review the relevant factorization theorem and evolution equations, and discuss the difficulties
associated with the two-dimensional nature of $\qt$. In \sec{gNu resummation}, we discuss the resummation of
the rapidity anomalous dimension via the solution of its RGE. We also briefly discuss how its intrinsic
nonperturbative contributions arise and can be handled in momentum space.
In \sec{soft func}, we derive the solution for the momentum-space evolution for
the soft and beam functions. We then discuss how all pieces get assembled into the final resummed $\qt$ spectrum in
\sec{implementation} and provide some comparisons to existing approaches and implementations in the literature in
\sec{comparison}. We conclude in \sec{conclusion}.
In the appendices we collect many relevant and useful definitions and relations for plus distributions.

\section{Scale setting in distribution space}
\label{sec:distr scale setting}

A key feature of observables $k$ that resolve the IR structure of soft and collinear emissions,
including the transverse momentum $q_T$, is that the logarithmic structure of the differential spectrum
in $k$ is given in terms of plus and delta distributions, which encode the cancellation
of the associated IR divergences between real and virtual contributions.

The correct scale setting to minimize logarithmic distributions is a crucial ingredient in the
resummation of the differential distributions directly in distribution space. In particular,
as we will see, for the $q_T$ spectrum this becomes a necessity to obtain well-defined predictions.

In this section, we discuss on general grounds several aspects of solving distributional
differential equations. In particular, we introduce a distributional scale setting
to correctly minimize distributional logarithms in the RGE boundary conditions.
We first discuss the simpler case of one-dimensional distributions, before generalizing it
to the two-dimensional case relevant for $q_T$ in \sec{distr scale setting 2d}.

\subsection{Toy example}
\label{sec:scale choice}

To illustrate the problem we consider a toy function $F(k, \mu)$, which contains
logarithms $\ln(k/\mu)$ and obeys the toy RGE
\begin{equation} \label{eq:toy RGE F}
\mu \frac{\del F(k,\mu)}{\del\mu} = -\as\, F(k, \mu)
\,.\end{equation}
(For the purposes of this toy example we neglect the running of $\as$ for simplicity.)
Solving \eq{toy RGE F} yields the formal solution
\begin{equation} \label{eq:toy solution F}
F(k,\mu) = F(k,\mu_0)\, U(\mu_0,\mu)
\,,\qquad
U(\mu_0,\mu) = \exp \Bigl(\as \ln\frac{\mu_0}{\mu} \Bigr)
\,,\end{equation}
where $U(\mu_0, \mu)$ is the evolution kernel with $U(\mu_0, \mu_0) = 1$ and $F(k, \mu_0)$
is the boundary condition. The evolution kernel resums logarithms $\ln(\mu_0/\mu)$ and
shifts the logarithms $\ln(k/\mu_0)$ of $F(k,\mu_0)$ into the logarithms $\ln(k/\mu)$ of $F(k,\mu)$.
Hence, if $F(k, \mu_0)$ is known ``exactly'' at the starting scale $\mu_0$, \eq{toy solution F}
gives $F$ at the different scale $\mu$ with all its logarithms $\ln(k/\mu)$ resummed.
More precisely, \eq{toy solution F} determines $F(k, \mu)$ to the logarithmic accuracy to which
both the boundary condition and the evolution kernel are known.

In some cases, the boundary condition is known (or assumed to be known) exactly.
For example, the nonperturbative parton distribution functions (PDFs) are extracted from the experimental data at some reference scale $\mu_0$ and are then evolved to an arbitrary scale $\mu$ using the DGLAP evolution equations.

In contrast, if the boundary condition $F(k, \mu_0)$ is calculated perturbatively,
it must be calculated including all logarithms (to the desired logarithmic accuracy).
Typically, this is achieved by choosing a particular starting scale $\mu_0$ for which
$F(k,\mu_0)$ is free of logarithms.
In the simplest case, $k$ is a scalar quantity such that $F$ is a regular function and the RGE \eq{toy RGE F} is multiplicative, as is the case for example for a hard matching coefficient in SCET.
The logarithmic structure of $F$ is then
\begin{equation}
F(k, \mu_0) = 1 + \as \ln\frac{k}{\mu_0} + \frac{1}{2} \as^2 \ln^2\frac{k}{\mu_0} + \cdots
\,,\end{equation}
where the ellipses denote higher-order terms $\sim\as^n \ln^n(k/\mu_0)$ as well as possible constant (nonlogarithmic) terms.
In this case, all logarithms vanish at $\mu_0 = k$, and $F(k,\mu_0=k)$
can be calculated in fixed-order perturbation theory,
\begin{equation}
F(k,\mu_0=k) = 1 + \cdots
\,,\end{equation}
where the ellipses now only contain possible higher-order constant terms.
Equation~\eqref{eq:toy solution F} then gives $F(k, \mu)$ at an arbitrary scale $\mu$ by
\begin{equation} \label{eq:mult_canonical}
F(k,\mu) = (1 + \cdots)\, U(\mu_0 = k, \mu)
\,,\end{equation}
where the logarithms $\ln(k/\mu)$ in $F(k, \mu)$
are now directly and \textit{fully} resummed by the evolution kernel $U(k, \mu)$.
Hence, the crucial ingredient that allows the
RGE to \textit{predict} the logarithms $\ln(k/\mu)$, as opposed to merely shifting
them from one scale to another, is the choice of $\mu_0$ that
eliminates all logarithms in the boundary condition $F(k,\mu_0)$.

In the more complicated cases we are interested in, $F(k, \mu)$ is a differential distribution
involving plus and delta distributions in the observable $k$. Let us first consider a toy
example where after transforming to an appropriate conjugate space, e.g.\ Fourier or Mellin space,
the RG \eq{toy RGE F} has an analogous form
\begin{equation} \label{eq:toy conjugate RGE}
\mu \frac{\del \tilde F(y,\mu)}{\del\mu} = -\as\, \tilde F(y, \mu)
\,,\qquad
\tilde F(y, \mu) = \tilde F(y, \mu_0)\, \tilde U(\mu_0, \mu)
\,.\end{equation}
Here, $y$ is the conjugate variable to $k$ and $\tilde F(y, \mu)$ is the corresponding
transformed function in conjugate space. We stress that as long as $\mu_0$ is kept symbolic,
the formal RGE solution is equivalent in any space,
i.e.\ it does not actually matter in which space the evolution kernel is determined,
it might just be easier to find a concrete solution in a particular space.
The differences arise entirely in how the boundary condition is chosen.
The above discussion for the multiplicative example now applies to $\tilde F(y, \mu)$,
which has the simple logarithmic structure
\begin{equation}
\tilde F(y, \mu_0) = 1 - \as \ln(y \mu_0) + \frac{1}{2} \as^2 \ln^2(y\, \mu_0) + \cdots
\,.\end{equation}
Choosing $\mu_0 = 1/y$, the conjugate boundary condition $\tilde F(y, \mu_0 = 1/y)$
is free of logarithms and can be calculated in fixed-order perturbation theory,
while all logarithms $\ln(y\mu)$ in the conjugate function
$\tilde F(y, \mu)$ are correctly predicted by $\tilde U(\mu_0 = 1/y, \mu)$.
This procedure of solving the RGE with setting the starting scale (or more generally
determining the boundary condition) intrinsically in conjugate space
fundamentally resums the conjugate logarithms and therefore we will refer
to it as \emph{evolution or resummation in conjugate space}.

For the reasons discussed before, we are interested in directly resuming the
(distributional) momentum-space logarithms of $k$, which we will refer to as
\emph{evolution/resummation in momentum (or distribution) space}.
The first attempt would be to use the canonical scale choice from momentum space, $\mu_0 = k$,
in the formal solution in conjugate space, \eq{toy conjugate RGE}. Doing so, the required boundary condition is
\begin{equation} \label{eq:toy wrong boundary}
\tilde F(y, \mu_0 = k) = 1 - \as \ln(yk) + \frac{1}{2} \as^2 \ln^2(yk) + \cdots
\,.\end{equation}
However, using fixed-order perturbation theory amounts
to truncating this series at some low order in $\as$ and neglecting all logarithmic terms
at higher orders in $\as$. This would only be sufficient if it were guaranteed
that the neglected logarithms in the boundary condition do not affect the desired logarithmic
accuracy in the final momentum-space distribution, which is far from obvious. Although naively, one
might think that the $y$-integration in the inverse transform should be dominated by $y\sim 1/k$, where the
neglected logarithms are small, the integration also includes the regions $y\to 0$ and $y\to \infty$, where the
neglected logarithms get arbitrarily large. As we will see, this is the reason
why this naive attempt in fact fails in case of $q_T$ (leading to the aforementioned spurious
divergences).

Instead, performing the evolution in momentum space requires to set the scales
\textit{and} determine the appropriate fixed-order boundary condition in momentum space.
Before discussing the analogous exponential toy example in distribution space, we can
illustrate the arising difficulties with the simple distribution
\begin{equation} \label{eq:D0example}
D_0(k, \mu)
= \delta(k) + \as \cL_0(k, \mu)
\,,\end{equation}
where
\begin{equation}
\cL_0(k, \mu)
\equiv \frac{1}{\mu} \cL_0\Bigl(\frac{k}{\mu}\Bigr) \equiv \biggl[ \frac{\theta(k)}{k} \biggr]_+^\mu
\,,\end{equation}
and $\cL_0(x) = [\theta(x)/x]_+$ is the standard plus distribution (see appendix \ref{app:plus distributions 1d} for more details). Since $\cL_0(k, \mu)$ encodes a logarithmic divergence, it counts as a single logarithm in logarithmic accuracy counting. The distribution $D_0(k, \mu)$ fulfills the differential equation
\begin{equation}
\mu \frac{\del D_0(k,\mu)}{\del\mu} = - \as \delta(k)
\,,\end{equation}
which has the general solution
\begin{equation}  \label{eq:D0solution}
D_0(k,\mu) = D_0(k,\mu_0) + \as \delta(k) \ln\frac{\mu_0}{\mu}
\,.\end{equation}
Using the full result for $D_0(k, \mu_0)$ from \eq{D0example}, together
with the distributional identity [see \eq{plus func rescaling 1d}]
\begin{equation} \label{eq:L0shift}
\delta(k) \ln\frac{\mu_0}{\mu}
= \cL_0(k, \mu) - \cL_0(k, \mu_0)
\equiv \biggl[ \frac{\theta(k)}{k} \biggr]_+^\mu - \biggl[ \frac{\theta(k)}{k} \biggr]_+^{\mu_0}
\,,\end{equation}
\eq{D0solution} gives
\begin{align} \label{eq:example minimization}
D_0(k,\mu)
&= \delta(k) + \as \cL_0(k, \mu_0) + \as \Bigl[ \cL_0(k, \mu) - \cL_0(k, \mu_0) \Bigr]
\nn \\
&= \delta(k) + \as \cL_0(k, \mu)
\,.\end{align}
Thus, the $\mu_0$-dependent terms cancel and we reproduce the correct result for $D_0(k, \mu)$.
This shows that the formal solution in \eq{D0solution} is sufficient to shift the $\mu$-dependence from $\mu_0$ to $\mu$.
However, obtaining the correct result for $D_0(k, \mu)$ from it like this crucially relies on the fact that we used the full result for the boundary condition $D_0(k, \mu_0)$.

In reality, calculating the boundary condition in fixed-order perturbation theory would amount
to truncating $D_0(k, \mu_0) = \delta(k)$ and neglecting the higher-order $\as \cL_0(k, \mu_0)$ term.%
\footnote{
This is one of the reasons for utilizing N$^n$LL$'$ counting in the resummation of distributions~\cite{Ligeti:2008ac, Abbate:2010xh, Berger:2010xi, Almeida:2014uva}. There, the boundary conditions are included to one higher fixed order, which remedies this issue
because the $\PlusLog{0}$-terms are now explicitly included in the boundary condition.}
Naively setting $\mu_0 = k$ to eliminate this term is clearly ill defined.
Since $\mu_0$ appears in the boundary condition of the distribution,
this would lead to an ill-defined expression $\delta(k)\ln k$ [as can been seen e.g.\ from \eq{L0shift}].
Instead, we need a distributional way of choosing ``$\mu_0 = k$'' that behaves analogously to the multiplicative case above. Namely, it should set the logarithmic distribution $\cL_0(k, \mu_0)$ in the boundary condition to zero so fixed-order perturbation theory can be used for the boundary condition without having to neglect any higher-order logarithmic terms, and such that the RGE directly predicts the correct $\cL_0(k, \mu)$.
One of the key results of this paper is a generic method to do so, which works for arbitrary distributions
and will be discussed next.

\subsection{Distributional scale setting}
\label{sec:cumulant prescription}

A general plus distributions of a function $g(k)$ is defined through the conditions (see appendix \ref{app:plus distributions 1d} for details)
\begin{align}
\Bigl[\theta(k)\, g(k, \mu) \Bigr]^\mu_+ &= \theta(k)\, g(k, \mu)
\qquad \text{for } k \neq 0
\,, \\
\int^\mu \del k\, \Bigl[\theta(k)\, g(k, \mu)\Bigr]^\mu_+ &= 0
\,.\end{align}
Here and in the following we always keep the lower integration limit implicit in the support of the distribution.
An important example are the logarithmic distributions
\begin{equation} \label{eq:PlusLog1d_def}
\cL_n(x) = \biggl[\frac{\theta(x) \ln^n\! x}{x}\biggr]_+
\,,\qquad
\cL_n(k, \mu) \equiv \frac{1}{\mu} \cL_n\Bigl(\frac{k}{\mu}\Bigr)
\equiv \biggl[\frac{\theta(k)}{k} \ln^n\!\frac{k}{\mu}\biggr]_+^\mu
\,.\end{equation}

For a general distribution $D(k, \mu)$ we define the distributional scale setting $\mu = k\dis$ as
\begin{equation}  \label{eq:distr scale setting 1d}
\boxed{
D(k, \mu = k\dis) \equiv \frac{\del}{\del k} \biggl[ \int^k\! \del k'\, D(k', \mu = k) \biggr]
\,.}
\end{equation}
Here, the derivative acts on everything inside the square brackets, and $\mu = k$ is set normally in the integrand, which is well defined since $D$ is evaluated at the integration variable $k'$.

The idea behind \eq{distr scale setting 1d} is that it turns logarithmic distributions into ordinary logarithms, which can then be eliminated by a normal scale choice. For example,
\begin{equation}
\cL_0(k,\mu = k\dis)
= \frac{\del}{\del k} \int^k\! \del k'\, \biggl[ \frac{\theta(k')}{k'} \biggr]_+^{\mu = k}
= \frac{\del}{\del k} \biggl[ \theta(k)\, \ln\frac{k}{\mu} \bigg\vert_{\mu = k} \biggr]
= \frac{\del}{\del k}\, 0
= 0
\,.\end{equation}
Similarly for the higher logarithmic distributions,
\begin{align}
\cL_n(k, \mu = k\dis)
&= \frac{\del}{\del k} \int^k\! \del k'\, \biggl[ \frac{\theta(k')}{k'} \ln^n\!\frac{k'}{\mu} \biggr]_+^{\mu} \bigg\vert_{\mu = k}
\nn \\
&= \frac{\del}{\del k} \biggl[ \frac{\theta(k)}{n+1} \ln^{n+1}\frac{k}{\mu} \bigg\vert_{\mu = k} \biggr]
= \frac{\del}{\del k}\, 0
= 0
\,.\end{align}
Note that although $\cL_n(k, \mu) = [\ln^n(k/\mu)/k]_+^\mu$ contains a pure logarithm $\ln(k/\mu)$, the appearance of $\mu$ in the boundary formally prohibits to simply set ``$\mu = k$'' to extract the boundary term.%
\footnote{Although this naive choice is mathematically ill-defined, it will actually yield the same result as the proper scale setting. It explicitly fails for $\PlusLog{0}$, which does not contain an explicit logarithm.}
It is also easy to see that this generalizes to any distribution $[\theta(k) g(k)]_+^\mu$,
\begin{align}
\Bigl[ g(k,\mu) \Bigr]_+^\mu \Big|_{\mu = k\dis}
&= \frac{\del}{\del k} \int^k\! \del k'\, \Bigl[\theta(k') g(k', k) \biggr]_+^{k}
= \frac{\del}{\del k}\, 0
= 0
\,.\end{align}
The cumulant integral exactly vanishes by definition of the distribution.

On the other hand, any $\mu$-independent constant terms (i.e.\ pure boundary terms) are unaffected since the operator $\frac{\del}{\del k}$ precisely inverts the cumulant $\int^k \del k'$. In particular,
\begin{equation}
\delta(k) \Big\vert_{\mu = k\dis}
= \frac{\del}{\del k} \int^k\! \del k'\, \delta(k')
= \frac{\del}{\del k} \theta(k)
= \delta(k)
\,.\end{equation}
Here it is crucial to keep track of any $\theta(k)$ appearing in the cumulant to properly recover the distributional structure.
For future convenience we use the notation
\begin{equation}
\boxed{
D[k] \equiv D(k,\mu) \Big|_{\mu = k\dis}
}\end{equation}
to denote the pure $\mu$-independent constant term, which will typically be $D[k] \sim \delta(k)$, but in general could also contain regular (integrable) functions of $k$.

Having a well-defined method for distributional scale setting allows us to easily solve distributional differential equations, as it allows us to essentialy treat distributions like ordinary logarithms.
To see that \eq{distr scale setting 1d} has the desired properties in more nontrivial cases, consider the simple examples
\begin{align} \label{eq:deq_prop1}
\delta(k)\, \ln^{n+1}\frac{\mu_0}{\mu} \bigg\vert_{\mu_0 = k\dis}
&= (n+1) \cL_n(k, \mu)
\qquad (n \geq 0)
\,, \nn \\
\qquad
(m+1) \cL_m(k, \mu)\, \ln^{n}\frac{\mu_0}{\mu} \bigg\vert_{\mu_0 = k\dis}
&= (n+m+1) \cL_{m+n}(k, \mu)
\qquad (n \geq 0)
\,, \nn \\
\qquad
\cL_m(k, \mu_0 )\, \ln^{n}\frac{\mu_0}{\mu} \bigg\vert_{\mu_0 = k\dis}
&= 0
\qquad (n \geq 0)
\,.\end{align}
These relations are quite intuitive in that the distributional scale setting essentially
moves any naively appearing $\ln k$ terms inside a suitably regulated logarithmic distribution.
The appearing prefactors count the order of the distributional logarithm in each equation.

More generally, we have
\begin{align} \label{eq:deq_prop2}
\qquad
\delta(k)\, g(\mu_0, \mu) \Big\vert_{\mu_0 = k\dis}
&= \delta(k)\, g(\mu, \mu) + \biggl[\theta(k) \frac{\del g(k, \mu)}{\del k} \biggr]_+^\mu
\,, \nn \\
\qquad
\Bigl[\theta(k)\, f(k, \mu)\Bigr]_+^\mu\, g(\mu_0, \mu) \Big\vert_{\mu_0 = k\dis}
&= \biggl[\theta(k) \frac{\del}{\del k} \biggl(g(k, \mu) \int_\mu^k \del k' f(k', \mu)\biggr) \biggr]_+^\mu
\,, \nn \\
\qquad
\Bigl[\theta(k)\, f(k, \mu_0)\Bigr]_+^{\mu_0} \, g(\mu_0, \mu) \Big\vert_{\mu_0 = k\dis}
&= 0
\,.\end{align}

\subsection{Integrating distributional differential equations}
\label{sec:rge solution 2}

First, consider again the simple example in \eq{D0example}. Setting $\mu_0 = k\dis$ now has the desired effect of being able to predict the complete logarithmic distribution from the general solution \eq{D0solution} by a scale choice,
\begin{equation}
D_0(k,\mu) = D_0(k,\mu_0) + \as \delta(k) \ln\frac{\mu_0}{\mu} \bigg\vert_{\mu_0 = k\dis}
= \bigl[\delta(k) + 0\bigr] + \as \cL_0(k, \mu)
\,.\end{equation}

Similarly, we can reproduce a $\cL_n(k, \mu)$ for $n \geq 1$ from its $\mu$ dependence. Consider a distribution $D_n(k, \mu)$, which satisfies the differential equation
\begin{equation}
\mu \frac{\del D_n(k, \mu)}{\del \mu}
= -n\,\as \cL_{n-1}(k, \mu)
= -n\,\as \biggl[\frac{\theta(k)}{k} \ln^{n-1}\frac{k}{\mu}\biggr]^\mu
\,.\end{equation}
Integrating this from $\mu_0$ to $\mu$, we get the general solution
\begin{align}
D_n(k, \mu)
&= D_n(k, \mu_0 ) - n\,\as \int_{\mu_0}^\mu\! \frac{\del\mu'}{\mu'} \biggl[ \frac{\theta(k)}{k} \ln^{n-1} \frac{k}{\mu'} \biggr]_+^{\mu'}
\nn \\
&= D_n(k, \mu_0 ) + \as\biggl[ \frac{\theta(k)}{k} \Bigl(\ln^n\!\frac{k}{\mu} - \ln^n\!\frac{k}{\mu_0}\Bigr) \biggr]_+^{\mu_0} + \frac{\delta(k)}{n+1}\,\as\ln^{n+1}\frac{\mu}{\mu_0}
\,.\end{align}
Here we used \eq{plus func rescaling 1d} to shift the boundary condition in the plus distribution from $\mu'$ to $\mu_0$, such that the integral can be pulled inside the plus distribution and performed.
(This step explicitly requires that $\mu_0$ does not depend on $k$, as otherwise the boundary condition of the plus distribution would be changed.)
Using \eq{deq_prop1} to set $\mu_0 = k\dis$, we thus find
\begin{equation}
D_n(k, \mu) = D_n[k] + \as\cL_{n}(k, \mu)
\,.\end{equation}

In practice, we can also specialize \eq{distr scale setting 1d} to define an integral over an arbitrary distribution $G$ with starting scale $\mu_0 = k\dis$ as
\begin{equation}  \label{eq:integral scale setting 1d}
\boxed{
\int_{k\dis}^\mu\! \frac{\del\mu'}{\mu'}\, G(k,\mu')
\equiv \frac{\del}{\del k} \int^k\! \del k'\, \int_k^\mu\! \frac{\del\mu'}{\mu'}\, G(k',\mu')
\,.}\end{equation}
This allows us to write a generic integral solution as
\begin{equation}
\mu\frac{\del D(k, \mu)}{\del\mu} = G(k, \mu)
\qquad\Rightarrow\qquad
D(k,\mu) = D[k] + \int_{k\dis}^\mu\! \frac{\del\mu'}{\mu'}\, G(k, \mu')
\,.\end{equation}
Although this looks like an ordinary integral solution, the important difference lies in the lower integration limit $k\dis$ which enforces the distributional scale setting.

\subsection{Toy example in distribution space}

We can now discuss the exponential toy example in distribution space. Consider a distribution with the logarithmic structure
\begin{equation} \label{eq:Ftoy}
F(k, \mu) = \delta(k) + \as \PlusLog{0}(k,\mu) + \as^2 \PlusLog{1}(k,\mu) + \frac{1}{2} \as^3 \PlusLog{2}(k,\mu) + \cdots
\,,\end{equation}
corresponding to an exponential in distribution space.
It satisfies the differential equation
\begin{equation} \label{eq:FtoyRGE}
\mu \frac{\del F(k,\mu)}{\del\mu} = - \as\, F(k,\mu)
\,,\end{equation}
whose general solution is easily seen to be
\begin{equation} \label{eq:FtoyRGEsol}
F(k,\mu) = F(k,\mu_0)\, U(\mu_0, \mu)
\,,\qquad
U(\mu_0, \mu) = \exp\Bigl(\as \ln\frac{\mu_0}{\mu} \Bigr)
\,.\end{equation}
Using the definition in \eq{distr scale setting 1d} to set $\mu_0 = k\dis$ we obtain
\begin{align} \label{eq:distr_canonical_explicit}
F(k,\mu)
&= F(k,\mu_0) \exp\biggl(\as \ln\frac{\mu_0}{\mu} \biggr) \bigg|_{\mu_0 = k\dis}
= \frac{\del}{\del k} \biggl[
\int^k\! \del k'\, F(k', k) \exp\biggl(\as \ln\frac{k}{\mu} \biggr) \biggr]
\nn \\
&= \frac{\del}{\del k} \biggl[
(1 + \dotsb) \theta(k) \exp\biggl(\as \ln\frac{k}{\mu} \biggr) \biggr]
\nn \\
&= \delta(k) \exp\Bigl(\as \ln\frac{\xi}{\mu} \biggr)
+ \biggl[ \theta(k) \frac{\del}{\del k} \exp\biggl(\as \ln\frac{k}{\mu} \biggr) \biggr]_+^\xi
\nn \\
&= \delta(k) + \as \biggl[\frac{\theta(k)}{k} \exp\biggl(\as \ln\frac{k}{\mu} \biggr) \biggr]_+^\mu
\,.\end{align}
The $k'$-integral in the first step only acts on $F(k')$ and eliminates all distributions leaving only constant boundary terms $1 + \dotsb$. In the second step we used \eq{thetaderivative} to take the derivative involving the $\theta(k)$, where $\xi$ is arbitrary and cancels between the two terms. In the last step we chose $\xi = \mu$. Expanding the exponential in $\as$, we can easily see that this reproduces the full distributional logarithmic structure of \eq{Ftoy}
\begin{align}
F(k,\mu)
&= \delta(k) + \as \biggl[ \frac{\theta(k)}{k} \Bigl( 1 + \as \ln\frac{k}{\mu} + \frac{1}{2} \as^2 \ln^2\frac{k}{\mu} + \cdots \Bigr) \biggr]_+^\mu
\nn \\
&= \delta(k) + \as \PlusLog{0}(k,\mu) + \as^2 \PlusLog{1}(k,\mu) + \frac{1}{2} \as^3 \PlusLog{2}(k,\mu) + \cdots
\,.\end{align}

With the general properties in \eqs{deq_prop1}{deq_prop2}, we can also simply use the fact that setting $\mu_0 = k\dis$ eliminates any contributions from distributional logarithms in the boundary condition $F(k, \mu_0)$. We can therefore directly plug in $F(k, \mu_0= k\dis) = F[k] = (1 + \dotsb)\delta(k)$ for the boundary condition to get
\begin{align} \label{eq:distr_canonical}
F(k,\mu)
&= F[k]\, U(\mu_0, \mu) \Big|_{\mu_0 = k\dis}
= (1 + \dotsb) \delta(k)\, U(\mu_0, \mu) \Big|_{\mu_0 = k\dis}
\,.\end{align}
Using \eq{deq_prop2} we can see that this is identical to what we got in \eq{distr_canonical_explicit},
\begin{align}
F(k,\mu) &= \delta(k)\, U(\mu_0, \mu) \Big|_{\mu_0 = k\dis}
\nn\\
&= \delta(k) + \Bigl[\theta(k) \frac{\del}{\del k} U(k, \mu) \Bigr]_+^\mu
= \delta(k) + \as \biggl[\frac{\theta(k)}{k} \exp\biggl(\as \ln\frac{k}{\mu} \biggr) \biggr]_+^\mu
\,.\end{align}

\paragraph{Alternative derivation}

Another way to derive the same solution without distributional scale setting is to start from the general Ansatz
\begin{equation}
F(k,\mu) = f_0(\mu)\, \delta(k) + \Bigl[\theta(k)\, f_1(k,\mu)\Bigr]_+^\mu
\,.\end{equation}
Plugging this back into \eq{FtoyRGE} and using \eq{plus func derivative 1d 2} to take the $\mu$ derivative,
\begin{align}
\mu \frac{\del F(k,\mu)}{\del\mu}
&= \delta(k) \biggl( \mu \frac{\del f_0(\mu)}{\del\mu} - \mu f_1(\mu,\mu) \biggr)
+ \biggl[\theta(k)\,  \mu \frac{\del f_1(k,\mu)}{\del\mu}\biggr]_+^\mu
\nn \\
&\stackrel{!}{=} -\as f_0(\mu)\, \delta(k) - \as \Bigl[\theta(k)\, f_1(k,\mu)\Bigr]_+^\mu
\,,\end{align}
yields a coupled system of differential equations for $f_0$ and $f_1$,
\begin{align}
\mu \frac{\del f_1(k,\mu)}{\del\mu} &= -\as f_1(k,\mu)
\,, \nn \\
\mu \frac{\del f_0(\mu)}{\del\mu} &= - \as f_0(\mu) + \mu f_1(\mu,\mu)
\,.\end{align}
The advantage is that these are now two ordinary RGEs, which can be solved straightforwardly without having to be careful about producing distributions from taking derivatives of $\theta(k)$. Solving for $f_1$ in terms of $f_0$,
\begin{align}
f_1(k,\mu)
&= f_1(k,k) \exp\biggl(\as \ln\frac{k}{\mu}\biggr)
= \biggl[ \frac{\del f_0(k)}{\del k} + f_0(k) \frac{\as}{k} \biggr] \exp\biggl(\as \ln\frac{k}{\mu}\biggr)
\nn \\
&= \frac{\del}{\del k} f_0(k) \exp\biggl(\as \ln\frac{k}{\mu}\biggr)
\,.\end{align}
Plugging back into the Ansatz, we obtain the result
\begin{equation}
 F(k,\mu) = f_0(\mu)\, \delta(k) + \biggl[\theta(k)\, \frac{\del}{\del k} f_0(k) \exp\biggl(\as \ln\frac{k}{\mu}\biggr) \biggr]_+^\mu \,,
\end{equation}
which for $f_0 = 1 + \cdots$ confirms the earlier result \eq{distr_canonical_explicit}.
This form suggests to interpret $f_0(\mu)$ as the boundary term, as it completely predicts the (logarithmic) structure of $F$.
Indeed, setting now $\mu = k\dis$ yields $F[k] = f_0(\xi)\, \delta(k) + [\theta(k) f_0'(k)]_+^\xi$,
confirming that the boundary term can be obtained as the coefficient $f_0(\mu)$ of $\delta(k)$.
Note that the essential element in this derivation is the same as above, namely that with the distributional canonical scale setting $F(k, \mu = k\dis) = F[k]$ reduces to a pure boundary term that is free of logarithmic distributions.
In actual applications, it would be calculated in a fixed-order expansion in $\as(\mu)$.

This simple toy example illustrates that with the distributional scale setting, using the canonical scale choice $\mu_0= k\dis$ as in \eq{distr_canonical} becomes exactly analogous to using the ordinary canonical scale choice in the multiplicative case in \eq{mult_canonical}. In particular, the pure boundary condition $F[k]$ can now be calculated in fixed-order perturbation theory, while the RGE solution predicts the complete distributional logarithmic structure.
It should be evident that this is true generically also for more complicated cases. We will encounter two more complicated examples when discussing the RGE for the rapidity anomalous dimension in \sec{gNu resummation} and for the soft function in \sec{soft func}.

\subsection{Comparison to evolution in conjugate space}
\label{sec:comparison rge tools}

It is interesting to compare the result of the momentum-space evolution
to the evolution in appropriate conjugate spaces,
i.e.~solving the RGEs in conjugate space with scale setting therein.
In general, using equivalent boundary conditions in different spaces can lead to different predictions in
physical space.
To illustrate this, we consider again the simple example distributions $D_n$, satisfying
\begin{equation} \label{eq:DnRGE}
\mu \frac{\del D_n(k,\mu)}{\del\mu} = - \as\, n \cL_{n-1}(k,\mu)
\qquad (n \geq 1)
\,,\end{equation}
and assume that only the LO boundary term is known from a fixed-order calculation,
\begin{equation} \label{eq:DnBC}
 D_n(k,\mu) = \delta(k) + \cdots
\,,\end{equation}
where the ellipses denote the unknown higher order terms.
Since all unknown logarithmic higher order terms vanish for $\mu = k\dis$,
the required LO boundary term to solve \eq{DnRGE} distributionally is $D[k] = \delta(k)$.
The solution then is (see \sec{rge solution 2})
\begin{equation} \label{eq:Dnexample}
D_n(k, \mu) = \delta(k) + \as \cL_{n}(k, \mu)
\,.\end{equation}
In the following we compare this momentum-space result to the resummation in cumulant and Fourier space.

\subsubsection{Cumulant space}

Taking the cumulant, \eqs{DnRGE}{DnBC} become
\begin{align} \label{eq:DnRGE cumulant}
 \mu \frac{\del \bar D_n(\kcut, \mu)}{\del \mu} &= -\as\, \theta(\kcut)\ln^{n}\frac{\kcut}{\mu}
\,, \\ \label{eq:DnBC cumulant}
 \bar D_n(\kcut,\mu) &= \theta(\kcut) + \cdots
\,,\end{align}
where the cumulant is defined as
\begin{equation}
 \bar D_n(\kcut,\mu) = \int^{\kcut}\! \del k'\, D_n(k', \mu)
\,.\end{equation}
The solution to \eq{DnRGE cumulant} is easily obtained as
\begin{align}
\bar D_n(\kcut, \mu)
= \bar D_n(\kcut, \mu_0) - \as\, \theta(\kcut) \int_{\mu_0}^\mu \frac{\del\mu'}{\mu'} \ln^n\!\frac{\kcut}{\mu'}
\,.\end{align}
The important point is that all distributions in momentum space correspond to logarithms $\ln(\kcut/\mu)$ in cumulant space, which can be fully resummed by choosing $\mu_0 = \kcut$.
With this choice, all logarithms in $\bar D_n(\kcut,\mu)$ are eliminated,
and from \eq{DnBC cumulant} it follows that the LO boundary condition is $\bar D_n(\kcut,\kcut) = \theta(\kcut)$.
The solution is thus given by
\begin{align}
\bar D_n(\kcut, \mu)
&= \theta(\kcut) + \frac{\as}{n+1} \theta(\kcut) \ln^{n+1}\frac{\kcut}{\mu}
\,.\end{align}
Transforming back to momentum space by taking the derivative with respect to $\kcut$ yields
\begin{equation}
D_n(k, \mu)
= \delta(k) + \as \biggl[ \frac{\theta(k)}{k} \ln^n\!\frac{k}{\mu} \biggr]_+^\mu
= \delta(k) + \as \cL_n(k,\mu)
\,,\end{equation}
which exactly reproduces the momentum-space solution in \eq{Dnexample}.
Hence, resummation in cumulant and momentum space to predict the logarithmic structure is equivalent for this example.
This is of course not very surprising, considering the intimate relation between
distribution and cumulant space.

\subsubsection{Fourier space}

Taking the Fourier transform of \eqs{DnRGE}{DnBC} using \eq{Ln to y}, we find
\begin{align} \label{eq:DnRGE Fourier}
\mu \frac{\del \tilde D_n(y, \mu)}{\del \mu}
&= -\as \sum_{k=0}^{n} (-1)^k \binom{n}{k} \ln^k\bigl(i y \mu e^{\gamma_E}\bigr) \dR[1]{n-k}
\,,\\ \label{eq:DnBC Fourier}
\tilde D_n(y,\mu) &= 1 + \cdots
\,,\end{align}
where the Fourier transform is defined as
\begin{equation}
\tilde D_n(y, \mu) = \int\!\del k\, e^{-i k y} D_n(k, \mu)
\,,\end{equation}
and the constant $\dR[1]{n}$ is defined as (see appendix \ref{app:Fourier trafo 1d} for more details)
\begin{equation}
\dR[1]{n} = \frac{\del^n}{\del a^n}\, e^{\gamma_E a} \Gamma(1+a) \bigg\rvert_{a=0}
\,.\end{equation}
The general solution is given by
\begin{equation}
\tilde D_n(y, \mu) = \tilde D_n(y, \mu_0) - \as \int_{\mu_0}^\mu \frac{\del\mu'}{\mu'} \sum_{k=0}^{n} (-1)^k \binom{n}{k} \ln^k\bigl(i y \mu' e^{\gamma_E}\bigr) \dR[1]{n-k}
\,.\end{equation}
Since all logarithms depend on $i y \mu e^{\gamma_E}$, they are fully resummed by choosing $\mu_0 = -i e^{-\gamma_E} / y$,%
\footnote{%
In principle, one is of course free to shift constants from the logarithms into subleading terms.
This convention for the canonical logarithm in conjugate space is the most natural one, since
at least at lowest order $n=0$ the pure plus distribution $\PlusLog{0}(k,\mu)$ corresponds to a pure logarithm $\ln(i y \mu e^{\gamma_E})$, see table \ref{tbl:Ln to y space} in appendix \ref{app:plus distributions 1d}.}
which eliminates all logarithms in the boundary term.
This allows us to use the known fixed-order boundary term \eq{DnBC Fourier}, $\tilde D_n(y, \mu_0) = 1$, and gives
\begin{equation}
\tilde D_n(y, \mu) = 1 - \as \frac{\dR[1]{n+1}}{n+1}  + \as \FT[\PlusLog{n}](y,\mu)
\,.\end{equation}
Transforming back to momentum space thus yields
\begin{equation} \label{eq:Dnsolution FT}
D_n(k, \mu) = \delta(k) + \as \biggl[\cL_n(k,\mu) - \frac{\dR[1]{n+1}}{n+1} \delta(k) \biggr]
\,.\end{equation}
While the plus distribution matches the one in \eq{Dnexample}, which is of course necessary to provide the correct $\mu$-dependence, the solution contains an additional term $-\delta(k)\dR[1]{n+1}/(n+1)$ at $\cO(\as)$.

This shows explicitly that the Fourier-space resummation, which fundamentally resums logarithms of the conjugate variable $y$, is not equivalent to the momentum-space resummation, as it induces an additional boundary term.
This discrepancy is due to the fact that pure plus distributions $\PlusLog{n}(k,\mu)$ do not correspond to pure powers of $\ln( i y \mu e^{\gamma_E})$ and vice versa.

In principle, the additional $\as \delta(k)$ term in \eq{Dnsolution FT} would be compensated for
if we were to include the boundary condition $\tilde D_n(y, \mu_0)$ in Fourier space to $\cO(\as)$.
However, for a general distribution, such spurious terms are generated to all orders in $\as$. (We will see this explicitly in \sec{gNu resummation} for the example of the rapidity anomalous dimension.)
Thus in practice for the purposes of resummation, where the boundary condition is only
calculated to some fixed order, the resummation in Fourier space intrinsically
induces additional subleading terms to all orders in $\as$, which is one the main motivations
for performing the resummation directly in momentum space.

\subsection{Implementation of scale variations and profile scales}
\label{sec:profiles}

For phenomenological applications it is necessary to also use noncanonical scale choices.
First, varying the scales away from their strict canonical values is a standard and convenient
way to probe the size of higher-order logarithms and thereby estimate perturbative uncertainties.
Furthermore, it is important to be able to dynamically turn off the resummation in the fixed-order region of the distribution. A standard way to achieve this is to employ profile scales $\mu_0(k)$~\cite{Ligeti:2008ac, Abbate:2010xh}, which smoothly interpolate as a function of $k$ between the strict canonical scale choice in the resummation region and a common fixed scale $\mu_{\rm FO}$ in the fixed-order region. Profile scales are also used to implement dynamical
scale variations to distinguish different sources of perturbative uncertainties,
for example resummation and fixed-order uncertainties~\cite{Berger:2010xi, Stewart:2011cf, Stewart:2013faa}.

We discuss the implementation of scale variations and more generally profile scales
within our framework using the exponential toy example in \eq{Ftoy} with the
associated RGE solution in \eq{FtoyRGEsol}:
\begin{equation}
 F(k,\mu) = F^\FO(k,\mu_0)  U(\mu_0,\mu)
\,,\qquad
U(\mu_0, \mu) = \exp\Bigl(\as \ln\frac{\mu_0}{\mu} \Bigr)
\,.\end{equation}
Here we have included the superscript ``FO'' to stress that the boundary condition is obtained from a fixed-order calculation, whereas $U$ is the evolution kernel.
In a typical application, one wants to perform the resummation using the canonical scale choice $\mu_0 = k\dis$ for $k\ll Q$, where the logarithms of $k$ are large and should be resummed. On the other hand, for $k \sim Q$, one wants to
turn off the resummation and recover the exact fixed-order result by taking $\mu_0 = \mu_\FO = \mu$.
Both requirements are fulfilled by choosing $\mu_0$ to be a profile function $\mu_0(k)$ behaving as
\begin{align}
 &\mu_0(k) \rightarrow k \,,\qquad\qquad k \ll Q \\
 &\mu_0(k) \rightarrow \mu_\FO = \mu \,,\quad k \sim Q \,.
\end{align}
The profile furthermore smoothly interpolates between the two regimes to capture the turning-off of the resummation.
Such a profile can be conveniently implemented by generalizing the distributional scale setting in \eq{distr scale setting 1d}, $\mu_0 = k\dis$, to a generic function $\mu_0 = \mu_0(k)\dis$,
\begin{equation} \label{eq:FtoyProfile}
 F(k,\mu) = \frac{\del}{\del k} \biggl[ \int^k\! \del k'\, F^\FO(k',\mu_0(k))  U(\mu_0(k),\mu) \biggr]
\,.\end{equation}
In the resummation regime, this reproduces exactly the canonical distributional scale setting $\mu_0 = k\dis$.
In the fixed-order regime, $U=1$ and $\mu_0$ is independent of $k$, such that the derivative simply inverts the integral to reproduce $F(k,\mu)=F^\FO(k,\mu)$.

Furthermore, the profile should allow scale variations to probe subleading logarithms.
To see in more detail how this works, consider the NLO boundary term
\begin{equation}
\label{eq:Ftoy_FO}
 F^\FO(k,\mu_0) = \delta(k) + \as \bigl[  f_1 \delta(k) + \cL_0(k,\mu_0) \bigr]
\,.\end{equation}
The solution with arbitrary profile $\mu_0(k)$ is then
\begin{align}
 F(k,\mu)
&= \frac{\del}{\del k} \biggl\{ \theta(k) \biggl(1 + \as\biggl[ f_1 + \ln\frac{k}{\mu_0(k)}\biggr] \biggr)U(\mu_0(k),\mu) \biggr\} \nn\\
&= \delta(k) \biggl(1 + \as\biggl[ f_1 + \ln\frac{\mu}{\mu_0(\mu)}\biggr] \biggr)U(\mu_0(\mu),\mu) \nn\\&\quad
  + \as \biggl[ \frac{\theta(k)}{k}  \biggl(
    1 + \as \biggl[f_1 + \ln\frac{k}{\mu_0(k)}\biggr]  \frac{\del\ln\mu_0(k)}{\del\ln k}
  \biggr) U(\mu_0(k),\mu)
  \biggr]_+^\mu
\,.\end{align}

The canonical scale choice to predict all logarithmic terms to this order is $\mu_0(k) = k$, such that we get
\begin{align}
 F(k,\mu)
&= \bigl(1 + \as f_1 \bigr)  \bigl[\delta(k) + \as \cL_0 + \as^2 \cL_1 + \cdots \bigr]
\,,\end{align}
which is exactly the expected structure.

More general, we can vary the scale $\mu_0$ around the canonical value $k$, $\mu_0(k) = a \cdot k$, to probe subleading logarithms in the resummation regime.
With this choice, we obtain
\begin{align}
 F(k,\mu)
&= \bigl[1 + \as (f_1 - \ln a) \bigr]  \bigl[\delta(k) + \as \cL_0 + \as^2 \cL_1 + \cdots \bigr] e^{\as \ln a}
\,.\end{align}
This clearly probes subleading terms through the induced exponential.
Note that the $-\ln a$ in the first term cancels the $\cO(\as)$-term of the exponential to correctly reproduce the fixed-order content.

Lastly, we can test the fixed-order structure by varying $\mu_0$ in the fixed-order regime.
To fully turn off the evolution kernel in \eq{FtoyProfile}, we choose $\mu_0(k) = \mu = a \cdot \mu_\FO$, yielding
\begin{align}
 F(k,a \mu_\FO)
&= \delta(k) + \as \bigl[ f_1 \delta(k) + \cL_0(k, a \mu_\FO) \bigr] = F^\FO(k, a \mu_\FO)  \\
&=\delta(k) + \as \bigl[ ( f_1 - \ln a) \delta(k) + \cL_0(k,\mu_\FO)  \bigr] = F^\FO(k,\mu_\FO) - \as \ln a\,\delta(k)
\nn
\,.\end{align}
The effect of such a scale variation is to induce the term $\as \ln(a)\delta(k)$, which precisely probes the $\cO(\as)$ boundary term.
For $a=1$, this obviously restores the FO content.

For a more general scale variation, one can also vary $\mu_0(k)$ and $\mu$ independently.
We consider the example $\mu_0(k) = a \cdot \mu_\FO$, $\mu = \mu_\FO$, where we only vary $\mu_0$.
This gives
\begin{align}
 F(k,\mu_\FO)
&= \delta(k) + \as \bigl[ ( f_1 - \ln a) \delta(k) + \cL_0(k,\mu_\FO)\bigr] e^{\as \ln a}
\,.\end{align}
The additional exponential probes higher order logarithms and is a remnant of the mismatch between the resummation scale $\mu_0$ and the common scale $\mu$.

Distributional scale setting thus allows for a straightforward implementation of profile scales,
which can be used to smoothly transition from resummation to fixed-order regime
and allow one to probe the typical size of subleading terms and thereby estimate perturbative
uncertainties in both regimes.
Of course, in actual applications one has to choose suitable profile functions and variations
such that they produces reasonable uncertainty estimates.

\subsection{Distributional scale setting in 2D}
\label{sec:distr scale setting 2d}

In this section we collect the corresponding formulas for distributional scale setting and solving distributional differential equations for the two-dimensional case. The derivations are almost identical to the one-dimensional case, and are not repeated here.

We consider the example of transverse momentum dependent functions, which will be used throughout the rest of this paper.
Such functions inherently contain divergences as $1/p_T^2$ in the limit of small transverse momenta $\pt$, which are regulated through plus distributions. These are defined through the conditions
\begin{align}
\Bigl[f(\pt, \mu) \Bigr]^\mu_+ &= f(\pt, \mu)
\qquad \text{for } |\pt| > 0
\,, \\
\int_{|\pt| \le \mu}\! \del^2\pt\, \Bigl[f(\pt, \mu)\Bigr]^\mu_+ &= 0
\,,\end{align}
see appendix \ref{app:plus distributions pt} for more details. Important examples are the logarithmic distributions
\begin{equation}
\PlusLog{n}(\pt, \mu)
\equiv \frac{1}{\pi \mu^2} \biggl[ \frac{\mu^2}{\pt^2} \ln^n\!\frac{\pt^2}{\mu^2} \biggr]_+^\mu
\equiv \frac{1}{\pi \mu^2} \PlusLog{n}\biggl(\frac{\pt^2}{\mu^2}\biggr)
\,,\end{equation}
which are the two-dimensional analog of the $\cL_n(k, \mu)$ in \eq{PlusLog1d_def}.
As in the one-dimensional case, the boundary condition of the $\PlusLog{n}(\pt, \mu)$ encodes a logarithmic dependence $\ln^n(p_T/\mu)$, which can be seen by shifting their boundary condition [see \eq{plus func rescaling}],
\begin{align} \label{eq:plus func rel}
 \cL_n(\pt,\mu) = \frac{1}{\pi\mu^2} \biggl[ \frac{\mu^2}{\pt^2} \ln^n\!\frac{\pt^2}{\mu^2} \biggr]_+^{\mu_0} + \frac{\delta(\pt)}{n+1} \ln^{n+1} \frac{\mu_0^2}{\mu^2}
\end{align}

For a general two-dimensional distribution $D(\pt, \mu)$ we define the distributional scale setting $\mu = p_T\dis$ as
\begin{equation}  \label{eq:distr scale setting}
\boxed{
D(\pt, \mu = p_T\dis)
\equiv \frac{1}{2\pi p_T} \frac{\del}{\del p_T} \biggl[\int_{|\kt| \le p_T}\! \del^2\kt \, D(\kt, \mu=p_T) \biggr]
\,.
}
\end{equation}
The derivative acts on everything inside square brackets, and $\mu = p_T$ can be set normally in the integrand since the integral runs over the auxiliary vector $\kt$. Like in the one-dimensional case, \eq{distr scale setting} sets purely distributional terms to zero,
\begin{equation} \label{eq:distr scale setting practical}
\cL_n(\pt, \mu = p_T\dis) = 0
\,,\qquad
\Bigl[f(\pt, \mu)\Bigr]_+^\mu \Bigl\vert_{\mu= p_T\dis} = 0
\,,\end{equation}
since the cumulant integral exactly vanishes by definition of the plus distribution, whereas any $\mu$-independent constant (boundary) terms are not affected at all. For convenience, we denote the pure $\mu$-independent boundary terms as
\begin{equation}
D[\pt]  \equiv D(\pt,\mu) \Big|_{\mu = p_T\dis}
\,.\end{equation}
Here, $D[\pt]$ can only depend on the quantity $\pt$. The simplest case is $D[\pt] \sim \delta(\pt)$,
but in general it can also contain regular (integrable) functions of $\pt$. We will see examples of more
general boundary terms in \secs{gNu resummation}{soft func}.

The distributional scale setting can be readily applied to solve differential equations with two-dimensional distributions.
The solution to the differential equation
\begin{equation}
 \mu \frac{\del D(\pt,\mu)}{\del\mu} = G(\pt,\mu)
\end{equation}
is given by
\begin{equation}
D(\pt,\mu) = D[\pt] + \int_{p_T\dis}^\mu \frac{\del\mu'}{\mu'} G(\pt,\mu')
\,,\end{equation}
where the integral over the distribution $G(\pt,\mu)$ starting at the canonical scale $\mu = p_T\dis$ is given by
\begin{equation}  \label{eq:integral scale setting}
\int_{p_T\dis}^\mu\! \frac{\del\mu'}{\mu'}\, G(\pt,\mu')
\equiv \frac{1}{2\pi p_T} \frac{\del}{\del p_T} \int_{|\kt| \le p_T}\! \del^2\kt\, \int_{p_T}^\mu\! \frac{\del\mu'}{\mu'}\, G(\kt,\mu')
\,,\end{equation}
A list of useful integrals is given in appendix \ref{app:cumulants}.

We conclude this section by comparing the resummation in distribution space with resummation in Fourier space, analogous to \sec{comparison rge tools}. Specifying to the example
\begin{equation}
 \mu \frac{\del D_n(\pt, \mu)}{\del \mu} = -2 n \as \PlusLog{n-1}(\pt,\mu) \qquad(n \ge 1)
\,,\end{equation}
and ignoring $\as$ running, the distributional solution for the simplest boundary term $D_n[\pt] = \delta(\pt)$ is given by
\begin{equation}
 D_n(\pt,\mu)
 = \delta(\pt) + \as \cL_n(\pt,\mu)
\,.\end{equation}
The solution obtained in cumulant space is again equivalent.

In contrast, solving the RGE with scale setting in Fourier space with the boundary condition $\tilde D_n(\bt, \mu_0=b_0/b_T) = 1$ yields
\begin{equation}
 D_n(\pt, \mu) = \delta(\pt) + \as \biggl[ \cL_n(\pt,\mu) - \frac{\dR{n+1}}{n+1} \biggr]
\,,\end{equation}
where the constant is given by \eq{R(n)},
\begin{equation}
\dR{n} = \frac{\del^n}{\del a^n} e^{2\gamma_E a} \frac{\Gamma(1+a)}{\Gamma(1-a)} \bigg\rvert_{a=0}
\,.\end{equation}

As for the one-dimensional case, we find that performing the resummation in Fourier space adds an additional boundary term compared to the momentum-space resummation.
In this specific example, it could be compensated for by including the boundary condition $\tilde D_n(\bt, \mu)$ in Fourier space to $\cO(\as)$.
However, for a general distribution, such spurious term are generated to all orders in $\as$.
Thus in practice, the resummation in Fourier space induces additional subleading term to all orders in $\as$ as well.

\section{Overview and complications in \texorpdfstring{$q_T$}{qT} resummation}
\label{sec:TMD}

Many of the difficulties in the resummation of the $q_T$ spectrum in momentum space
are due to the intrinsic two-dimensional nature of $\qt$ and the involved convolutions,
and are absent for one-dimensional variables like transverse energy or thrust.
In this section, we explore in detail the underlying reason for this.
After briefly reviewing $q_T$ factorization and the relevant associated RG equations
in \sec{TMD factorization}, we argue in \sec{TMD convolutions} that the appearing
two-dimensional convolutions requires very careful scale setting,
which turns out to be the crucial complication of $q_T$ resummation in momentum space.
In particular, we will see that setting the boundary scales to the overall $q_T$ does not
correctly resum all logarithms, as one might naively expect.
This will be illustrated in \secs{naive illustration}{TMD illustration} by reproducing a spurious divergence,
which is well-known in the literature and as we will show originates from this incorrect scale setting,
or more generally from an incorrect treatment of the boundary condition.
In the following \secs{gNu resummation}{soft func} we then show how the $q_T$ resummation via
RG evolution in momentum space can be carried out using the distributional scale setting
of \sec{distr scale setting}.

\subsection{Review of \texorpdfstring{$q_T$}{qT} factorization}
\label{sec:TMD factorization}

We briefly review the factorization theorem for the transverse-momentum distribution and set up our notation.
We follow the rapidity renormalization formalism of refs.~\cite{Chiu:2011qc,Chiu:2012ir} using soft-collinear effective theory (SCET) \cite{Bauer:2000ew, Bauer:2000yr, Bauer:2001ct, Bauer:2001yt}. There are several other formalisms known
in the literature, in particular the original formulation by Collins, Soper, and Sterman (CSS)~\cite{Collins:1981uk, Collins:1981va, Collins:1984kg} and its applications and extensions~\cite{Collins:1350496, Balazs:1997xd, Balazs:2000wv, Balazs:2006cc, Wang:2012xs, Catani:2000vq, deFlorian:2000pr, deFlorian:2001zd, Catani:2010pd, Catani:2013tia, Bozzi:2010xn, Catani:2015vma, Bozzi:2007pn, deFlorian:2012mx, Cieri:2015rqa}, as well as other work in SCET \cite{Becher:2010tm, Becher:2011xn, Becher:2012yn, GarciaEchevarria:2011rb, Echevarria:2012js, Echevarria:2014rua}.
A recent review of transverse-momentum dependent factorization and evolution can be found in \mycite{Rogers:2015sqa}.

We consider the measurement of the transverse momentum $\qt$ of a color-singlet system $X$ with invariant mass $Q$ and total rapidity $Y$.
In the limit of very small transverse momentum, $q_T \ll Q$, the cross section is dominated by soft and collinear gluon emissions from the incoming partons that recoil against the hard system $X$.
The emissions cause large logarithms $\ln^m(q_T^2/Q^2)$, with a power $m \le 2n-1$ at order $\as^n$, which we like to resum to all orders to retain predictive power in the perturbative series at small $q_T$.
In the limit $q_T \ll Q$, the cross section can be factorized as~\cite{Chiu:2012ir}
\begin{align} \label{eq:factorization thm}
\frac{\del\sigma}{\del Q^2 \del Y \del \qt}
\equiv \frac{1}{q_T} \frac{\del\sigma}{\del Q^2 \del Y \del q_T \del\phi}
&= \sigma_0\, H(Q,\mu) \int\! \del^2\ki{a}\, \del^2\ki{b}\, \del^2\ki{s} \, \delta(\qt - \ki{a} - \ki{b} - \ki{s})
\nn \\ & \quad
\times B_a(\omega_a, \ki{a},\mu,\nu)\,  B_b(\omega_b, \ki{b},\mu,\nu)\, S(\ki{s},\mu,\nu)
\,,\end{align}
where
\begin{equation}
\omega_a = Q e^Y
\,,\qquad
\omega_b = Q e^{-Y}
\,,\end{equation}
and for simplicitly we have kept the sum over partonic channels (and helicities in case of incoming gluons) implicit.
Equation~\eqref{eq:factorization thm} is valid up to power corrections in $q_T/Q$.
We are only interested in the leading-power cross section, which contains all singular logarithms,
and drop the power correction.

Here, $\sigma_0$ denotes the Born cross section and $H(Q,\mu)$ is the hard function containing virtual corrections to the partonic process. Following refs.~\cite{Stewart:2009yx, Stewart:2010qs}, the bare beam functions $B_i$ are defined in SCET in terms of forward proton matrix elements of collinear quark and gluon operators,
\begin{align} \label{eq:beam func def}
B^\text{bare}_q(\omega, \pt)
&= \theta(\omega) \bra{ P(p_n) } \bar\chi_n \frac{{\slashed{\bar n}}}{2} \delta(\omega-\overline{\mathcal P}_{n})\, \delta(\pt-\mathcal{P}_{\perp n}) \chi_n \ket{P(p_n)}
\,, \nn \\
B^{\mu\nu\,\text{bare}}_g(\omega, \pt)
&= \omega \theta(\omega) \bra{ P(p_n) }\mathrm{Tr}\{\cB_{n,\perp}^{\mu}\, \delta(\omega - \overline{\mathcal P}_{n})\, \delta(\pt-\mathcal{P}_{\perp n})\, \cB_{n,\perp}^{\nu}\} \ket{P(p_n)}
\,.\end{align}
They encode the effects of collinear initial-state radiation and are equivalent to transverse-momentum dependent parton distributions. They depend on the flavor $i$ and light-cone momentum $\omega = x p_n^-$ of the parton that enters the hard interaction, where $p_n$ is the proton momentum. They also depend on the total transverse momentum $\pt$ of collinear initial-state radiation that was emitted prior to the hard interaction.
The soft function $S$ measures the total transverse momentum originating from soft emissions and is defined as the vacuum matrix element
\begin{align} \label{eq:soft func def}
S_{gg}^\text{bare}(\pt)
&= \frac{1}{N_c^2-1} \bra{0}\mathrm{Tr}\{\bar T[S^\dagger_{n\perp} S_{ \bar n\perp} ]\, \delta(\pt - \mathcal{P}_{\perp s})\, T[S^\dagger_{\bar n\perp} S_{ n\perp}]\}\ket{0}
\,, \\
S_{qq}^\text{bare}(\pt)
&= \frac{1}{N_c} \bra{0}\mathrm{Tr}\{\bar T[S^\dagger_{n\perp} S_{ \bar n\perp}]\, \delta(\pt - \mathcal{P}_{\perp s})\, T[S^\dagger_{\bar n\perp} S_{ n\perp}]\}\ket{0}
\,.\end{align}
The $\cB^\mu_{n,\perp}$ and $\chi_n$ are collinear gluon and quark fields in SCET, $\mathcal{P}$ is the momentum label operator, and the $S_{n,\perp}$ are soft Wilson lines along the light-cone direction $n$. For more details see refs.~\cite{Chiu:2012ir, Luebbert:2016itl}.

The bare hard, beam, and soft functions are divergent quantities and require renormalization.
The UV divergences are regulated as usual by dimensional regularization. The resulting
renormalized functions, which appear in \eq{factorization thm}, satisfy the renormalization group equations
\begin{align}
\label{eq:RGE muH}
 \mu \frac{\del H(Q,\mu)}{\del\mu} &= \gMuH(Q,\mu)\, H(Q, \mu) \,, \\
\label{eq:RGE muB}
 \mu \frac{\del B(\omega, \pt, \mu, \nu)}{\del\mu} &= \gMuB(\omega, \mu,\nu)\, B(\omega, \pt, \mu, \nu) \,, \\
\label{eq:RGE muS}
 \mu \frac{\del S(\pt, \mu, \nu)}{\del\mu} &= \gMuS(\mu,\nu)\, S(\pt, \mu, \nu)
\,,\end{align}
where the anomalous dimensions have the all-order structure
\begin{align}
\label{eq:gMuH}
 \gMuH(Q,\mu) &= 4 \GammaC[\as(\mu)] \ln\frac{Q}{\mu} + \gMuH[\as(\mu)] \,, \\
\label{eq:gMuB}
 \gMuB(\omega,\mu,\nu) &= 2 \GammaC[\as(\mu)] \ln\frac{\nu}{\omega} + \gMuB[\as(\mu)] \,, \\
\label{eq:gMuS}
 \gMuS(\mu,\nu)    &= 4 \GammaC[\as(\mu)] \ln\frac{\mu}{\nu} + \gMuS[\as(\mu)]
\,.\end{align}
The $\mu$-independence of the cross section implies the RG consistency relation
\begin{equation}
\gMuH(\as) + 2\gMuB(\as) + \gMuS(\as) = 0
\,.\end{equation}

The beam and soft functions furthermore depend on a rapidity scale $\nu$, associated with an additional regulator required to regulate rapidity divergences. These arise because both soft and beam modes describe modes of virtuality $\mu^2 \sim q_T^2$, leading to an ``overlap'' of soft and collinear momentum regions which are not resolved by dimensional regularization
There have been a variety of rapidity regulators suggested in the literature to deal with these divergences.
In the original CSS approach a non-lightlike axial gauge was employed~\cite{Collins:1981uk}, whereas in recent work Wilson lines off the light cone are used \cite{Collins:1350496, Ji:2004wu}.
In the context of SCET, the utilized regulators include the analytic regulator acting on eikonal propagators~\cite{Beneke:2003pa, Chiu:2007yn, Becher:2011dz}, the $\eta$-regulator inserted into Wilson lines~\cite{Chiu:2011qc, Chiu:2012ir}, the $\delta$-regulator adding a mass to eikonal propagators~\cite{Chiu:2009yx, GarciaEchevarria:2011rb}, and the exponential regulator acting on the phase space~\cite{Li:2016axz}.
For all of these approaches, the beam and soft functions are currently known to NNLO \cite{Catani:2011kr,Catani:2012qa,Gehrmann:2014yya,Luebbert:2016itl,Echevarria:2015byo,Echevarria:2016scs}.
These fixed-order ingredients are not necessary for the purpose of this paper, which only relies on the RGE structure of beam and soft functions.

For concreteness we employ the $\eta$-regulator together with the rapidity renormalization group~\cite{Chiu:2011qc, Chiu:2012ir}.
Our discussion can be applied similarly to other regulators.
The crucial point is that the additional rapidity regulator induces an additional rapidity scale, here denoted as $\nu$.
In particular, $\nu$ is analogous to the scale $\zeta$ in the original CSS formulation~\mycite{Collins:1981uk},
see also \mycite{Rogers:2015sqa}. (In some formalisms or applications the rapidity scale is kept implicit and/or
fixed to canonical values.)

The rapidity renormalization group equations are given by
\begin{align}
\label{eq:RGE nuB}
 \nu \frac{\del B(\omega, \pt, \mu, \nu)}{\del\nu}
 &= \int\! \del^2\kt\, \gamma_{\nu,B}(\kt,\mu)\, B(\omega, \pt - \kt, \mu, \nu)
\,, \\
\label{eq:RGE nuS}
 \nu \frac{\del S(\pt, \mu, \nu)}{\del\nu}
 &= \int\!\del^2\kt\, \gamma_{\nu,S}(\kt,\mu)\, S(\pt - \kt, \mu, \nu)
\,.\end{align}
The overall $\nu$-independence of the cross section implies the consistency relation
\begin{equation} \label{eq:gNu}
\gNu(\pt,\mu) \equiv \gamma_{\nu,S}(\pt,\mu) = - 2 \gamma_{\nu,B}(\pt,\mu)
\,,\end{equation}
which means that there is only one independent rapidity anomalous dimension, which we denote as $\gNu$.
The crucial difference to the $\mu$-RGE is that the $\nu$-RGE is intrinsically a convolution.
The commutativity of $\del/\del\nu$ and $\del/\del\mu$ relates the $\mu$ and $\nu$ anomalous dimensions through
\begin{equation} \label{eq:consistency}
\mu \frac{\del}{\del\mu} \gNu(\kt,\mu) = \nu \frac{\del}{\del\nu} \gMuS(\mu,\nu) \delta(\kt) = - 4 \GammaC[\as(\mu)] \delta(\kt)
\,.\end{equation}

We stress that the singular logarithmic structure of the $q_T$ spectrum is fully encoded in the RGE equations eqs.~\eqref{eq:RGE muH}--\eqref{eq:consistency}. While their precise definitions and derivation depend to some extent
on the employed effective-field theory framework, equivalent evolution equations with the same momentum structure
exist in all approaches. In particular, the original CSS formulation for $q_T$ resummation is based on analogous evolution equations~\cite{Collins:1981uk, Collins:1981va, Collins:1984kg}.
Also, the corresponding system of differential equations in Fourier space, which is often considered,
is completely equivalent.
The SCET framework with rapidity renormalization we use is convenient in that it makes
the structure maximally general and explicit.

The nontrivial task at hand is to solve the RG equations with appropriate momentum-space boundary conditions
in order to perform the resummation in momentum space. In other words, we want
to predict the all-order distributional logarithmic structure in $q_T$ from the RG evolution.
As we will see, carrying out the RG evolution in momentum space is quite complicated due to the distributional nature coupled with the two-dimensional convolutions in $\kt$.
The correct solution for the momentum-space RG evolution is the main purpose of this paper.

An important comment concerns the definition of the formal resummation accuracy.
In problems with Sudakov double logarithms, the cross section exponentiates into
the form $\sigma \sim \exp[\as^n \ln^{n+1} + \as^n \ln^n + \as^n \ln^{n-1} + \dotsb]$. Including
the first set of terms in the exponent $\sim \as^n \ln^{n+1}$ corresponds to the leading-logarithmic (LL) order, including
the next set of terms $\sim \as^n \ln^n$ corresponds to the next-to-leading logarithmic (NLL) order, and so forth.
Alternatively, one can consider the logarithm of the cross section and count the corresponding terms in its $\as$ expansion.
One issue with this way of counting logarithms is that it is intrinsically ambiguous, i.e., it is always
possible to have slightly different definitions which agree to a given order, but differ by contributions
that in one or the other definition are formally subleading.
As we will see the $q_T$ spectrum in momentum space does not exponentiate into a simple exponential,
but rather it will have a generalized exponential structure in distribution space, which makes this
counting of explicit logarithms even less well-defined. (We will see that this can even lead to producing
spurious divergences in the resummed result from formally subleading terms.)

The exponential structure of the resummed $q_T$ spectrum is fully encoded in its evolution equations, or equivalently
the anomalous dimenions are in essence the generalized ``logarithm'' of the resummed cross section.
Therefore, an easy, fully consistent, and unambiguous way to define the resummation order is strictly by the
pure fixed-order expansions of the cusp and noncusp anomalous dimensions together
with the fixed-order expansions of the \emph{pure} boundary terms for the hard, beam, and soft functions.
Since these fixed-order series are the fundamental input to the RG evolution, with everything else
following from them, it makes sense to define the fundamental resummation accuracy
solely by the perturbative accuracy at which these inputs are included. In particular, with this strict definition
one is not allowed to disregard seemingly subleading logarithmic terms at intermediate stages. This also
means that the various RG consistency relations should always be exactly fulfilled by the resummed result.
Our goal is to derive the solution for the evolution and resummation in momentum space, i.e.,
with anomalous dimensions and boundary terms defined in momentum space, valid to in principle any order in this strict definition.
Note that given the resummed result at strict LL, NLL, etc. one can of course consider
additional choices or further approximations to simplify the result and study their numerical impact,
which however is not our concern here.

\paragraph{Notation:}
In the remainder of this paper, we denote the final transverse momentum of the produced color-singlet state $X$ by $\qt$,
while the transverse-momentum argument of a specific function is typically denoted by $\pt$, and the integration momenta in the convolution integrals are usually denoted by $\ki{i}$.
Convolutions are often abbreviated as
\begin{equation}
(f \otimes g)(\pt, \dots) \equiv \int\!\del^2\kt\, f(\pt - \kt, \dots)\, g(\kt, \dots)
\,,\end{equation}
where the ellipses denote possible additional variables.
Multiple convolutions are abbreviated as
\begin{equation} \label{eq:mult conv 2d}
 (f \otimes^n)(\pt)
 \equiv \int\del^2\ki{1} \dots \del^2\ki{n}\, f(\ki{1}) \dots f(\ki{n})\, \delta(\pt - \ki{1} - \cdots - \ki{n}) \,.
\end{equation}
These formulas are also summarized in appendix \ref{app:notation}.

\subsection{Implications of two-dimensional convolutions}
\label{sec:TMD convolutions}

The transverse momentum spectrum is generically given in terms of plus distributions, which are necessary to regulate its $1/q_T^2$ divergences.
Furthermore, the factorized cross section and its RGEs involve two-dimensional convolutions, see \eqs{factorization thm}{RGE  nuS}.
While the issue of scale setting with distributions has been addressed in \sec{distr scale setting},
we now discuss the implications of the two-dimensional convolutions.

To illustrate the arising subtleties, it is sufficient to focus on the rapidity RGE of the soft function,
as evolving it to the beam scale $\nu = \nu_B \sim Q$ eliminates all rapidity logarithms in the beam functions entering \eq{factorization thm}. A formal solution to \eq{RGE nuS} is readily derived to be
\begin{align}  \label{eq:naive solution S}
 S(\pt, \mu, \nu_B) &= \int\! \del^2\kt\, V(\pt - \kt,\mu, \nu_B, \nu_S)\, S(\kt, \mu, \nu_S)
\,,\end{align}
where the rapidity evolution kernel $V$ is given by
\begin{align} \label{eq:naive V}
V(\pt,\mu,\nu_B,\nu_S)
&= \delta(\pt)  + \sum_{n=1}^\infty \frac{1}{n!} \ln^n\!\frac{\nu_B}{\nu_S}\, (\gNu \otimes^n)(\pt,\mu)
\nn \\
&= \delta(\pt)
+  \ln\frac{\nu_B}{\nu_S} \gNu(\pt, \mu)
\nn \\ & \quad
+ \frac{1}{2} \ln^2\frac{\nu_B}{\nu_S} \int\!\del^2\ki{1} \int\!\del^2\ki{2} \,\gNu(\ki{1}, \mu)\, \gNu(\ki{2 }, \mu)\,
\delta(\ki{1} + \ki{2} - \pt)
\nn \\ & \quad
+ \dotsb
\,.\end{align}
Here $(\gNu \otimes^n)$ denotes $n$ convolutions of $\gNu$ with itself, see \eq{mult conv 2d}.
$V$ is an exponential in convolution space and \eq{naive V} can be derived equivalently in
convolution space or Fourier space (see \sec{naive illustration} below).
Taking the derivative with respect to $\nu$, one can easily verify
that it provides a solution for \eq{RGE nuS}.

The evolution kernel \eq{naive V} has a simple physical interpretation.
Each factor of $\gNu(\ki{i}, \mu)$ corresponds to a real soft emission with momentum $\ki{i}$.
The convolution integrals are the remaining transverse phase-space integrals, which are constrained such
that the transverse momenta of all emissions sum up to the total $\pt$.
Each emission is dressed with a rapidity logarithm $\ln(\nu_B/\nu_S)$
to evolve in rapidity from the soft scale $\nu_S$ to the beam scale $\nu_B$, which corresponds
to the effective range in rapidity over which the soft emission has been integrated.
The $n$ emission term then scales with $\ln^n(\nu_B/\nu_S)$, which precisely builds up
an exponential in convolution space.

\begin{figure}[h]
\centering
\includegraphics[width=0.5\textwidth]{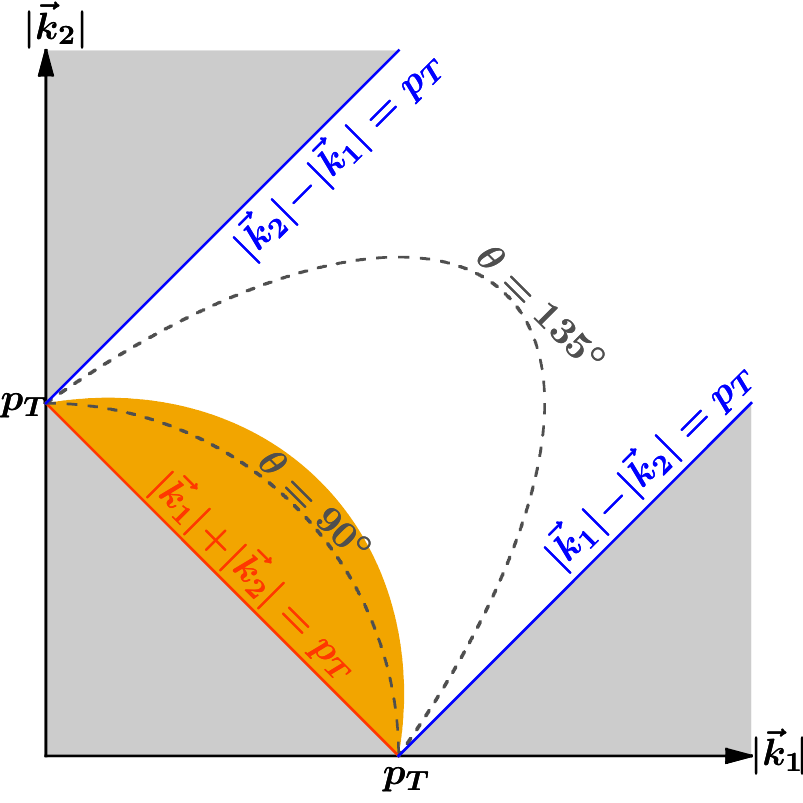}
\caption{Illustration of the transverse momentum regions contributing to the convolution $(\gNu \otimes \gNu)(\pt,\mu)$.
The region between the solid blue and orange lines contributes to the convolution integral.
The dashed lines correspond to a fixed angle $\theta = \angle(\ki{1},\ki{2}) = 90,\,135^\circ$ between the two emissions.
Canonical scaling $|\ki{1}| \sim |\ki{2}| \sim p_T$ is only fulfilled in the shaded orange region.}
\label{fig:convolution}
\end{figure}

To investigate the structure of $V$ in more detail, we focus on the first nontrivial convolution.
It involves integrating over two real emissions with momenta $\ki{1,2}$ such that $\ki{1}+\ki{2} = \pt$.
Figure~\ref{fig:convolution} illustrates the momentum regions in $|\ki{1,2}|$ contributing to this integral.
The region between the solid lines is allowed, while the gray region outside cannot fulfill the measurement constraint.
The dashed lines correspond to a fixed angle between the two emissions of $\angle(\ki{1},\ki{2}) = 90^\circ, 135^\circ$.
The larger this angle is, the larger the allowed magnitudes $|\ki{1,2}|$ are.
In the limit where the emissions are back-to-back, the magnitudes $|\ki{1,2}|$ can become infinitely large, as long as their difference still gives $\pt$. This is the limit given by the two blue lines.
Hence, the convolution integrals in \eq{naive V} in principle receive contributions from infinitely large momenta. Physically, the limit of both emissions having large $|\ki{1,2}|\sim Q$ should be power-suppressed in $q_T/Q$ and hence not affect the singular logarithmic structure. However, in this limit the emissions are not correctly described anymore by the underlying soft expansion, which assumes $|\ki{1,2}| \sim p_T \ll Q$. Therefore, we can in principle receive spurious contributions to the integral from this region. On the other hand, there can also be relevant physical contributions from any intermediate region $p_T \ll |\ki{1,2}| \ll Q$ that must be correctly taken into account.
It was already argued in ref.~\cite{Parisi:1979se} (see also ref.~\cite{Monni:2016ktx}) that the very small $\qt$ region can be influenced (or even be dominated) by such kinematic cancellations of harder emissions.

One might now be worried that the factorization theorem is intrinsically ill-defined,
as it contains effects of arbitrarily hard emissions.
This is unavoidable, as the soft approximation eliminates the phase-space constraints
that would normally cut off such emissions, as already noticed in \mycite{Frixione:1998dw}.
However, all large rapidity logarithms should arise from the soft region
where all emissions are of the order of the final $q_T$, $|\ki{i}| \sim q_T$.
From the above observations, it is clear that the rapidity evolution kernel
\eq{naive V} violates this requirement. This is perfectly fine because a priori
\eqs{naive solution S}{naive V} only shift logarithms $\ln(p_T/\nu_S)$ in $S(\pt,\mu,\nu_S)$
into logarithms $\ln(p_T/\nu_B)$ in $S(\pt,\mu,\nu_B)$.
In particular, if $S(\pt,\mu,\nu_S)$ is already resummed,
\eqs{naive solution S}{naive V} are valid solutions of the rapidity RGE.

The situation changes if we want to use the rapidity RGE to predict all logarithms of the soft function.
Then we would like to start from a boundary condition for the soft function without any real emissions,
and subsequently let the evolution add real emissions by convolving with $\gNu$.
Each such emission with momentum $\ki{i}$ should scale with a rapidity logarithm $\sim \ln(\nu_B / |\ki{i}|)$
to evolve in rapidity from its own emission scale to the beam scale.
For the example of two emissions, we thus expect a contribution of the form
\begin{equation} \label{eq:conv illustration}
 \sim \int\!\del^2\ki{1} \int\! \del^2\ki{2}\,
 \ln\frac{\nu_B}{|\ki{1}|} \gNu(\ki{1},\mu)  \ln\frac{\nu_B}{|\ki{2}|} \gNu(\ki{2},\mu) \delta(\pt - \ki{1} - \ki{2})
\,.\end{equation}
(For the moment we ignore that the rapidity logarithms need to be properly
included into plus distributions, which will be taken care of in the
final solution in \sec{soft func}.)
If the $\ki{1,2}$ integrals were dominated by $|\ki{1,2}| \sim p_T$,
indicated by the shaded orange region in figure~\ref{fig:convolution},
both rapidity logarithms could be approximated as $\sim \ln(\nu_B/p_T)$,
and a double logarithm $\ln^2(\nu_B/p_T)$ could be pulled out of the convolution integrals.
This is precisely what happens in \eq{naive V}.
However, as explained above, the convolution also gets contributions from energetic emissions
$|\ki{1,2}| \gg p_T$, which happen to be back-to-back such that $\ki{1} + \ki{2} = \pt$.
In this region, the rapidity logarithms $\ln(\nu_B/|\ki{1,2}|)$ in \eq{conv illustration} get smaller
and eventually become irrelevant for $|\ki{1,2}| \sim Q$. Instead, the approximation $|\ki{1,2}| \sim p_T$
would result in a spuriously large rapidity logarithm $\ln^2(\nu_B/p_T)$, which would
artificially enhance this otherwise suppressed phase-space region.
To properly treat these emissions, it is necessary to keep the
correct rapidity logarithms $\sim \ln(\nu_B / |\ki{i}|)$ in the convolutions.
This is obviously not possible by simply choosing a fixed value for $\nu_S$ in
\eqs{naive solution S}{naive V}.
In particular, if we were to use the naive canonical choice $\nu_S\sim p_T$,
we would artificially enhance unphysical contributions from energetic emissions.
In \sec{TMD illustration}, this point will be stressed by showing that doing so
actually produces a spurious divergence in the $q_T$-spectrum.
The correct RG evolution to predict the resummed soft function is significantly
more complicated than \eq{naive V}, and in particular will rely on distributional scale setting.
It will be derived in \sec{soft func}.

So far we have focused on the role of the rapidity scale.
The two-dimensional convolution structure also has implications for the scale $\mu$ entering
the rapidity anomalous dimension and soft function.
For this purpose, it suffices to consider \eqs{naive solution S}{naive V},
which correctly evolve the (already resummed) soft function between two scales $\nu_S$ and $\nu_B$.
The final results in \eqs{naive solution S}{naive V} after all integrations can only depend on $p_T$
and thus can only contain logarithms $\ln(p_T/\mu)$.
One might therefore expect that when performing the rapidity evolution
at $\mu \sim p_T$, it should be sufficient to evaluate $\gNu$ at fixed order.
However, what is important is that the rapidity anomalous dimension contains logarithms $\ln(|\ki{i}|/\mu)$ (in distributional form). These are only minimized by $\mu \sim p_T$ if the convolutions were dominated by $|\ki{i}| \sim p_T$. However, as argued above, since the convolution intrinsically probes momenta $|\ki{i}| \gg p_T$, setting $\mu \sim p_T$ can induce spurious logarithms $\ln(|\ki{i}|/p_T)$. Hence, the rapidity anomalous dimension $\gNu$ entering the convolutions should always be resummed in order to correctly describe emissions at any $|\ki{i}| \lesssim Q$.
As we will see in \sec{gNu resummation}, the main effect of resumming $\gNu(\ki{i},\mu)$ is
to evaluate it  at $\as(|\ki{i}|)$ rather than $\as(\mu)$.
This suppresses the amplitude of energetic emissions,
which is particularly important because there is no phase-space suppression
due to the soft approximations, as discussed above.

At this point we can also discuss why these complications do not arise for scalar quantities like thrust or transverse energy. Taking transverse energy as the closest example%
\footnote{The factorization of $E_T$ is known to be affected by Glauber modes \cite{Gaunt:2014ska}, but the $E_T$-distribution can still serve as an example for the different mathematical structure of the renormalization group equations.}
to $q_T$, the measurement constraint (for two emissions) changes to
\begin{equation}
 \delta(\ki{1} + \ki{2} - \pt) \quad\rightarrow\quad \delta(|\ki{1}| + |\ki{2}| - E_T)
\,.\end{equation}
This is clearly a much stronger constraint and forces both momenta to be of the order of the total transverse energy, $|\ki{i}| \sim E_T$. In figure~\ref{fig:convolution}, this corresponds to forcing the momenta to lie on the orange line, and it does not allow any contributions from the large momentum region.
In particular, the rapidity logarithms $\ln(\nu_B/|\ki{i}|)$ are now well approximated by $\sim \ln(\nu_B/E_T)$.
As we will see in the next subsection, this is precisely the reason why in the one-dimensional case the convolution structure of the RGE does not lead to a spurious singularity.

\subsection{Rapidity evolution in Fourier space}
\label{sec:naive illustration}

To illustrate that the problems observed in the previous section are not
an artifact of the momentum space approach, we now consider the resummation
of the soft function in Fourier space.
There, the rapidity RGE in \eq{RGE nuS} becomes multiplicative,
\begin{equation} \label{eq:RGE nuS FT}
 \tilde S(\bt,\mu,\nu) = \tgNu(\bt,\mu)\, \tilde S(\bt,\mu,\nu)
\,,\end{equation}
where $\bt$ is Fourier conjugate to $\pt$ and the tilde denotes the
Fourier transformed  quantities.
Equation~\eqref{eq:RGE nuS FT} is easily solved by
\begin{equation}
 \tilde S(\bt,\mu,\nu) = \tilde S(\bt,\mu,\nu_0) \exp\biggl[ \ln\frac{\nu}{\nu_0} \tgNu(\bt,\mu)\biggr]
\,,\end{equation}
from which we recover the momentum space solution
\begin{equation}
S(\pt,\mu,\nu) = \int\!\frac{\del^2\bt}{(2\pi)^2}\, e^{i \pt \cdot \bt} \tilde S(\bt,\mu,\nu_0) \exp\biggl[ \ln\frac{\nu}{\nu_0} \tgNu(\bt,\mu)\biggr]
\,.\end{equation}
For \emph{arbitrary} $\nu_0$, this is exactly equal to \eqs{naive solution S}{naive V},
and allows one to correctly shift logarithms from $\nu_0$ to $\nu$.

However, to instead predict the soft function including its logarithms, one needs
to choose $\nu_0$ suitably.
Since the final $S(\pt,\mu,\nu)$ result is known to contain logarithms $\ln(p_T/\nu)$,
one might be tempted to set $\nu_0 = p_T$ and use the fixed-order boundary condition $\tilde S(\bt,\mu,\nu_0) = 1 + \dotsb$.
However, we know that $\tilde S$ contains logarithms $\ln(\nu_0 b_T)$, and these are of course
not eliminated by $\nu_0 \sim p_T$. Since the Fourier integral runs over all $\bt$,
they can in principle become relevant at small and large $\bt$, in particular wherever the simple scaling
$b_T \sim p_T^{-1}$ is violated. (This is the same situation as discussed in \eq{toy wrong boundary}.)
This is analogous to setting $\nu_S$ in \eq{naive solution S} to the overall
$q_T$ and using the pure fixed-order boundary condition for $S(\kt, \mu, \nu_S)$.

We will explicitly see in the next section that the $b_T \to 0$ region causes
troubles if one were to set $\tilde S(\bt,\mu,\nu_0) = 1$.
In Fourier space, the problem is easily overcome by choosing $\nu_0 \sim 1/b_T$,
which would allow one to evaluate the boundary condition in a pure fixed-order expansion.
However, this corresponds to an intrinsic scale setting and thus resummation in Fourier space.
The correct momentum-space analog corresponds to the discussion in \eq{conv illustration}
and will be derived in \sec{soft func}.

\subsection{Illustration: effects from energetic emissions}
\label{sec:TMD illustration}

In the \sec{TMD convolutions} we argued that the two-dimensional convolutions of $\gNu$ are
intrinsically sensitive to large transverse momenta. As a result
the formal solution \eq{naive V} does not allow one to correctly predict the all-order
soft function because these energetic emissions are artificially enhanced
by large rapidity logarithms. We will now explicitly demonstrate that this incorrect
treatment is what causes a well-known spurious singularity in the evolution kernel.

We first calculate the rapidity evolution kernel $V$ in \eq{naive V} using the fixed
leading-order expression for $\gNu$, which is given by
\begin{equation} \label{eq:gnuLO}
\gNu(\pt, \mu) = 2 \GammaC[\as(\mu)]\, \PlusLog{0}(\pt, \mu)
\,.\end{equation}
Inserting this into \eq{naive V}, the evolution kernel can be calculated either
through Fourier transformation or by explicitly calculating $\PlusLog{0} \otimes^n$ [see \eq{Plus0^n}].
Both techniques yield
\begin{align}  \label{eq:V naive sol}
V(\pt,\mu,\nu, \nu_0)
&= \Bigl[ \delta(\pt) + \omega_s \PlusPol{\omega_s}(\pt,\mu) \Bigr]
 e^{-2\gamma_E \omega_s} \frac{\Gamma(1-\omega_s)}{\Gamma(1+\omega_s)}
\,,\end{align}
where
\begin{equation} \label{eq:omegas}
\omega_s =  2 \GammaC[\as(\mu)] \ln\frac{\nu}{\nu_0}
\,,\end{equation}
and the plus distribution is defined as [see \eq{plus func pol}]
\begin{align}
\PlusPol{a}(\pt, \mu)
&\equiv \frac{1}{\pi \mu^2} \biggl[ \biggl(\frac{\pt^2}{\mu^2}\biggr)^{a-1} \biggr]_+^\mu
\equiv \frac{1}{\pi \mu^2} \PlusPol{a}(\pt^2 / \mu^2)
\,.\end{align}
The result in \eq{V naive sol} contains an explicit divergence at $\omega_s = 1$, which has
been encountered before~\cite{Frixione:1998dw,Becher:2010tm,Becher:2011xn,Chiu:2012ir,Monni:2016ktx}.
Using what would seem to be the canonical scale choices, $\mu = \nu_0 = q_T$ and $\nu = Q$, the divergence at $\omega_s = 1$ occurs when
\begin{equation}
\GammaC[\as(q_T)] = \ln^{-1}\frac{Q^2}{q_T^2}
\,.\end{equation}
For illustration, for gluon-induced processes like $gg\to H$, where $\GammaC \sim C_A$,
this happens at $q_T \approx 8\,\GeV$ for $Q = 125\,\GeV$
and $q_T \approx 27~\GeV$ for $Q = 1\,\TeV$.
For quark-induced process like Drell-Yan, where $\GammaC \sim C_F$,
it occurs at $q_T \approx 2\,\GeV$ for $Q = m_Z$
and $q_T \approx 4\,\GeV$ for $Q = 1\,\TeV$.

Clearly, the all-order $q_T$ spectrum cannot contain such a singularity, especially since for large $Q$
it happens at purely perturbative $q_T$.
Its appearance for small enough $q_T/Q$ respectively large enough $\ln(q_T/Q)$ indicates
that the above naive attempt simply does not properly treat the resummation of logarithms.
As we have shown, \eq{V naive sol} can be derived working entirely in momentum space and
without any reference to Fourier space, which means that it is not related to a possibly
ill-defined inverse Fourier transformation.

To show that the origin of this divergence is indeed due to contributions from large transverse momenta
in the convolution, as argued earlier, we regulate the LO anomalous dimension in \eq{gnuLO} by introducing an
explicit cut off $\Lambda$ in momentum space,
\begin{equation}  \label{eq:gNu regulated}
\gNu(\pt, \mu) \rightarrow \gNu^{(\Lambda)}(\pt,\mu) \equiv \gNu(\pt, \mu) \theta(\Lambda - p_T)
\,.\end{equation}
Since $\gNu(\pt,\mu)$ corresponds to a single real emission with momentum $\pt$, this is equivalent
to cutting off emissions with large transverse momentum $|\pt| > \Lambda$.
Using \eq{gNu regulated} in \eq{naive V}, the rapidity evolution kernel becomes rather complicated and is most easily evaluated in Fourier space, where the regulated anomalous dimension \eq{gNu regulated} reads
\begin{equation}
\tilde{\gNu}^{(\Lambda)}(\bt,\mu)
= 2 \GammaC[\as(\mu)] \biggl[ \ln\frac{\Lambda^2}{\mu^2}
- \frac{b_T^2\Lambda^2}{4} {_2F_3}\biggl(- \frac{b_T^2\Lambda^2}{4}\biggr) \biggr]
\,,\end{equation}
where $_2F_3(x) = {_2F_3}( 1, 1; 2, 2, 2; x)$ is a generalized hypergeometric function.
The rapidity evolution kernel then is
\begin{align} \label{eq:V regulated}
V(\pt, \mu, \nu, \nu_0)
&= \int\!\frac{\del^2\bt}{(2\pi)^2}\, e^{i \bt \cdot \pt} \exp\biggl[\ln\frac{\nu}{\nu_0} \tgNu^{(\Lambda)}(\bt,\mu)\biggr]
\nn \\
&= \frac{1}{2 \pi} \biggl(\frac{\Lambda^2}{\mu^2}\biggr)^{\omega_s} \int_0^\infty\! \del b_T\,b_T J_0(b_T p_T)
\exp\biggl[-\omega_s \frac{b_T^2 \Lambda^2}{4} {_2F_3} \biggl(-\frac{b_T^2\Lambda^2}{4}\biggr) \biggr]
\nn \\
&= \frac{2}{\pi\mu^2} \biggl(\frac{\Lambda^2}{\mu^2}\biggr)^{\omega_s-1} \int_0^\infty\! \del x\,x
J_0\biggl(\frac{2x p_T}{\Lambda}\biggr) \exp\Bigl[ - \omega_s\, x^2 {_2F_3}(-x^2)\Bigr]
\,.\end{align}
This is perfectly finite for $\omega_s = 1$.
It shows that by cutting off energetic emissions the divergence disappears, and
confirms our observation that it is precisely the large momentum region that is
being treated incorrectly.
Taking the $\Lambda\to\infty$ limit in \eq{V regulated} is slightly nontrivial but reproduces \eq{V naive sol}.
In particular, from the expression in the last line of \eq{V regulated} it is easy to see
that for $\omega_s = 1$ the prefactor diverges for $\Lambda \to \infty$,
while it does not for $\omega_s < 1$.

Interestingly, by choosing $\Lambda = p_T$, meaning that no single emission is allowed to
be harder than the total $\pt$, one obtains
\begin{align}
V(\pt, \mu, \nu, \nu_0)
&= \frac{2}{\pi \mu^2} \biggl(\frac{p_T^2}{\mu^2}\biggr)^{\omega_s-1} \int_0^\infty\! \del x\,x J_0(2x)
\exp\biggl[ - \omega_s\, x^2 {_2F_3}(-x^2) \biggr]
\,,\end{align}
where the $p_T$-dependence is exactly the same as in the unregulated solution \eq{V naive sol},
while the remaining integral is a function of $\omega_s$ that is finite for any $\omega_s$
(except for $\omega_s=0$, for which the whole expression has to reproduce the boundary condition $V(\pt,\mu,\nu, \nu) = (2\pi)^2\delta(\pt)$).
This stresses again that emissions much harder than the actual final state $\pt$ cause the divergence.

To see more explicitly how the divergence in \eq{V naive sol} arises, we can focus on its $\delta(\pt)$ term.
Using the expression for $\cL_0\otimes^n$ from \eq{Plus0^n}, we get
\begin{align}
V(\pt,\mu, \nu, \nu_0)
&= \delta(\pt) + \sum_{n=1}^\infty \frac{\omega_s^n }{n!} (\cL_0\otimes^n)(\pt, \mu)
= \delta(\pt) \sum_{n=0}^\infty \frac{\omega_s^n }{n!} (-1)^n \dR[2]{n} + \dotsb
\nn \\
&\sim \delta(\pt) \sum_{n=0}^\infty \frac{\omega_s^n }{n!} \times n! + \dotsb
\,,\end{align}
where the ellipses denote the remaining distributions $\cL_n$ that built up the $\cL^{\omega_s}$ term in \eq{V naive sol}.
The $(-1)^n\dR[2]{n}$ is exactly the $n$th derivative of the $\omega_s$-dependent factor in \eq{V naive sol}, $e^{-2\gamma_E \omega_s} \Gamma(1-\omega_s)/\Gamma(1+\omega_s)$, see \eq{R(n)}.
The last sum in the first line hence precisely leads to the $\delta(\pt)$ piece of \eq{V naive sol},
including its divergence at $\omega_s = 1$.
In the last step we inserted the asymptotic behaviour $\dR[2]{n} \sim (-1)^n n!$ [see \eq{R2n_largen}].
Each factor $\omega_s$ corresponds to a single $\gNu$ and hence a single real emission.
Therefore, the $n!$ growth of the contribution from $n$ emissions, which leads to the divergence for $\omega_s = 1$,
simply reflects the number of combinations of $n$ individual emissions $\ki{i}$ to yield an overall
$\pt = \sum_i \ki{i}$. Hence it is of a kinematic origin and intrinsic to the two-dimensional
nature of $\pt$.

To further illustrate this, we can show that this divergence does not appear for the one-dimensional case,
where the convolution momentum is strictly limited to $k' \leq k$. The analogous one-dimensional
problem (which would be relevant for $E_T$) is given by
\begin{align}
 \nu \frac{\del F(k,\mu,\nu)}{\del\nu} &= \int\!\del k'\, \gamma_{\nu,F}(k-k',\mu) F(k',\mu,\nu)
\,,\quad \nn\\
 \gamma_{\nu,F}(k,\mu) &= 2 \GammaC[\as(\mu)] \cL_0(k,\mu)
\,.\end{align}
The corresponding evolution kernel is given by,
\begin{align}
 U(k,\mu,\nu,\nu_0)
 &= \delta(k) + \sum_{n=1}^\infty \frac{\omega_s^n}{n!} (\cL_0 \otimes^n)(k,\mu)
 = \Bigl[ \delta(k) + \omega_s \PlusPol{\omega_s}(k,\mu) \Bigr] \frac{e^{-\gamma_E \omega_s}}{\Gamma(1+\omega_s)}
\nn \\
 &= \delta(k) \sum_{n=0}^\infty \frac{\omega_s^n}{n!} (-1)^n \dRc[1]{n} + \cdots
= \delta(k) \frac{e^{-\gamma_E \omega_s}}{\Gamma(1+\omega_s)} + \cdots
\,,\end{align}
where the ellipses in the second line denote plus distributions $\cL_n(k,\mu)$, which we omitted for simplicity to focus on the $\delta(k)$ term.
As before, each power of $\omega_s$ corresponds to a single emission.
The $n$-emission term comes with a coefficient $(-1)^n\dRc[1]{n}$, which is the $n$th derivative of $e^{-\gamma_E \omega_s}/\Gamma(1+\omega_s)$, see \eq{dR1c}.
This function is well defined for all $\omega_s$, and in particular its derivatives do not show the factorial growth
that is present in the two-dimensional case $\dR{n} \sim (-1)^n n!$.

Finally, for completeness we show how the divergence arises from the Fourier space calculation.
In this case, the evolution kernel is given by the inverse Fourier transform
\begin{align}
V(\pt,\mu, \nu, \nu_0)
&= \int\!\frac{\del^2\bt}{(2\pi)^2}\,e^{i \bt \cdot \pt} \exp\biggl[ \ln\frac{\nu}{\nu_0} \tgNu(b_T,\mu) \biggr]
\,.\end{align}
The Fourier transform of the LO expression for $\gNu$ in \eq{gnuLO} is given by
\begin{equation}
 \tgNu(b_T,\mu)
 = \int\!\del^2\pt\, e^{-i \bt \cdot \pt} \gNu(\pt,\mu)
 = -2 \GammaC[\as(\mu)] \ln\frac{b_T^2 \mu^2}{b_0^2}
\,,\end{equation}
such that
\begin{align}
V(\pt,\mu, \nu, \nu_0)
&=  \frac{1}{2 \pi} \int_0^\infty\! \del b_T\,b_T\,J_0(b_T p_T) \exp\biggl(-\omega_s \ln\frac{b_T^2 \mu^2}{b_0^2} \biggr)
\,,\end{align}
with $\omega_s$ as in \eq{omegas}.
The integral converges at large $b_T$. At small $b_T$ the integrand behaves as
\begin{equation}
b_T\, J_0(b_T p_T) \exp\biggl(-\omega_s \ln\frac{b_T^2 \mu^2}{b_0^2}\biggr)  \sim b_T^{1-2\omega_s}
\,,\end{equation}
where we approximated $J_0(b_T p_T) \approx 1$ for $b_T \to 0$.
Hence we find a logarithmic divergence from the small $b_T$ limit when $\omega_s \to 1$.
This is consistent with the previous calculation, since roughly $b_T \sim p_T^{-1}$,
and thus we can again conclude that energetic emission cause the divergent term.
To identify the factorial growth, we expand the exponential in the integrand and perform the integration in the small $b_T$ region, where $J_0(b_T p_T) \sim 1$,
\begin{align}
V(\pt,\mu, \nu, \nu_0)
&= \frac{1}{2 \pi} \int_0^\infty\! \del b_T\,b_T\,J_0(b_T p_T) \exp\biggl(-\omega_s \ln\frac{b_T^2 \mu^2}{b_0^2} \biggr)
\nn \\
&\sim  \frac{1}{2 \pi} \sum_{n=0}^\infty \frac{1}{n!} (-\omega_s)^n \int\!\del b_T\,b_T \ln^n\!\frac{b_T^2 \mu^2}{b_0^2}
\nn\\
&\sim \frac{ b_0^2}{4\pi\mu^2}  \sum_{n=0}^\infty \frac{1}{n!} \omega_s^n n!
 = \frac{ b_0^2}{4\pi\mu^2}  \frac{1}{1-\omega_s}
\,.\end{align}
Hence, the small-$b_T$ region of the Fourier integral reproduces the $n!$ growth
responsible for the first pole in $\omega_s = 1$ in \eq{V naive sol}.
Note that expanding the Bessel function $J_0$ to higher orders similarly produces
the other poles at $\omega_s = 2, 3, \cdots$ of \eq{V naive sol}.

In conclusion, the two-dimensional nature of the convolutions appearing in transverse momentum
distributions pose a significant complication to performing the associated renormalization group evolution
and resummation.
The main problem is that these convolutions can intrinsically probe emissions at arbitrarily large momentum via kinematic cancellations. In principle, such kinematic cancellations among several hard emissions to produce a small value of $\qt$ is a physical effect which will be present in the full all-order result for the spectrum,
However, in the above naive approach, the large-momentum emissions get artificially enhanced by
fake rapidity logarithms.
By explicitly cutting off such emissions, we have shown that they indeed produces unphysical contributions and are the origin of a spurious singularity in the rapidity evolution kernel.
As discussed in \sec{TMD convolutions}, the appearance of these energetic emissions
requires a careful scale setting when performing the RG evolution for $q_T$-spectrum in
momentum space, which will be the focus of the remainder of this paper.

\section{Resummation of the rapidity anomalous dimension}
\label{sec:gNu resummation}

In this section, we carry out the resummation of the rapidity anomalous dimension $\gNu(\pt, \mu)$ in
momentum space by solving its differential equation [see \eq{consistency}],
\begin{equation}  \label{eq:gNu RGE}
 \mu \frac{\del\gNu(\pt,\mu)}{\del\mu} = -4\GammaC[\as(\mu)]\, \delta(\pt)
\,,\end{equation}
using the techniques introduced in \sec{distr scale setting}.
Since \eq{gNu RGE} encodes the consistency (i.e.\ exact path independence) between the $\mu$ and $\nu$
evolutions~\cite{Chiu:2011qc, Chiu:2012ir}, the solution of \eq{gNu RGE} is
an important ingredient in the full momentum-space resummation.
In particular, as was discussed in \sec{TMD}, since the two-dimensional convolutions are intrinsically
sensitive to emissions at all momentum scales, one cannot naively use a fixed-order approximation
for $\gNu$ even when the rapidity RGE is performed at $\mu \sim q_T$.

We will also use the resummed result for $\gNu(\pt, \mu)$ as an example
to study the differences to carrying out the resummation in Fourier space,
the appearance of nonperturbative effects in the $q_T$ spectrum,
as well as the implementation of a profile scale and how it allows to probe subleading logarithms.

\subsection{Resummation of \texorpdfstring{$\gNu$}{gammaNu} in closed form}

We can solve \eq{gNu RGE} distributionally by integrating it from arbitrary $\mu_0$ to $\mu$
and then setting $\mu_0 = p_T\dis$ using \eq{distr scale setting},
\begin{align}
 \gNu(\pt,\mu)
 &= \biggl[\gNu(\pt,\mu_0) - \delta(\pt) \int_{\mu_0}^\mu\! \frac{\del\mu'}{\mu'}\, 4 \GammaC[\as(\mu')] \biggr]_{\mu_0 = p_T\dis}  \nn \\
 &= \frac{1}{2 \pi p_T} \frac{\del}{\del p_T} \biggl\{ \theta(p_T)\gNu[\as(p_T)] - \theta(p_T) \int_{p_T}^\mu\! \frac{\del\mu'}{\mu'}\, 4 \GammaC[\as(\mu')] \biggr\}
\,.\end{align}
Here we used that the boundary term can only depend on $\as(p_T)$ and that by virtue of the cumulant must be proportional to $\theta(p_T)$, to write it as $\theta(p_T) \gNu[\as(p_T)]$.
Evaluating the derivative [see \eq{cumulant inverse}] yields
\begin{align} \label{eq:gNu solution}
\gNu(\pt,\mu)
&= \biggl[ \frac{4\GammaC[\as(p_T)]}{2 \pi p_T^2}  \biggr]_+^\mu
 + \biggl[ \frac{1}{2 \pi p_T^2} \frac{\del \gNu[\as(p_T)]}{\del \ln p_T} \biggr]_+^\xi + \delta(\pt) \gNu[\as(\xi)]
\,.\end{align}
The $\mu$ dependence is carried by the first distribution, which can easily be seen to fulfill \eq{gNu RGE}.
The last two terms arise from the derivative acting on $\theta(p_T) \gNu[\as(p_T)]$ and are hence independent of any scale except $p_T$. The $\xi$ dependence exactly cancels between the two terms, but is necessary to introduce a plus distribution to regulate the $1/p_T^2$ divergence. The same result is also obtained by directly solving \eq{gNu RGE} in cumulant space.

Let us briefly discuss the form of \eq{gNu solution}.
Recall that $\gNu(\pt,\mu)$ corresponds to a single real emission with transverse momentum $\pt$.
The factor $1/p_T^2$ inside the plus distributions corresponds to the propagator associated with such an emission.
The associated infrared singularity at $\pt\to0$ has to cancel against virtual corrections,
which is encoded by the plus distribution that regulates the divergence.
The effect of solving the RGE of $\gNu$ is to evaluate the anomalous dimension at $\as(p_T)$ rather than $\as(\mu)$.
This is not very surprising, as we would expect the $\mu$-RGE to resum virtual corrections to this single emission,
which naturally pushes $\as$ to be evaluated at the emission scale rather than the overall scale $\mu$.

The boundary term $\gNu[\as(\xi)]$ can be extracted by integrating \eq{gNu solution} up to $p_T \le \mu$,
\begin{equation} \label{eq:gNu boundary}
 \theta(\mu) \gNu[\as(\mu)] = \int_{|\pt| \le \mu}\!\del^2\pt\,\gNu(\pt,\mu)
\,.\end{equation}
It corresponds precisely to the noncusp piece of the anomalous dimensions. We expand it as
\begin{equation}
 \gNu(\as) = \sum_{n=0}^\infty \gNuConst{n} \AS^{n+1}
\,,\end{equation}
and the constants $\gNuConst{n}$ are the coefficients of $\delta(\pt)$ in the fixed-order calculation of $\gNu(\pt,\mu)$.

\subsection{Iterative resummation of \texorpdfstring{$\gNu$}{gammaNu}}
\label{sec:gNu iterative}

It is instructive to see how the resummed result for $\gNu(\pt, \mu)$ arises order-by-order in perturbation theory by using a recurrence relation. To obtain the recurrence relation, we expand
\begin{equation}
\gNu(\pt,\mu) = \sum_{n=0}^\infty \gNu^{(n)}(\pt,\mu) \ASmu^{n+1}
\,.\end{equation}
Differentiating this with respect to $\mu$ and plugging it back into \eq{gNu RGE} gives the relation
\begin{equation} \label{eq:gNu recurrence}
\mu \frac{\del\gNuC{n}(\pt,\mu)}{\del\mu} = - 4 \Gamma_n \delta(\pt) + \sum_{m=0}^{n-1} 2 (m+1) \beta_{n-m-1} \gNuC{m}(\pt,\mu)
\,,\end{equation}
where $\Gamma_n$ and $\beta_n$ are the coefficients of the cusp anomalous dimensions and beta function, see \eq{Gammabetacoeffs}.
Applying the integration rule in \eq{integral scale setting} yields the solution
\begin{align} \label{eq:gNu recurrence solution}
 \gNuC{n}(\pt,\mu)
 &= 2 \Gamma_n \PlusLog{0}(\pt,\mu) + \sum_{m=0}^{n-1} 2 (m+1) \beta_{n-m-1} \int\limits_{\mu_0 = p_T\dis}^\mu \hspace{-2ex} \frac{\del\mu'}{\mu'} \gNuC{m}(\pt,\mu')
+ \gNuConst{n} \delta(\pt)
\,.\end{align}
The first two terms follow from integrating the right hand side of \eq{gNu recurrence}, leaving only a pure number times $\delta(\pt)$ as possible boundary term.
This shows explicitly that the noncusp coefficients $\gNuConst{n}$ determine the boundary condition,
which can (and must) be determined in fixed-order perturbation theory, while all plus distributions arise from the $\mu$-evolution. Note that the remaining integral has to be evaluated with distributional scale setting according to \eq{integral scale setting}.

The first few terms following from \eq{gNu recurrence solution} are given by
\begin{align} \label{eq:gNu 3loop}
\gNuC{0}(\pt, \mu)
&= 2 \Gamma_0\, \PlusLog{0}(\pt,\mu)
\,, \nn \\
\gNuC{1}(\pt, \mu)
&= -2 \beta_0 \Gamma_0\, \PlusLog{1}(\pt,\mu) + 2 \Gamma_1\, \PlusLog{0}(\pt,\mu) + \gNuCNC{1}\, \delta(\pt)
\,, \nn \\
\gNuC{2}(\pt, \mu)
&= 2 \beta_0^2 \Gamma_0\, \PlusLog{2}(\pt,\mu)
  - 2 (\beta_1 \Gamma_0 + 2 \beta_0 \Gamma_1 )\, \PlusLog{1}(\pt,\mu)
\nn \\ & \quad
+ 2 (\Gamma_2 - \beta_0 \gNuCNC{1})\, \PlusLog{0}(\pt,\mu) + \gNuCNC{2}\, \delta(\pt)
\,, \nn \\
\gNuC{3}(\pt, \mu)
&= -2 \beta_0^3 \Gamma_0\, \cL_3(\pt,\mu)
 + ( 5 \beta_0 \beta_1 \Gamma_0 + 6 \beta_0^2 \Gamma_1 )\, \cL_2(\pt,\mu)
\nn\\&\quad
 - 2 ( \beta_2 \Gamma_0 + 2 \beta_1 \Gamma_1 + 3 \beta_0 \Gamma_2 - 3 \beta_0^2 \gNuCNC{1} )\, \cL_1(\pt,\mu)
\nn\\&\quad
 + ( 2\Gamma_3 - 2 \beta_1 \gNuCNC{1} - 3\beta_0 \gNuCNC{2} )\, \cL_0(\pt,\mu)
 + \gNuCNC{3}\, \delta(\pt)
\,.\end{align}
Here we already used $\gNuCNC{0} = 0$ to simplify the expressions.
One can easily check that the same result is obtained by expanding the full resummed result in \eq{gNu solution}.

\subsection{Comparison to resummation in Fourier space}
\label{sec:gNu comparison}

The transverse momentum resummation is typically performed in Fourier (impact parameter) space.
The rapidity anomalous dimension in Fourier space satisfies the corresponding differential equation
\begin{equation}
\mu \frac{\del\tgNu(\bt,\mu)}{\del\mu}  = -4 \GammaC[\as(\mu)]
\,.\end{equation}
Here, $\tgNu(\bt,\mu)$ is the Fourier transform of $\gNu(\pt,\mu)$.
It naturally depends on $\ln(b_T^2 \mu^2 / b_0^2)$ with $b_0 = 2 e^{-\gamma_E}$,
such that the resummation with canonical scale choice $\mu_0 = b_0/b_T$ is
\begin{equation}  \label{eq:gNu solution b}
\tgNu(\bt,\mu) = \tgNu[\as(b_0/b_T)] - \int_{b_0/b_T}^\mu\! \frac{\del\mu'}{\mu'}\, 4 \GammaC[\as(\mu')]
\,,\end{equation}
where $\tgNu[\as(b_0/b_T)]$ is the boundary term.

To compare \eq{gNu solution b} to the momentum space solution in \eq{gNu solution}, we need to take the inverse
Fourier transform, which involves integrating $\tgNu(\bt,\mu)$ over the nonperturbative region $1/b_T \lesssim \LQCD$.
It is hence easier to compare at the level of the perturbative reexpansion of $\gNu$.
A recurrence relation similar to \eq{gNu recurrence solution} is easily derived in Fourier space,
\begin{align}  \label{eq:gNu recurrence b solution}
{\tilde\gamma}_\nu^{(n)}(\bt, \mu)
&= -2 \Gamma_n \ln\frac{b_T^2\mu^2}{b_0^2} + 2 \sum_{m=0}^{n-1}(m+1) \beta_{n-m-1} \int_{b_0/b_T}^\mu\! \frac{\del \mu'}{\mu'}\, \gNuC{m}(\bt,\mu') + \gNuConstB{n}
\,.\end{align}
Calculating the first few terms and transforming them to momentum space, we obtain
\allowdisplaybreaks
\begin{align}
\gNuC{0}(\pt, \mu) &= 2 \Gamma_0 \PlusLog{0}(\pt,\mu)
\,, \nn \\
\gNuC{1}(\pt, \mu) &= -2 \beta_0 \Gamma_0 \PlusLog{1}(\pt,\mu) + 2 \Gamma_1 \PlusLog{0}(\pt,\mu) + \gNuConstB{1} \delta(\pt)
\,, \nn \\
\gNuC{2}(\pt, \mu)
&= 2 \beta_0^2 \Gamma_0 \PlusLog{2}(\pt,\mu)
  - 2 (\beta_1 \Gamma_0 + 2 \beta_0 \Gamma_1 )\, \PlusLog{1}(\pt,\mu)
\nn \\ & \quad
+ 2 ( \Gamma_2 - \beta_0 \gNuConstB{1} )\, \PlusLog{0}(\pt,\mu)
  +   \Bigl(\gNuConstB{2} + \frac{8}{3} \zeta_3 \beta_0^2 \Gamma_0 \Bigr) \delta(\pt)
\,, \nn \\
\gNuC{3}(\pt, \mu)
&= -2 \beta_0^3 \Gamma_0\, \cL_3(\pt,\mu)
 + ( 5 \beta_0 \beta_1 \Gamma_0 + 6 \beta_0^2 \Gamma_1 )\, \cL_2(\pt,\mu)
\nn\\&\quad
 - 2 ( \beta_2 \Gamma_0 + 2 \beta_1 \Gamma_1 + 3 \beta_0 \Gamma_2 - 3 \beta_0^2 \gNuConstB{1} )\, \cL_1(\pt,\mu)
\nn\\&\quad
+ ( 2\Gamma_3 - 2 \beta_1 \gNuConstB{1} - 3\beta_0 \gNuConstB{2} - 8 \zeta_3 \beta_0^3 \Gamma_0 )\, \cL_0(\pt,\mu)
\nn\\&\quad
+ \Bigl( \gNuConstB{3} + \frac{20}{3} \zeta_3 \beta_0 \beta_1 \Gamma_0  + 8 \zeta_3 \beta_0^2 \Gamma_1  \Bigr)\, \delta(\pt)
\,.\end{align}
\allowdisplaybreaks[0]
The first two terms agree with \eq{gNu 3loop}, and hence to this order we find
identical noncusp constants, $\gNuConstB{1} = \gNuCNC{1}$.
The differences compared to \eq{gNu 3loop} start at $\cO(\as^3)$, where the $b$-space resummation induces additional terms. This means that the 3-loop and 4-loop noncusp constants are related by
\begin{align}
\gNuConst{2} &= \gNuConstB{2} + \frac{8}{3} \zeta_3 \beta_0^2 \Gamma_0
\,, \nn \\
\gNuConst{3} &= \gNuConstB{3} + \frac{20}{3} \zeta_3 \beta_0 \beta_1 \Gamma_0  + 8 \zeta_3 \beta_0^2 \Gamma_1
\,.\end{align}
In addition, the $\cL_0$ term in $\gNuC{3}$ differs by a contribution $8 \zeta_3 \beta_0^3 \Gamma_0 $,
which is induced by the different boundary term $\gNuConstB{2}$,
which feeds into the logarithmic distribution at higher orders.

Similar additional terms are induced at each higher order.
The reason is that pure $b$-space logarithms $\ln^n(b^2 \mu^2 / b_0^2)$ do not correspond
to pure plus distributions $\cL_{n-1}(\pt,\mu)$, but also induce $\delta(\pt)$ terms
when Fourier transformed back to momentum space.
The implications of this were discussed already in \secs{comparison rge tools}{distr scale setting 2d}.
If $\gNu$ is calculated in full fixed order to $\as^n$, the boundary terms can be extracted up to this order, which also takes into account the differences between $\gNuConst{n}$ and $\gNuConstB{n}$ up to this order.
However, for the unknown higher order terms one would set $\gNuConstB{n} = 0$ in the $b$-space resummation (i.e.\ by setting the boundary condition in $b$-space) and $\gNuConst{n} = 0$ in the momentum-space resummation (i.e.\ by setting the boundary condition in momentum space). This means if one were to use the $b$-space resummation to obtain the momentum-space result this would induce additional boundary terms which then lead to additional subleading terms to all orders compared to the momentum-space resummation, and vice versa.

Even though the inverse Fourier transformation over the full $b$-space result is not possible,
we can express the differences in closed form by relating \eqs{gNu solution}{gNu solution b} through
\begin{align} \label{eq:gNu from bspace}
 \gNu(\pt, \mu) &= \FT^{-1}[\tgNu(\bt, \mu)](\pt) - \biggl[\frac{1}{2 \pi p_T^2} \frac{\del \Delta\gNu[\as(p_T)]}{\del\ln p_T} \biggr]_+^\mu - \delta(\pt)\, \Delta\gNu[\as(\mu)]
\,,\end{align}
where
\begin{align} \label{eq:Delta gNu}
\Delta\gNu[\as(\mu)] &= \int_0^\infty\! \del b_T\,\mu\, J_1(b_T \mu)\, \tgNu(|\bt|, \mu) - \gNu[\as(\mu)]
\,.\end{align}
Here we used that $\tgNu$ only depends on the magnitude of $\bt$, $\tgNu(\bt, \mu) \equiv \tgNu(|\bt|, \mu)$.
The relation is derived by relating the pure $\delta(\pt)$ pieces via \eq{gNu boundary} to each other.
It holds in the sense that the fixed-order expansions of both sides match to all orders.
The $\Delta\gNu$ contains the differences in the boundary condition and the second term in \eq{gNu from bspace} explicitly shows the subleading logarithmic terms contained in the inverse Fourier transform of the $b$-space solution.

\subsection{Turning off resummation using profiles}
\label{sec:gNu profiled}

Having obtained the resummed rapidity anomalous dimension,
we can use it as an illustrative example for the implementation of a profile scale
to smoothly transition between resummation and fixed-order regimes, as discussed
in \sec{profiles}.
Starting from the formal solution
\begin{equation}
 \gNu(\pt,\mu) = \gNu(\pt,\mu_0) - \delta(\pt) \int_{\mu_0}^\mu\! \frac{\del\mu'}{\mu'}\, 4 \GammaC[\as(\mu')] \,,
\end{equation}
the aim is to choose a profile scale $\mu_0(p_T)$ that smoothly transitions from the canonical scale choice $\mu_0 = p_T\dis$ in the resummation regime to the fixed scale $\mu_0 = \mu$ that turns off the resummation.
Note that the profile function can also depend on further variables such as $Q$ or the final $\qt$. Since these dependences do not require distributional scale setting, we suppress them.

As discussed in \sec{profiles}, the $p_T$-dependence of $\mu_0(p_T)$ requires distributional scale setting using \eq{distr scale setting} generalized to a function $\mu_0(p_T)$,
\begin{align} \label{eq:gNu profile 1}
\gNu(\pt,\mu)
= \frac{1}{2\pi p_T} \frac{\del}{\del p_T} \int_{|\kt|\le p_T} \! \del^2\kt \biggl\{
  \gNu^\FO(\kt,\mu_0(p_T)) - \delta(\kt) \int_{\mu_0(p_T)}^\mu\! \frac{\del\mu'}{\mu'}\, 4 \GammaC[\as(\mu')]
 \biggr\}
\,.\end{align}
Here the superscript ``FO'' makes explicit that the boundary term is obtained from a fixed-order calculation.
Focusing for the moment on the second term, which predicts the logarithmic terms at higher orders in $\as$, the derivative can be evaluated [see \eq{cumulant inverse}],
\begin{align} \label{eq:gNu profile}
\gNu(\pt,\mu)
&\supset \biggl[ \frac{1}{2\pi p_T^2} \frac{\del\ln\mu_0(p_T)}{\del\ln p_T} 4 \GammaC[\mu_0(p_T)] \biggr]_+^\mu - \delta(\pt)\int_{\mu_0(\mu)}^\mu\! \frac{\del\mu'}{\mu'}\, 4 \GammaC[\as(\mu')]
\,.\end{align}
This has to be compared to the corresponding term in the canonical solution in \eq{gNu solution},
\begin{align}
\gNu(\pt,\mu)
&\supset \biggl[ \frac{4 \GammaC[\as(p_T)]}{2 \pi p_T^2}  \biggr]_+^\mu
\,.\end{align}
This result is recovered using the canonical scale $\mu_0(p_T) = p_T$, for which the integral in \eq{gNu profile} vanishes. On the other hand, the resummation is turned off by choosing $\mu_0(p_T) = \mu$, for which both terms in \eq{gNu profile} vanish and only the fixed-order term in \eq{gNu profile 1} survives.
In between, the shape of the derivative $\frac{\del\ln\mu_0(p_T)}{\del\ln p_T} \ne 1$ entering in the plus distribution controls how the plus distribution, which is entirely generated by the resummation, is being turned off until it vanishes in the fixed-order regime.

Another purpose of the profile scale is to probe the size of higher-order terms through scale variations. To show concretely how this works, consider choosing $\mu_0(p_T) = 2 p_T$, for which \eq{gNu profile} becomes
\begin{align}
\gNu(\pt,\mu)
&\supset \biggl[ \frac{4 \GammaC[\as(2 p_T)]}{2\pi p_T^2} \biggr]_+^\mu
- \delta(\pt)\int_{2 \mu}^\mu\! \frac{\del\mu'}{\mu'}\, 4 \GammaC[\as(\mu')]
\,.\end{align}
We observe two effects.
On the one hand, the plus distribution now has a different shape in $\pt$ due to $\GammaC[\as(2p_T)]$.
This is reasonable, as all higher-order logarithms in $\gNu$ are resummed into the scale entering the cusp anomalous dimension $\GammaC$, so we expect a variation there.
On the other hand, the $\delta(\pt)$ term does not vanish anymore, and hence probes the constant fixed-order boundary terms.

In conclusion, the distributional scale setting allows one to use profile scales as usual, allowing to smoothly connect the resummation and fixed-order regimes as well as to implement profile scale variations.

\paragraph{Example at lowest order}

As an illustrative example, we consider resumming $\gNu$ at lowest order (i.e.\ LL including $\beta_0$ and $\Gamma_0$)
with a profile scale in conjunction with the full $\cO(\as)$ boundary term,
\begin{equation}
\gNu^\FO(\pt, \mu)
= 2 \Gamma_0 \frac{\as(\mu)}{4\pi} \PlusLog{0}(\pt,\mu)
\,.\end{equation}
(This combination corresponds to a partial NLL result and would typically not arise in practice, but it is useful for illustration.)
From \eq{gNu profile 1}, we obtain
\begin{align}
\gNu^\mathrm{(N)LL}(\pt,\mu)
&= \frac{\Gamma_0}{\pi} \biggl[
\frac{1}{2\pi p_T} \frac{\del}{\del p_T}
\biggl\{ \as[\mu_0(p_T)] \ln\frac{p_T}{\mu_0(p_T)}
 - \int_{\mu_0(p_T)}^\mu \frac{\del\mu'}{\mu'} \as(\mu') \biggr\} \biggr]_+^\mu
\nn\\ & \quad
 + \frac{\Gamma_0}{\pi} \delta(\pt) \biggl\{ \as[\mu_0(\mu)] \ln\frac{\mu}{\mu_0(\mu)} - \int_{\mu_0(\mu)}^\mu \frac{\del\mu'}{\mu'} \as(\mu')\biggr\}
\,.\end{align}
The first term in each of the curly brackets is from the fixed-order boundary condition and the second term is from the LL $\mu$ evolution. Expanding the result in $\as(\mu)$, we obtain
\begin{align}
\gamma_\nu^{\mathrm{(N)LL} (0)}(\pt, \mu)
&= 2 \Gamma_0\, \PlusLog{0}(\pt, \mu) = \gNuC{0}(\pt, \mu)
\,, \nn \\
\gamma_\nu^{\mathrm{(N)LL} (1)}(\pt, \mu)
 &= - 2 \beta_0 \Gamma_0 \frac{1}{\pi\mu^2} \biggl[ \frac{\mu^2}{\pt^2} \ln\frac{\mu_0(p_T)^2}{\mu^2} \biggr]_+^\mu
 + 2 \beta_0 \Gamma_0 \frac{1}{\pi\mu^2} \biggl[ \frac{\mu^2}{\pt^2} \ln\frac{\mu_0(p_T)^2}{p_T^2} \frac{\del\ln\mu_0(p_T)}{\del\ln p_T} \biggr]_+^\mu
 \nn \\&\quad + 4 \Gamma_0 \beta_0 \ln^2\frac{\mu}{\mu_0(\mu)} \delta(\pt)
\,.\end{align}
As discussed in generality above, by including the full $\cO(\as)$ boundary condition, the $\cO(\as)$ result $\gNuC{0}$ is exactly reproduced independently of the choice of $\mu_0(p_T)$. The two-loop term
has to be compared to the full fixed-order result from \eq{gNu 3loop},
\begin{align}
\gNuC{1}(\pt, \mu)
&= -2 \beta_0 \Gamma_0 \frac{1}{\pi\mu^2} \biggl[ \frac{\mu^2}{\pt^2} \ln\frac{p_T^2}{\mu^2} \biggr]_+^\mu + 2 \Gamma_1 \frac{1}{\pi\mu^2} \biggl[ \frac{\mu^2}{\pt^2} \biggr]_+^\mu + \gNuCNC{1} \delta(\pt)
\,.\end{align}
With the fixed-order scale $\mu_0(p_T)=\mu$, the two-loop term $\gamma_\nu^{\mathrm{(N)LL} (0)}$ completely vanishes, leaving only the fixed-order term $\gamma_\nu^{(0)}$.
With the canonical resummation scale $\mu_0(p_T)=p_T$, the first term in $\gamma_\nu^{\mathrm{(N)LL} (0)}$ exactly reproduces the leading-logarithmic two-loop term $\sim \beta_0\Gamma_0\PlusLog{1}$, while the other terms vanish. Varying $\mu_0(p_T)$ induces higher-order terms with precisely the structure of the formally higher-order $\Gamma_1$ and $\gNuCNC{1}$ terms. Furthermore, the $p_T$-dependence of the variation allows one to distinguish and separately probe the size of the higher-order logarithmic and higher fixed-order boundary pieces.

\subsection{Nonperturbative modeling with the moment expansion}
\label{sec:gNu nonpert}

It is well known from the resummation in Fourier space that the rapidity evolution
kernel becomes intrinsically nonperturbative at $1/b \ll \LQCD$~\cite{Collins:1981uk, Collins:1981va, Collins:1984kg}.
In momentum space this corresponds to the fact that the resummed result for
$\gNu(\pt,\mu)$ in \eq{gNu solution} explicitly depends on $\as(p_T)$, which means
that it becomes nonperturbative for $p_T \lesssim \LQCD$.
This nonperturbative region is necessarily probed
in the rapidity RGE of the soft function, see e.g.~\eqs{RGE nuS}{naive V}, since
$\gNu(\kt, \mu)$ is integrated over all $\kt$.

Although we will show in \sec{perturbativity} that the final result for the
momentum convolutions is actually perturbative up to power corrections $\cO(\LQCD^2/p_T^2)$,
it is still important to properly handle and isolate the nonperturbative contributions
to $\gNu$. To do so, we can write the true all-order result for $\gNu$ as
\begin{align} \label{eq:gNunonpert}
\gNu(\pt,\mu)
&= \biggl[ \frac{4 \GammaC[\as(\mu_0(p_T))]}{2 \pi p_T^2} \biggr]_+^\mu
 + \biggl[ \frac{1}{2 \pi p_T^2} \frac{\del \gNu[\as(\mu_0(p_T))]}{\del\ln p_T} \biggr]_+^\xi
 + \delta(\pt)\, \gNu[\as(\mu_0(\xi))]
\nn \\  &\quad
+ \gNuNP(\pt)
\,.\end{align}
Here, we have replaced the fixed canonical scale $\mu_0 = p_T\dis$ in the resummed result
in \eq{gNu solution} by a cutoff function $\mu_0(p_T)$, which is equal to
$\mu_0(p_T) = p_T$ for $p_T \gtrsim \LQCD$ but cuts off and approaches some constant (perturbative) value for $p_T \lesssim \LQCD$.
The difference to the true result is absorbed into the nonperturbative contribution $\gNuNP(\pt)$.
Since the resummed perturbative result is now only ever evaluated at perturbative scales, all nonperturbative
contributions must be contained in $\gNuNP(\pt)$.
The precise choice of $\mu_0(p_T)$ here corresponds to a perturbative scheme dependence
in the definition of $\gNuNP(\pt)$, which cancels on the right-hand sice in \eq{gNunonpert}
(and which we suppress in our notation). At the same time, $\gNuNP(\pt)$ is $\mu$ independent
because the (perturbative) $\mu$ dependence is by definition carried by the resummed perturbative
contributions.

Letting $\mu_0(p_T) = p_T$ in the perturbative regime, $\gNuNP(\pt)$ will have support of order $\LQCD$.
Hence we can expand it in its moments,
\begin{equation} \label{eq:moment expansion}
 \gNuNP(\pt) = \biggl( \sum_{n=1}^\infty \Omega_n \Delta_{\pt}^n \biggr)\, \delta(\pt)
 \equiv \gNuNP[\Delta_{\pt}]\, \delta(\pt)
\,,\end{equation}
where $\Delta_{\pt} = \partial^2/\partial^2 p_x + \partial^2/\partial^2 p_y$ is the Laplace operator.
Since $\gNu$ is known to be azimuthaly symmetric, no other operators, such as e.g.\ linear operators, may arise.
The associated $\Omega_n$ coefficients are nonperturbative parameters, which have to be extracted from data.
Like $\gNuNP(\pt)$ itself, they are $\mu$ independent but scheme dependent.
This is reminiscent of the moment expansions to treat the nonperturbative corrections
in one-dimensional soft functions~\cite{Hoang:2007vb, Ligeti:2008ac}.

The moment expansion \eq{moment expansion} becomes more intuitive upon Fourier transformation.
In impact parameter space, it turns into a simple polynomial in $-b_T^2$,
\begin{equation}
 {\tilde \gamma}_\nu^\text{(np)}(\bt) = \sum_{n=1}^\infty \Omega_n (-b_T^2)^n = \gNuNP[-b_T^2]
\,.\end{equation}
The analysis of \mycite{Scimemi:2016ffw} found that the perturbative result for
${\tilde \gamma}_\nu(\bt, \mu)$ has a leading renormalon contribution which scales as $b_T^2$.
This is consistent with the fact that the first nonperturbative moment $\Omega_1$ scales with $\Delta_{\pt}$
or $b_T^2$, and the renormalon in the perturbative series should be precisely cancelled by
a corresponding renormalon in $\Omega_1$.
We expect that the above cutoff definition can be used to provide a renormalon-free scheme definition
for $\Omega_1$ in momentum space and presumably $\gNuNP(\pt)$ as a whole.
It would be interesting to investigate this in more detail in the future.

To compare the above moment expansion to the literature, consider the formal solution of the rapidity RGE in \eq{RGE nuS}
in impact parameter space,
\begin{equation}
\tilde S(\bt, \mu, \nu) = \tilde S(\bt, \mu, \nu_0) \exp\biggl[\ln\frac{\nu}{\nu_0} \tgNu(b_T, \mu) \biggr]
\,.\end{equation}
Including both the perturbative and nonperturbative contributions to $\tgNu$, this yields
\begin{equation}
\tilde S(\bt, \mu, \nu)
= \tilde S(\bt, \mu, \nu_0) \exp\Bigl[\ln\frac{\nu}{\nu_0} \tgNu^\text{(pert)} (b_T, \mu)\Bigr]
\exp\Bigl[-\Omega_1\ln\frac{\nu}{\nu_0} b_T^2  + \cdots \Bigr]
\,,\end{equation}
where $\tgNu^\text{(pert)} (b_T, \mu)$ is the Fourier transform of the resummed perturbative contribution.
Hence, the momentum-space resummation reproduces and confirms the often-used procedure in the literature to model nonperturbative effects using a Gaussian factor in impact parameter space, which is also motivated by renormalon analyses~\cite{Korchemsky:1994is, Scimemi:2016ffw}.
In practice, a variety of nonperturbative models have been suggested, see e.g.~refs.~\cite{Echevarria:2012pw,Collins:2014jpa} for recent studies.
The above also confirms that the nonperturbative correction scales with a rapidity logarithm~\cite{Collins:1981va, Becher:2013iya}.

\section{Resummation of soft and beam functions}
\label{sec:soft func}

In the previous section we have solved the RGE for the rapidity anomalous dimension $\gNu$,
which already illustrated some of the key features of the distributional scale setting in a
nontrivial setting. In this section, we will solve the RGEs for both beam and soft function
and derive the correct momentum-space evolution to predict their complete distributional logarithmic
structure.
The implications on the full $q_T$ spectrum are discussed in \sec{implementation}.

\subsection{Soft function}

The soft function obeys the coupled system of renormalization group equations, \eqs{RGE muS}{RGE nuS},
\begin{align}
 \mu \frac{\del S(\pt,\mu,\nu)}{\del\mu} &= \gMuS(\mu,\nu) S(\pt,\mu,\nu)
\,, \\
 \nu \frac{\del S(\pt,\mu,\nu)}{\del\nu} &= \int\! \del^2\kt\,\gNu(\kt,\mu)\, S(\pt - \kt,\mu,\nu)
\,,\end{align}
where the $\nu$-RGE involves a two-dimensional convolution, which
is the main source of complications, as discussed in \sec{TMD convolutions}.

\subsubsection{Iterative solution}
We first derive an iterative solution order by order in $\as$.
This requires the expansions
\begin{align}
 S(\pt,\mu,\nu) &= \sum_{n=0}^\infty S^{(n)}(\pt,\mu,\nu) \ASmu^n
\,, \\
 \gMuS(\mu,\nu) &= \sum_{n=0}^\infty \Bigl( 4 \Gamma_n \ln\frac{\mu}{\nu} + \gMuSCNC{n} \Bigr) \ASmu^{n+1}
\,, \\
 \gNu(\pt,\mu)  &= \sum_{n=0}^\infty \gNuC{n}(\pt,\mu) \ASmu^{n+1}
\,,\end{align}
where the $\gNuC{n}$ were derived in \sec{gNu iterative}.
The two RGEs then become
\begin{align}
\label{eq:mu RGE soft iter}
 \mu \frac{\del S^{(n)}(\pt,\mu,\nu)}{\del\mu} &= \sum_{m=0}^{n-1} \Bigl(  4 \Gamma_{n-m-1} \ln\frac{\mu}{\nu} + \gMuSCNC{n-m-1} + 2 m \beta_{n-m-1} \Bigr) S^{(m)}(\pt,\mu,\nu) \,, \\
\label{eq:nu RGE soft iter}
 \nu \frac{\del S^{(n)}(\pt,\mu,\nu)}{\del\nu} &= \sum_{m=0}^{n-1} \bigl(\gNuC{n-m-1} \otimes S^{(m)}\bigr)(\pt,\mu,\nu) \,.
\end{align}
These equations allow to determine the $\as^n$ coefficient $S^{(n)}$ from the lower order terms $S^{(m)}$ with $m < n$.
Solving first the $\nu$-RGE in \eq{nu RGE soft iter}, we obtain the formal solution
\begin{align}
 S^{(n)}(\pt,\mu,\nu) = S^{(n)}(\pt,\mu,\nu_0) + \int_{\nu_0}^\nu\! \frac{\del\nu'}{\nu'}\, \sum_{m=0}^{n-1} \gNuC{n-m-1}(\pt,\mu) \otimes S^{(m)}(\pt,\mu,\nu') \,.
\end{align}
The missing boundary term $S^{(n)}(\pt,\mu,\nu_0)$ is deduced from the $\mu$-RGE in \eq{mu RGE soft iter},
\begin{align}
S^{(n)}(\pt,\mu,\nu_0)
 &= S^{(n)}(\pt,\mu_0,\nu_0)  \\&\quad+ \int_{\mu_0}^\mu\! \frac{\del\mu'}{\mu'} \sum_{m=0}^{n-1}  \left(  4 \Gamma_{n-m-1} \ln\frac{\mu'}{\nu_0} + \gMuSCNC{n-m-1} + 2 m \beta_{n-m-1} \right) S^{(m)}(\pt,\mu',\nu_0) \,. \nn
\end{align}
The remaining boundary term $S^{(n)}(\pt,\mu_0,\nu_0)$ can only contain logarithms $\ln(\mu_0/p_T)$ or $\ln(\nu_0/p_T)$ or any combination of these. In particular, these logarithms can be hidden inside the boundary condition of plus distributions $\PlusLog{n}(\pt,\mu), \PlusLog{n}(\pt,\nu)$.
To eliminate them, we apply \eq{distr scale setting} to set $\mu_0 = \nu_0 = p_T\dis$.
The remaining pure fixed-order boundary term must then be proportional to $\delta(\pt)$, and we obtain
\begin{align}
 &S^{(n)}(\pt,\mu,\nu)
 = S_n \delta(\pt) \nn \\
 &+ \int_{p_T\dis}^\mu\! \frac{\del\mu'}{\mu'} \sum_{m=0}^{n-1} \Bigl(  4 \Gamma_{n-m-1} \ln\frac{\mu'}{\nu_0} + \gMuSCNC{n-m-1} + 2 m \beta_{n-m-1} \Bigr)   S^{(m)}(\pt,\mu',\nu_0) \biggr|_{\nu_0 = p_T\dis} \nn \\
 &+ \int_{p_T\dis}^\nu\! \frac{\del\nu'}{\nu'} \sum_{m=0}^{n-1} \bigl(\gNuC{n-m-1}\otimes S^{(m)} \bigr)(\pt,\mu,\nu') \,.
\label{eq:soft iterative solution}
\end{align}
In the $\mu$-integral, both distributional scale settings should be performed in one step, $\mu_0 = \nu_0 = p_T\dis$.
By construction, this solution fulfills both RGEs and fully resums all logarithmic distributions.
We have explicitly verified this solution by predicting the structure of the soft function through $\cO(\as^6)$ and comparing with the results obtained from impact parameter space,
where the distributions become simple functions and hence the complication of distributional scale setting does not arise.
As for $\gNu$, the constant pieces differ in general, i.e.\ $S_n \ne \tilde S_n$,
thereby also inducing different logarithms at higher orders.
Upon making the appropriate identifications between $\tilde S_n$ and $S_n$, exact agreement between
both iterative solutions is obtained.
This shows that the distributional scale setting is well-defined for the soft function, even though it involves two distinct scales.
For illustration, the fixed-order expansion of the soft function is given through
$\cO(\as^3)$ in appendix \ref{app:soft func}, including the relation between the
boundary terms $S_n$ and $\tilde S_n$.

For completeness, we also give the iterative solution when first solving the $\mu$-RGE and then the $\nu$-RGE:
\begin{align}
 &S^{(n)}(\pt,\mu,\nu) = S_n \delta(\pt) \nn \\
 &\quad+ \int_{p_T\dis}^\mu\! \frac{\del\mu'}{\mu'} \sum_{m=0}^{n-1}  \Bigl(  4 \Gamma_{n-m-1} \ln\frac{\mu'}{\nu} + \gMuSCNC{n-m-1} + 2 m \beta_{n-m-1} \Bigr) S^{(m)}(\pt,\mu',\nu) \nn\\
 &\quad+ \int_{p_T\dis}^\nu\! \frac{\del\nu'}{\nu'} \sum_{m=0}^{n-1} \bigl(\gNuC{n-m-1} \otimes S^{(m)} \bigr)(\pt,p_T\dis,\nu') \,.
\label{eq:soft iterative solution 2}
\end{align}

Equations~\eqref{eq:soft iterative solution}~and~\eqref{eq:soft iterative solution 2} are useful to illustrate again the correct scale setting for distributions.
Firstly, the lower order terms $S^{(m)}$ feeding into $S^{(n)}$ are themselves distributions with boundaries $\mu$ or $\nu$.
Hence integrating over $\mu$ and $\nu$ with starting scale $p_T$ requires distributional scale setting, which we apply according to \eq{distr scale setting}.
Secondly, we see that it is crucial to set $\nu_0 = p_T\dis$ at each order rather than keeping $\nu_0$ arbitrary and setting $\nu_0 = p_T$ only at the very end.
The reason lies in the convolution term in \eqs{soft iterative solution}{soft iterative solution 2}, $(\gNuC{n-m-1} \otimes S^{(m)})(\pt,\mu,\nu')$:
The convolution depends on whether the integrand contains a formal scale $\nu_0$ or whether it is set to the convolution variable $\kt$.
In contrast, multiplicative RGEs are not sensitive to this, such that the scale can be set at the very end.

\subsubsection{Solution in closed form}
\label{sec:soft func all order}

We now proceed to derive a closed form of the solution. We start by solving the
$\mu$-RGE in \eqs{RGE muS}{gMuS}
at $\nu_0=p_T\dis$.
The solution is straightforwardly obtained using the technique of \sec{rge solution 2}.
The formal solution
\begin{equation}
 S(\pt,\mu,\nu_0) = S(\pt,\mu_0,\nu_0) \exp\biggl[ \int_{\mu_0}^\mu \frac{\del\mu'}{\mu'} \gMuS(\mu',\nu_0) \biggr]
\end{equation}
only requires to set $\mu_0 = \nu_0 = p_T\dis$, yielding%
\footnote{To allow scale variations, one could of course choose profile functions $\mu_0 = \mu_0(p_T)\dis, \nu_0 = \nu_0(p_T)\dis$.}
\begin{align} \label{eq:soft mu evolution}
 S(\pt,\mu,\nu_0 = p_T\dis)
 &= \frac{1}{2\pi p_T} \frac{\del}{\del p_T} \theta(p_T) \cS[\as(p_T)] \exp\biggl[ \int_{p_T}^\mu \frac{\del\mu'}{\mu'} \gMuS(\mu',p_T) \biggr]
\\
 &= \delta(\pt) \cS[\as(\mu)]+ \biggl[ \frac{1}{2\pi p_T} \frac{\del}{\del p_T} \cS[\as(p_T)] \exp\biggl\{ \int_{p_T}^\mu\! \frac{\del\mu'}{\mu'}\, \gMuS(\mu',p_T) \biggr\} \biggr]_+^\mu
\,.\nn\end{align}
The plus distribution contains the exponentiated Sudakov double logarithm at $\nu_0 = p_T\dis$ as dictated by the RGE.
The pure boundary term, which can only depend on $\as$, is defined as
\begin{equation}
 \theta(p_T) \cS[\as(p_T)] = \int_{|\kt| \le p_T} \!\del^2\kt\, S(\kt, \mu = p_T, \nu = p_T)
\,.\end{equation}
and has the perturbative expansion
\begin{equation}
 \cS(\as) = \sum_{n=0}^\infty S_n \AS^{n}
\,.\end{equation}
The constants $S_n$ can be obtained as coefficients of $\delta(\pt)$ in the fixed-order calculation.
More generally, the soft function at the canonical distributional scales is given in terms of this boundary term by
\begin{align}
 S(\pt,p_T\dis,p_T\dis) &= \frac{1}{2\pi p_T} \frac{\del}{\del p_T} \theta(p_T) \cS[\as(p_T)]
\nn \\
 &= \delta(\pt) \cS[\as(\xi)] + \biggl[\frac{1}{2\pi p_T} \frac{\del}{\del p_T} \cS[\as(p_T)] \biggr]_+^\xi
\,,\end{align}
where the $\xi$ dependence exactly cancels.

In the next step, the $\nu$-RGE in \eq{RGE nuS} has to be solved, which is more complicated.
Inspired by the iterative solution, we expand the soft function as
\begin{equation}
 S(\pt,\mu,\nu) = \sum_{n=0}^\infty S^{[n]}(\pt,\mu,\nu)
\,,\end{equation}
where the $S^{[n]}$ correspond to the soft function including $n$ contributions from $\gNu$.
More intuitively, $S^{[n]}$ is the piece of the soft function originating from $n$ real emissions.
Since $\gNu$ describes a single real emission, the rapidity RGE becomes
\begin{align}
 \nu \frac{\del S(\pt,\mu,\nu)}{\del\nu}
&= \sum_{n=0}^\infty \nu\frac{\del}{\del\nu} S^{[n]}(\pt,\mu,\nu)
= \sum_{n=0}^\infty (\gNu \otimes S^{[n]})(\pt,\mu,\nu) \nn \\
&= \sum_{n=1}^\infty (\gNu \otimes S^{[n-1]})(\pt,\mu,\nu)
\,.\end{align}
Hence the $S^{[n]}$ are determined through the recursive RGE
\begin{align}
 \nu \frac{\del}{\del\nu} S^{[n]}(\pt,\mu,\nu) &= (\gNu \otimes  S^{[n-1]})(\pt,\mu,\nu)
\,, \quad
 \nu \frac{\del}{\del\nu} S^{[0]}(\pt,\mu,\nu) = 0
\,.\end{align}
The solution at order $n$ is
\begin{equation}
 S^{[n]}(\pt,\mu,\nu) = S^{[n]}(\pt,\mu,p_T\dis) + \int_{p_T\dis}^\nu \!\frac{\del\nu'}{\nu'} (\gNu \otimes S^{[n-1]})(\pt,\mu,\nu')
\,,\end{equation}
where it is now quite intuitive that $\nu_0$ should be set to $p_T\dis$ and not to the overall $q_T$.
Iterating this, the only leftover boundary term is precisely $S(\pt,\mu,p_T\dis) = \sum_n S^{[n]}(\pt,\mu,p_T\dis)$,
and we can write the formal all-order solution as
\begin{align} \label{eq:soft nu evolution}
 S(\pt,\mu,\nu)
&= S(\pt,\mu,p_T\dis) + \int_{p_T\dis}^\nu\! \frac{\del\nu_1}{\nu_1} \int\!\del^2 \ki{1}\, \gNu(\pt - \ki{1},\mu) S(\ki{1},\mu,k_1\dis)
\nn \\ &\quad
+ \int_{p_T\dis}^\nu\! \frac{\del\nu_1}{\nu_1} \!\int\!\del^2 \ki{1}\, \gNu(\pt - \ki{1},\mu) \!\int_{k_1\dis}^{\nu_1}\! \frac{\del\nu_2}{\nu_2} \!\int\!\del^2 \ki{2}\, \gNu(\ki{1} - \ki{2},\mu) S(\ki{2},\mu,k_2\dis)
\nn \\ &\quad
+ \cdots
\nn \\
&= S(\pt,\mu,p_T\dis) + \sum_{n=1}^\infty \Biggl[\prod_{i=1}^n \int_{k_{i-1}\dis}^{\nu_{i-1}}\! \frac{\del\nu_i}{\nu_i} \int\!\del^2 \ki{i}\, \gNu(\ki{i-1} - \ki{i},\mu) \Biggr] S(\ki{n},\mu,k_n\dis)
\,,\end{align}
where in the last line $k_0 \equiv p_T$ and $\nu_{0} \equiv \nu$.
All $\nu_i$ integrals have to be understood according to \eq{integral scale setting}.

From the explicit form of the first few terms, it is easy to see that \eq{soft nu evolution}
fulfills the rapity RGE and that by choosing the overall $\nu = p_T\dis$ reproduces
the ($\mu$-evolved) boundary condition $S(\pt,\mu,p_T\dis)$.
Note that although both $\gNu$ and $S$ will be probed at nonperturbative $\as$ in the convolutions of \eq{soft nu evolution}, the result is actually perturbative up to corrections $\cO(\LQCD^2/p_T^2)$, as will be discussed in \sec{perturbativity}.
For practical purposes it is nevertheless necessary to use a nonperturbative modelling, as already discussed for $\gNu$ in \sec{gNu nonpert}.
To allow scale variations, one would replace all $k_i\dis$ by a profile functions $\nu_0(k_i)\dis$.

Together, \eqs{soft mu evolution}{soft nu evolution} completely predict
the logarithmic structure of the soft function to all orders.
Crucially, the canonical-scale boundary condition for the solution is $S(\pt,p_T\dis,p_T\dis)$,
which can be reliably calculated in a pure fixed-order calculation.
The iterative structure of the rapidity evolution \eq{soft nu evolution}
ensures that the correct rapidity logarithms $\ln(\nu/|\ki{i}|)$ evolving
each emission to the overall rapidity scale are resummed.

Once the soft function has been evolved to some scale $\nu_S$ using this method,
it can be evolved further to a different overall scale $\nu$ using the simple kernel of \eq{naive V},
\begin{align} \label{eq:soft nu shift}
 S(\pt, \mu, \nu) &= \int\! \del^2\kt\, V_S(\pt - \kt,\mu, \nu, \nu_S)\, S(\kt, \mu, \nu_S)
\,,\nn\\
 V_S(\pt,\mu,\nu,\nu_S) &= \delta(\pt)  + \sum_{n=1}^\infty \frac{1}{n!} \ln^n\!\frac{\nu}{\nu_S}\, (\gNu \otimes^n)(\pt,\mu)
\,,\end{align}
where we introduce the notation $V_S$ for the evolution kernel when used in this way for continuing the evolution.
This simple structure immediately emerges from \eq{soft nu evolution} when replacing all
$\nu_0 = k_i\dis$ by a common scale $\nu_S$, such that all $\nu$-integrals factor out of
the convolutions. This shows that \eq{soft nu shift} is indeed sufficient to continue the
$\nu$ evolution in \eq{soft nu evolution} from the overall scale $\nu_S$ to a new overall scale $\nu$.

\subsubsection{Comparison to ``naive'' scale setting}
\label{sec:soft exact vs naive}

Having obtained a solution correctly minimizing the boundary term distributionally,
we can turn back to the discussion of \sec{TMD convolutions}.
There we had argued that the naive solution, \eqs{naive solution S}{naive V}, leads to wrong
predictions, because energetic emissions are artificially enhanced by unphysical logarithms
$\ln(Q/q_T)$. Instead, each convolution should be dressed with its proper rapidity logarithm
$\ln(Q/|\ki{i}|)$, which is precisely what is happening in \eq{soft nu evolution}.

To see in more detail the different treatment of these rapidity logarithms,
we now compare the iterative solution to the naive evolution kernel \eq{naive V}.
To do this, we simply assume the trivial boundary condition $S(\kt,\mu,k_T\dis) = \delta(\kt)$.
This of course means that the following result is not the properly resummed soft function,
but it allows to disentangle effects from rapidity and $\mu$ evolution,
since the latter are now neglected in the boundary term.
Furthermore, we neglect running of $\as$ for simplicity to keep the following formula
as compact as possible, and work only to LL accuracy.
In this toy model, the rapidity anomalous dimension is given by
\begin{equation}
 \gNu(\pt,\mu) = 2 \GammaC \cL_0(\pt,\mu)
\,,\end{equation}
where at LL we can ignore all constant pieces.
To keep track of only the first two emissions, we evaluate the resummed toy soft function
only up to $\cO(\GammaC^2)$. We then obtain from \eq{soft nu evolution}
\begin{align} \label{eq:soft predicted 1}
 S^\text{(toy)}(\pt,\mu,\nu)
 &= \delta(\pt) + \int_{p_T\dis}^\nu\! \frac{\del\nu_1}{\nu_1} \gNu(\pt,\mu) \nn \\ &\quad
 + \int_{p_T\dis}^\nu\! \frac{\del\nu_1}{\nu_1} \int\del^2 \ki{1}\, \gNu(\pt - \ki{1},\mu) \int_{k_1\dis}^{\nu_1}\! \frac{\del\nu_2}{\nu_2} \gNu(\ki{1},\mu)
 +\cdots
\nn \\
 &=
 \delta(\pt)
 -2 \frac{\GammaC}{\pi \mu^2} \biggl[ \frac{\mu^2}{\pt^2} \ln\frac{\pt^2}{\mu\nu} \biggr]_+^\mu
 + \frac{2 \GammaC^2}{\pi \mu^2} \biggl[ \frac{\mu^2}{\pt^2} \ln\frac{\pt^2}{\mu^2} \ln\frac{\pt^2}{\nu^2} \ln\frac{\pt^2}{\mu\nu} \biggr]_+^\mu \nn \\ &\quad
 + 4 \zeta_3 \GammaC^2 \cL_0(\pt,\nu)
 +\cO(\GammaC^3)
\,.\end{align}

For comparison, we calculate the soft function using the naive resummation \eqs{naive solution S}{naive V}, where the starting scale $\nu_0$ is kept arbitrary.
Similarly to above, we set the boundary term to $S(\pt,\mu,\nu_0) = \delta(\pt)$,
and evaluate the resummed soft function only at LL without $\as$ running.
We find
\begin{align} \label{eq:soft predicted 2}
 S^\text{(toy)}(\pt, \mu, \nu)
 &= \delta(\pt) + \ln\!\frac{\nu}{\nu_0} \gNu(\pt,\mu) + \frac{1}{2} \ln^2\!\frac{\nu}{\nu_0} \int\! \del^2\kt\, \gNu(\pt-\kt,\mu) \gNu(\kt,\mu)
\nn \\
 &= \delta(\pt) + \frac{\GammaC}{\pi \mu^2} \ln\frac{\nu^2}{\nu_0^2} \biggl[ \frac{\mu^2}{\pt^2} \biggr]_+^\mu
 + \frac{\GammaC^2}{\pi \mu^2} \ln^2\frac{\nu^2}{\nu_0^2} \biggl[ \frac{\mu^2}{\pt^2} \ln\frac{\pt^2}{\mu^2} \biggr]_+^\mu
\,.\end{align}
Comparing the two results \eqs{soft predicted 1}{soft predicted 2},
we observe a very different logarithmic structure.
In particular, in the solution
\eq{soft predicted 1} derived from the known exact solution, all rapidity logs
are necessarily sensitive to the soft function $\pt$. This becomes crucial when
the soft function is inserted into further convolutions, for example to predict
the higher order terms of $S^\text{(toy)}$ itself.

\subsection{Beam functions}
\label{sec:beam func}

The beam function RGE in \eqs{RGE muB}{RGE nuB}
\begin{align}
 \mu \frac{\del B(\omega, \pt,\mu,\nu)}{\del\mu} &= \gMuB(\omega, \mu,\nu)\, B(\omega, \pt, \mu, \nu)
\,,\\
 \nu \frac{\del B(\omega, \pt, \mu, \nu)}{\del\nu} &= - \frac{1}{2} \int\! \del^2\kt\, \gNu(\kt,\mu)\, B(\omega, \pt - \kt, \mu, \nu)
\end{align}
are very similar to the soft function RGEs,
and hence both iterative solution \eqs{soft iterative solution}{soft iterative solution 2}
as well as the all-order solutions \eqs{soft mu evolution}{soft nu evolution}
can be applied to the beam function upon proper replacement of anomalous dimensions and boundary terms.

The main difference to the soft function is that the canonical rapidity scale is
$\nu_B = \omega$ rather than $\nu_B = p_T\dis$,
and hence does not require distributional scale setting.
The solution of the $\mu$-RGE at the canonical $\nu$-scale is
\begin{align} \label{eq:beam mu evolution}
B(\omega,\pt,\mu,\nu_B = \omega)
&= \delta(\pt)\, \cB(\omega,\mu)
\\&\quad
+ \biggl[ \frac{1}{2\pi p_T} \frac{\del}{\del p_T} \cB(\omega,p_T)
\exp\biggl\{\int_{p_T}^\mu\! \frac{\del\mu'}{\mu'}\, \gMuB(\omega,\mu',\nu_B=\omega) \biggr\} \biggr]_+^\mu
\,.\nn\end{align}
The boundary term is defined as
\begin{equation} \label{eq:beam boundary}
 \theta(p_T) \cB(\omega, p_T) = \int_{|\kt| \le p_T} \!\del^2\kt\, B(\kt, \mu=p_T, \nu_B = \omega)
\end{equation}
and is expanded as
\begin{equation}
 \cB(\omega,\mu) = \sum_{n=0}^\infty B_n(\omega,\mu) \ASmu^{n}
\,.\end{equation}
It is more complicated than for the soft function, because the beam functions are further matched onto PDFs by an operator product expansion (see \mycite{Luebbert:2016itl} for more details),
\begin{equation} \label{eq:beam OPE}
 B_a(\omega, \pt,\mu,\nu) = \sum_i \int\! \frac{\del z}{z}\, \cI_{ai}(z, \pt,\mu,\nu)\, f_i\Bigl(\frac{\omega/E_{\rm cm}}{z}, \mu\Bigr)
\,.\end{equation}
Here $B_a$ is the beam function for flavor $a$ and $f_i$ is the PDF for flavor $i$.
(We ignore for the moment that gluon beam functions furthermore have a helicity structure.)
While applying \eq{beam boundary} to \eq{beam OPE} eliminates all distributions in $\pt$ in the $\cI_{ai}$ matching kernels,
the boundary term still involves a convolution in $z$ with the PDF.

The rapidity evolution kernel is obtained from \eq{soft nu evolution} by replacing
all $\nu_0 = k_i\dis$ by $\nu_0 = \nu_B \sim \omega$. Since this scale is independent of $p_T$,
the $\nu$-integrals can be pulled out as
\begin{align} \label{eq:beam nu evolution}
 B(\omega,\pt,\mu,\nu) &=  \int\del^2\kt\, V_B(\kt,\nu,\nu_B)\, B(\omega,\pt-\kt,\mu,\nu_B)
\,,\nn\\
 V_B(\kt,\nu,\nu_B) &=  \delta(\kt) + \sum_{n=0}^\infty \frac{1}{n!} \biggl(-\frac{1}{2}  \ln\frac{\nu}{\nu_B}\biggr)^n (\gNu \otimes^n)(\kt,\mu)
\,,\end{align}
where the $-1/2$ arises from $\gamma_{\nu,B} = -\gNu/2$.
Hence, for the beam function we recover the simple exponential in convolution space
with the fixed-order boundary condition $B(\omega,\pt,\mu,\nu_B)$.
This allows for a simple scale variation by choosing $\nu_B$ to be a suitable profile.

\subsection{Perturbativity of convolutions}
\label{sec:perturbativity}

\begin{figure}[t!]
 \centering
 \includegraphics[width=0.5\textwidth]{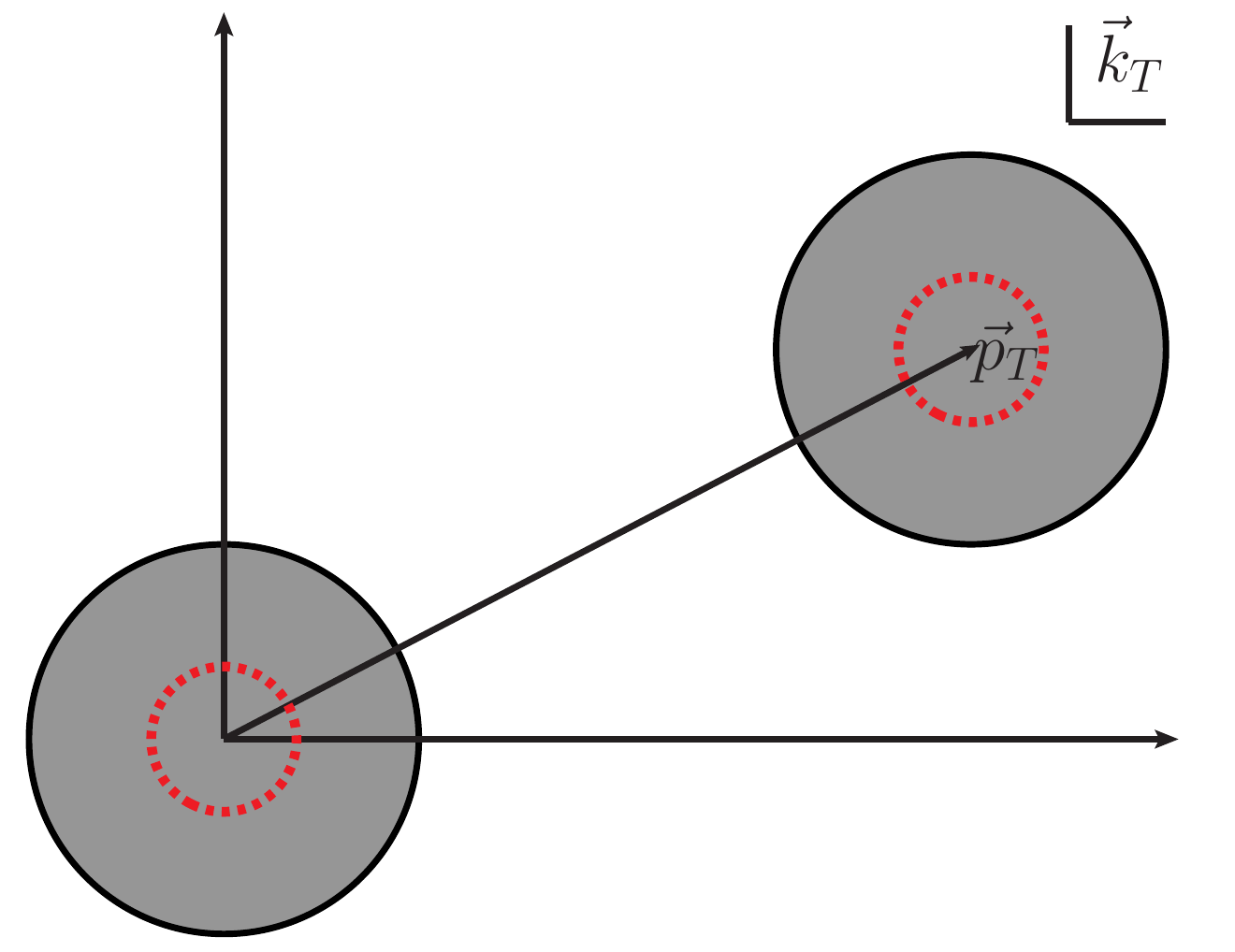}
 \caption{Illustration of division of the integration range of $\kt$ in discs around the two singularities at $\vec 0$ and $\vec p_T$.
          The grey discs represent the disc radius $p_T \gg \Lambda \gg \LQCD$.
          The red, dashed line illustrates the location of the Landau poles at $|\kt| = \LQCD$ and $|\pt - \kt| = \LQCD$.}
 \label{fig:conv perturbativity}
\end{figure}

The solution \eq{soft nu evolution} of the rapidity RGE contains multiple convolutions
of the form $(\gNu \otimes S)(\pt)$ or $(\gNu \otimes B)(\pt)$.
These convolutions naturally probe $\as(|\kt|)$ in the nonperturbative regime $|\kt| \lesssim \LQCD$.
Fortunately, nonperturbative effects turn out to be suppressed by $\cO(\LQCD^2/p_T^2)$,
which ultimately means that the $q_T$ spectrum is also perturbative up to corrections $\cO(\LQCD^2/q_T^2)$,
as one would naively expect.

To show this, we consider the convolution $(f_1 \otimes f_2)(\pt,\mu)$ of two generic distributions
\begin{equation}
\Bigl[ f_1(|\kt|) \Bigr]_+^\mu
\,,\qquad
 \Bigl[ f_2(|\kt|) \Bigr]_+^\mu
\,,\end{equation}
which only depend on the magnitude $|\kt|$, which is the typical case we are interested in.
In general, the distributions can depend on further parameters which we suppress,
as they do not affect the following calculation.
We also do not include $\delta(\kt)$ terms, as they trivially factor out of the convolution
$f_1 \otimes f_2$.

Assuming perturbative $|\pt|,\mu \gg \LQCD$, we introduce an auxiliary parameter $\Lambda$ fulfilling $\LQCD \ll \Lambda \ll p_T$ to split the integration range of the convolution
\begin{align}
 (f_1 \otimes f_2)(\pt,\mu) = \int \del^2\kt\, \Bigl[ f_1(|\pt - \kt|) \Bigr]_+^\mu \Bigl[ f_2(|\kt|) \Bigr]_+^\mu
\end{align}
into three pieces (see \fig{conv perturbativity}),
two discs around $\vec 0$ and $\pt$ of radius $\Lambda$, and the rest of the plane.

The convolution integral then splits into three pieces,
\begin{align} \label{eq:conv split}
 (f_1 \otimes f_2)(\pt, \mu) &=
  \!\!\int_{ B_\Lambda(\vec 0)}\!\! \del^2\kt\, f_1(|\pt - \kt|) \Bigl[f_2(|\kt|)\Bigr]_+^\mu +
  \!\!\int_{ B_\Lambda(\pt)}\!\! \del^2\kt\, \Bigl[f_1(|\pt - \kt|)\Bigr]_+^\mu f_2(|\kt|)
\nn \\
  &+ \int_{ \mathbb{R}^2 \backslash (B_\Lambda(\vec 0) \cup B_\Lambda(\vec p_T)) } \del^2\kt\, f_1(|\pt - \kt|) f_2(|\kt|) \,,
\end{align}
where $B_\Lambda(\pt)$ denotes the ball of radius $\Lambda$ around the origin $\pt$.
In the first integral, the range is restricted to be close the origin, and hence the plus prescription of $f_1$ can be dropped, and vice versa for the second integral.
The third integral is separated from both poles by $\Lambda \gg \LQCD$ such that both distributions can be dropped.
Furthermore, it also avoids both Landau poles at $|\kt| = \LQCD$ and  $|\pt - \kt| = \LQCD$, and hence is completely perturbative.
This isolates all the nonperturbative contributions into the first two integrals.

We now focus on the first integral.
Since by construction $|\kt| \le \Lambda \ll |\pt|$ , we can approximate $|\pt - \kt|^2 = |\pt|^2 \bigl(1 + \cO( \kt^2 / \pt^2 ) \bigr)$.
Assuming that the functions only depend on the squared magnitudes, we find
\begin{align}
 \int_{ B_\Lambda(\vec 0)} \del^2\kt\,
    f_1|(\pt - \kt|) \Bigl[f_2(|\kt|)\Bigr]_+^\mu
&= \int_{ B_\Lambda(\vec 0)} \del^2\kt\,
    f_1(|\pt|) \biggr[1 + \cO\biggr(\frac{\kt^2}{\pt^2}\biggr)\biggr] \Bigl[f_2(|\kt|)\Bigr]_+^\mu
\nn \\
&= f_1(|\pt|)  \biggr[1 + \cO\biggr(\frac{\Lambda^2}{\pt^2}\biggr)\biggr]
    \int_{ |\kt| \le \Lambda} \del^2\kt\, \Bigl[f_2(|\kt|)\Bigr]_+^\mu
\nn \\
&= f_1(|\pt|)  \biggr[1 + \cO\biggr(\frac{\Lambda^2}{\pt^2}\biggr)\biggr]
    2 \pi \int_{\mu}^{\Lambda} \del k_T\, k_T f_2(k_T) \,.
\end{align}
Here we used that the plus distribution by definition removes all contributions for $|\kt| \leq \mu$ and therefore
its integral is only sensitive to the upper integration limit $\Lambda$. Hence, we find that the leading nonperturbative effect is suppressed as $\cO(\Lambda^2/\pt^2)$.

The same approximation can be applied to the second integral in \eq{conv split},
such that \eq{conv split} becomes
\begin{align} \label{eq:conv nonpert}
 (f_1 \otimes f_2)(\pt, \mu) &=  \biggr[1 + \cO\biggr(\frac{\Lambda^2}{p_T^2}\biggr)\biggr] 2 \pi \int_{\mu}^{\Lambda} \del k_T\, k_T \Bigl[ f_1(p_T) f_2(k_T) + f_1(k_T) f_2(p_T)\Bigr] + \nn \\
 &\quad+ \int_{ \mathbb{R}^2 \backslash (B_\Lambda(\vec 0) \cup B_\Lambda(\vec p_T)) } \del^2\kt\, f_1(\pt - \kt) f_2(\kt)
\,.\end{align}

In conclusion, we find that convolutions $(f_1 \otimes f_2)(\pt,\mu)$ are well behaved with
nonperturbative contributions suppressed  as $\cO(\Lambda^2/\pt^2)$,
as long as both $|\pt|$ and $\mu$ themselves are perturbative, $|\pt|, \mu \gg \LQCD$.

\paragraph{Illustration:}
The above result can be illustrated by an explicit example, which also shows a different strategy
to evaluate such convolutions.
We consider the convolution $(f \otimes f)(\pt,\mu)$ of
\begin{equation}
 f(\pt,\mu) = \biggl[ \frac{\as(|\pt|)}{\pt^2} \biggr]_+^\mu
\,,\end{equation}
which is the prototype of nonperturbative convolutions appearing in the $q_T$ spectrum.
Even without an explicit calculation, its generic form is known from its $\mu$-dependence.
From
\begin{equation}
 \mu \frac{\del f(\pt,\mu)}{\del\mu} = -2 \pi \as(\mu) \delta(\pt)
\,,\end{equation}
it follows that
\begin{equation}
 \mu \frac{\del (f\otimes f)(\pt,\mu)}{\del \mu} = -4 \pi \as(\mu) \biggl[ \frac{\as(|\pt|)}{\pt^2} \biggr]_+^\mu \,.
\end{equation}
Integrating this according to the prescription \eq{integral scale setting}, we find
\begin{align}
 (f\otimes f)(\pt,\mu)
 &= (f\otimes f)(\pt,\mu_0 = p_T\dis) -  4\pi \int_{\mu_0= p_T\dis}^\mu \frac{\del\mu'}{\mu'} \as(\mu') \biggl[ \frac{\as(|\pt|)}{\pt^2} \biggr]_+^{\mu'} \nn \\
 &= (f\otimes f)(\pt,\mu_0= p_T\dis) - \Biggl[ \frac{4\pi \as(|\pt|)}{\pt^2} \int_{p_T}^\mu \frac{\del\mu'}{\mu'} \as(\mu') \Biggr]_+^\mu  \,.
\label{eq:ff nonpert}
\end{align}
The plus distribution is perturbative as long as both $|\pt|, \mu \gg \LQCD$,
which is precisely the regime we have considered in the above general derivation.
The structure of the boundary term is not known, but its most generic structure to be $\mu_0$-independent is
\begin{equation}
 (f\otimes f)(\pt,\mu_0= p_T\dis) = \biggl[ \frac{-\LQCD/|\pt| \cdot F'(\LQCD/|\pt|)}{2\pi \pt^2} \biggr]_+^\xi + F(\LQCD/\xi) \delta(\pt)
\,,\end{equation}
where the dependence on the arbitrary parameter $\xi$ cancels between the two terms.
Since $(f\otimes f)$ scales as $1/\pt^2$, $F$ must be a scalar function.
It can hence only depend on $\LQCD/|\pt|$,
as no other scales can be combined into a scaleless number.
To be any physically reasonable function, $F$ should vanish
(or at most become constant) for $\LQCD\to0$.
Furthermore we expect it to only depend on the magnitude $\pt^2$,
and hence the boundary term should scale as $\cO(\LQCD^2/\pt^2)$,
just as expected from the general calculation.

\section{The resummed transverse-momentum spectrum}
\label{sec:implementation}

We now discuss in more detail the final result for the resummed transverse momentum spectrum
\eq{factorization thm},
\begin{align} \label{eq:dsigma}
\frac{\del\sigma}{\del Q^2 \del Y \del \qt}
&= \sigma_0\, H(Q,\mu) \int\! \del^2\ki{a}\, \del^2\ki{b}\, \del^2\ki{s} \, \delta(\qt - \ki{a} - \ki{b} - \ki{s})
\nn \\ & \hspace{3cm}
\times B_a(\omega_a, \ki{a},\mu,\nu)\,  B_b(\omega_b, \ki{b},\mu,\nu)\, S(\ki{s},\mu,\nu)
\,,\end{align}
where $\omega_{a,b} = Q^{\pm Y}$.
All distributional logarithms in the $q_T$ spectrum are fully resummed in momentum space
by using the RG-evolved hard, beam, and soft functions in \eq{dsigma}.

To simplify the complicated structure of the $\nu$-evolution, we use \eq{beam nu evolution} to evolve
the beam functions $B_{a,b}$ from their natural rapidity scales $\nu_{a,b} \sim \omega_{a,b}$ to $\nu$.
Since both kernels are simple exponentials in convolution space, they can be combined into
a single evolution kernel, which due to $2 \gamma_{\nu,B} = -\gamma_{\nu,S} = -\gNu$
precisely yields the kernel used to shift the soft function, \eq{soft nu shift},
\begin{align}
(B_a \otimes B_b)(\pt,\mu,\nu)
&= \int\!\del^2\kt\,\del^2\ka\,\del^2\kb\,\delta(\pt - \kt - \ka - \kb)
\nn\\&\hspace{1.5cm}\times
  V_S(\kt, \mu,\nu_B,\nu)\, B_a(\omega_a, \ka,\mu,\nu_a)\, B_b(\omega_b, \kb,\mu,\nu_b)
\,,\end{align}
where $\nu_B = \sqrt{\nu_1 \nu_2}$.
Since beam functions inside the final convolution are evaluated at their natural scales $\nu_{a,b} \sim \omega_{a,b}$,
they are free of large rapidity logarithms.

In addition to soft and beam evolution, we also need the RG evolution of the hard function,
which is straightforwardly obtained from \eq{RGE muH},
\begin{equation}
 H(Q,\mu) = H(Q,\mu_H) \exp\biggl[ \int_{\mu_H}^{\mu}\! \frac{\del\mu'}{\mu'}\, \gMuH(Q,\mu') \biggr]
\,,\end{equation}
where $\mu_H \sim Q$ ensures that $H(Q,\mu_H)$ is free of large logarithms.

To assemble the cross section \eq{dsigma}, we set the arbitrary scale $\mu$ to the
total transverse momentum $q_T$. As usual, this requires distributional scale setting,
$\mu = q_T\dis \equiv \mu_T\dis$.\footnote{%
Alternatively, one could choose $\mu = \mu_H$, but then the final structure is less intuitive.}
We keep the symbolic notation $\mu_T$ rather than $q_T$ to make the origin
of all factors of $q_T$ and $\mu_T$ in the final formula clear.
We then find
\begin{align} \label{eq:dsigma 2}
\frac{\del\sigma}{\del Q^2 \del Y \del \qt}
&= \sigma_0\, H(Q,\mu_H) \frac{1}{2\pi q_T} \frac{\del}{\del q_T} \int\limits_{|\pt| \le q_T} \hspace{-2ex} \del^2\pt \nn\\&\quad \times
\exp\biggl[ \int_{\mu_H}^{\mu_T}\! \frac{\del\mu'}{\mu'}\, \gMuH(Q,\mu') \biggr]
\int\!\del^2\ka\,\del^2\kb\,\del^2\ki{s} \, \delta(\pt - \ka - \kb - \ki{s})
\int\!\del^2\ks'
\nn \\ & \quad \times
\biggl[
 \delta(\ks - \ks') + \sum_{n=1}^\infty \prod_{i=1}^n \int_{k_{i-1}\dis}^{\nu_{i-1}}\!\frac{\del\nu_i}{\nu_i} \int\! \del^2\ki{i}\, \gNu(\ki{i-1} - \ki{i},\mu_T)\,
 \delta\Bigl(\ki{s} - \ki{s}' - \sum_i \ki{i} \Bigr)
\biggr]
\nn \\ & \quad \times
B_a(\omega_a, \ka,\mu_T,\nu_a)\, B_b(\omega_b, \kb,\mu_T,\nu_b)\, S(\ki{s}',\mu_T,k'_s\dis)
\,.\end{align}
The first line contains the $\mu_T$-independent pieces and the cumulant integral from the distributional scale setting $\mu = \mu_T\dis$.
The second line makes explicit that the $\mu$-evolution resums logarithms $\ln(\mu_T/\mu_H) \sim \ln(q_T/Q)$.
The third line contains the $\nu$-evolution kernel, \eq{soft nu evolution},
to evolve the soft function to $\nu_0 \equiv \nu_B = \sqrt{\nu_a \nu_b}$.
The last line contains beam and soft functions evaluated at their natural $\nu$-scales,
and hence all rapidity logarithms in \eq{dsigma 2} are fully resummed.

Note that \eq{dsigma 2} could be the starting point for a numerical evaluation.
Although the infinite number of convolutions cannot be calculated in closed form,
one could for example evaluate the result iteratively and truncate the sum in \eq{dsigma 2}
once a desired numerical accuracy is reached.

One of the original motivations to carry out the momentum-space resummation
(see \mycites{Ellis:1997ii,Kulesza:1999gm,Frixione:1998dw}) was to
avoid the intrinsic nonperturbative sensitivity at large $q_T \gg \LQCD$
arising in the Fourier-space resummation (see \sec{lit CSS} for more details).
From \eq{dsigma 2} it is clear that the analogous sensitivity is still present in the
momentum-space resummation, namely the cross section is intrinsically sensitive to
the nonperturbative contributions from the rapidity anomalous dimension,
since the convolutions probe $|\ki{i}'| \lesssim \LQCD$.
Fortunately, in either case these effects turn out to be suppressed by $\LQCD^2/q_T^2$, as shown in \sec{perturbativity}.

The last line of \eq{dsigma 2} contains the $\mu$-evolved beam and soft functions given by
\eqs{soft mu evolution}{beam mu evolution}. To investigate its structure in more detail,
we omit the pure $\delta$-terms in \eqs{soft mu evolution}{beam mu evolution}, which trivially factor out of the convolution, and for simplicity write everything into a single plus distribution.
This gives
\begin{align} \label{eq:dsigma BBS}
 &B_a(\omega_a, \ka,\mu_T,\nu_a)\, B_b(\omega_b, \kb,\mu_T,\nu_b)\, S(\ki{s},\mu_T,k_s\dis)
\\ & \qquad
 \supset \biggl[
  \frac{1}{2\pi k_a} \frac{\del}{\del k_a}\,
  \frac{1}{2\pi k_b} \frac{\del}{\del k_b}\,
  \frac{1}{2\pi k_s} \frac{\del}{\del k_s}\,
  \cB(\omega_a, k_a)\, \cB(\omega_b, k_b)\, \cS[\as(k_s)]
\nn\\& \quad \qquad\times
  \exp\biggl\{
   \int_{k_a}^{\mu_T}\! \frac{\del\mu'}{\mu'}\, \gMuB(\omega_a,\mu',\omega_a) +
   \int_{k_b}^{\mu_T}\! \frac{\del\mu'}{\mu'}\,\gMuB(\omega_b,\mu',\omega_b) +
   \int_{k_s}^{\mu_T}\! \frac{\del\mu'}{\mu'}\, \gMuS(\mu',k_s)
  \biggr\}
 \biggr]_+^{\mu_T}
\,.\nn\end{align}
The first line of the plus distribution contains the pure fixed-order boundary terms,
which are free of any logarithmic distributions.
The exponential explicitly contains $\mu$-evolutions from the convolution momenta
to the scale $\mu_T$. These are required to properly satisfy the exact path independence
of the $\mu$- and $\nu$-evolution.
This can be checked explicitly by verifying the $\mu_T$-independence of \eq{dsigma 2}.
For practical purposes, it might be sufficient to use the fixed-order expansion for \eq{dsigma BBS},
but since this would deviate from the strict resummation order,
it would have to be verified numerically.

\subsection{Illustration at LL}
To illustrate \eq{dsigma 2}, we evaluate it to leading-logarithmic (LL) order, which
is strictly defined by keeping the LO boundary terms for hard, beam, and soft function,
keeping $\Gamma_0$ and $\beta_0$ in the anomalous dimensions, and dropping all noncusp
anomalous dimensions. In this limit, the $\mu$-evolved beam function boundary term becomes
\begin{align} \label{eq:beam func LL}
B(\omega, \pt, \mu_T,\nu=\omega)
&= \delta(\pt) f(\omega, \mu_T) + \biggl[ \frac{1}{2\pi p_T} \frac{\del}{\del p_T} f(\omega, p_T) \biggr]_+^{\mu_T}
 \approx \delta(\pt) f(\omega, \mu_T)
\,,\end{align}
where the $\mu$-evolution drops out because $\gMuB(\mu,\nu=\omega)$ vanishes at LL. We can also drop the PDF evolution,
since it is a subleading logarithmic effect (i.e.\ the PDF anomalous dimensions are counted as noncusp contributions).
The $\mu$-evolved soft function boundary term at LL is
\begin{align}
 S(\ki{s},\mu_T,k_s\dis)
&= \delta(\ki{s}) + \biggl[ \frac{1}{2\pi k_s} \frac{\del}{\del k_s} \exp\biggl\{ \int_{k_s}^{\mu_T}\! \frac{\del\mu'}{\mu'}\, 4 \GammaC[\as(\mu')] \ln\frac{\mu'}{k_s} \biggr\} \biggr]_+^{\mu_T}
\,.\end{align}
With strict canonical scales, $\nu_B = Q$ and $\mu_T = q_T$, \eq{dsigma 2} is then given by
\begin{align} \label{eq:dsigma LL}
\frac{\del\sigma^{\rm LL}}{\del Q^2 \del Y \del \qt}
&= \sigma_0 \frac{1}{2\pi q_T} \frac{\del}{\del q_T}
f_a(\omega_a,\mu_T)\, f_b(\omega_b,\mu_T) \int\limits_{|\pt| \le q_T} \hspace{-2ex} \del^2\pt\, \exp\biggl[ \int_{\mu_H}^{\mu_T}\! \frac{\del\mu'}{\mu'}\, \gMuH(Q,\mu') \biggr]
\int\!\del^2\ks
\nn \\ & \quad \times
\biggl[
 \delta(\pt - \ks
 ) + \sum_{n=1}^\infty \prod_{i=1}^n \int_{k_{i-1}\dis}^{\nu_{i-1}}\!\frac{\del\nu_i}{\nu_i} \int\! \del^2\ki{i}\, \gNu(\ki{i-1} - \ki{i},\mu_T) \, \delta\Bigl(\pt - \ki{s} - \sum_i \ki{i}\Bigr)
\biggr]
\nn \\ & \quad \times
\biggl( \delta(\ki{s}) + \biggl[ \frac{1}{2\pi k_s} \frac{\del}{\del k_s} \exp\biggl\{ \int_{k_s}^{\mu_T}\! \frac{\del\mu'}{\mu'}\, 4 \GammaC[\as(\mu')] \ln\frac{\mu'}{k_s} \biggr\} \biggr]_+^{\mu_T} \biggl)
\,.\end{align}
In this form, it is somewhat reminiscent of the form suggested by Dokshitzer, Dyakonov, and Troyan
for the LL cross section in \mycites{Dokshitzer:1978hw, Dokshitzer:1978yd},
\begin{align} \label{eq:dsigma DDT}
\frac{\del\sigma^\text{(DDT)}}{\del Q^2 \del Y \del \qt}
&= \frac{\sigma_0}{\pi} \frac{\del}{\del q_T^2} f_a(\omega_a,q_T)\, f_b(\omega_b,q_T)\, e^{\bS(Q,q_T)}
\,.\end{align}
However, comparing this to \eq{dsigma LL}, $e^{\bS}$ would be given by
the full cumulant integral of \eq{dsigma LL}, which has an exponential structure
but is by no means a simple exponential. Only if one were to neglect all convolutions,
i.e.\ keep only the first $\delta(\pt-\ks)$ term in the rapidity evolution in the second line of \eq{dsigma LL},
one would find the expected Sudakov factor
$\bS = \int_{Q}^{q_T} \frac{\del\mu'}{\mu'} \gMuH(Q,\mu')$.
In a full LL resummation, the rapidity evolution forbids such a simple relation,
and as a consequence the simple DDT form \eq{dsigma DDT} cannot hold for the $q_T$ spectrum.
(At higher orders, the effect of PDF running neglected in \eq{beam func LL} would similarly spoil this.)
In fact, the DDT formula was never derived as LL solution of a factorization theorem,
but from a summation of ladder diagrams relevant at LL \cite{Dokshitzer:1978hw}.

An interesting feature of \eq{dsigma LL} is that after carrying out all convolutions,
the result can always be written in the form $\delta(\pt) + \sum c_n \cL_n(\pt,\mu_T) + \sum d_m \cL_m(\pt,\nu_B)$.
With strict canonical scale setting, $\mu_T = q_T\dis$, the first sum vanishes due to $\cL_n(\qt,q_T\dis) = 0$.
In contrast, the second sum is converted into $\cL_m(\pt,\nu_B) \to \cL_m(\qt,\nu_B)$.
Since $\nu_B \sim Q$, this precisely yields the rapidity logarithms $\ln(Q/q_T)$,
which however do not have a simple exponential structure any more.
For noncanonical scales, the $\cL_n(\pt,\mu_T)$ would not exactly vanish,
but yield small corrections that probe the all-order logarithmic structure.

As an illustrative example, we carry out the first few convolutions,
but ignore $\as$-running for simplicity. This is analogous to the simple
study in \sec{soft exact vs naive}. We find
\begin{align} \label{eq:dsigma toy dis}
\frac{\del\sigma}{\del Q^2 \del Y \del \qt}
&= \sigma_0 \frac{1}{2\pi q_T} \frac{\del}{\del q_T} \theta(q_T)
f_a(\omega_a,q_T)\, f_b(\omega_b,q_T) \exp\biggl[ - \frac{\GammaC}{2} \ln^2\frac{Q^2}{q_T^2} \biggr]
\nn \\ & \quad \times
\biggl[
 1 - 2\GammaC^2 \zeta_3 \ln\frac{Q^2}{q_T^2} + \GammaC^3 \biggl( \frac{2 \zeta_3}{3} \ln^3\frac{Q^2}{q_T^2} + 6 \zeta_5 \ln\frac{Q^2}{q_T^2} \biggr) \nn\\&\qquad
 + \GammaC^4 \biggl( - 4 \zeta_5 \ln^3\frac{Q^2}{q_T^2} + 10 \zeta_3^2 \ln^2\frac{Q^2}{q_T^2} - 30 \zeta_7 \ln\frac{Q^2}{q_T^2}  \biggr)
 + \cO(\GammaC^5)
\biggr]
\,.\end{align}
We have not evaluated the derivative, as in this form the terms in square bracket
are exactly the result of the rapidity evolution. It clearly induces apparent
subleading terms in the cross section, which however are part of the strict LL order
that is based on the expansion of the anomalous dimensions.

For comparison, the Fourier-resummed spectrum (see \sec{soft exact vs naive} for more details)
with its canonical scale choices $\nu_S = \mu = b_0/b_T$, neglecting again $\as$-running, is given by
\begin{align} \label{eq:dsigma toy FT 1}
\frac{\del\sigma}{\del Q^2 \del Y \del \qt}
&= \sigma_0\, f_a(\omega_a,\mu)\, f_b(\omega_b,\mu) \int\!\frac{\del^2\bt}{(2\pi)^2} e^{i \bt \cdot \qt}
 \exp\biggl[ - \frac{\GammaC}{2} \ln^2\frac{Q^2 b_T^2}{b_0^2} \biggr]
\,.\end{align}
As before, we neglect running in the PDFs for simplicity to evaluate them at the random scale $\mu$.
The Sudakov double logarithm is clearly a well-behaved function that drops fast for both $\bt \to 0$
and $\bt \to \infty$, and hence the inverse Fourier integral must be fine as well.
To compare this expression to \eq{dsigma toy dis}, we split $\ln(Q b_T / b_0) = \ln(Q/\mu) + \ln(\mu b_T / b_0)$,
with $\mu$ arbitrary. Expanding the integrand in $\GammaC$, we only need to Fourier-transform powers of logarithms
$\ln^n(\mu b_T / b_0)$, which is well known (see appendix \ref{app:plus distributions ips}).
Setting $\mu = q_T\dis$ gives
\begin{align} \label{eq:dsigma toy FT 2}
\frac{\del\sigma}{\del Q^2 \del Y \del \qt}
&= \sigma_0 \frac{1}{2\pi q_T} \frac{\del}{\del q_T} \theta(q_T)
f_a(\omega_a,q_T)\, f_b(\omega_b,q_T) \exp\biggl[ - \frac{\GammaC}{2} \ln^2\frac{Q^2}{q_T^2} \biggr] \nn \\ & \quad
\times \biggl[
 1 - 2 \GammaC^2 \zeta_3 \ln\frac{Q^2}{q_T^2} + \GammaC^3 \biggl( \frac{2 \zeta_3}{3} \ln^3\frac{Q^2}{q_T^2} + 6 \zeta_5 \ln\frac{Q^2}{q_T^2} - \frac{10}{3} \zeta_3^2 \biggr)
\\ \nn &\qquad
 + \GammaC^4 \biggl( - 4 \zeta_5 \ln^3\frac{Q^2}{q_T^2} + 10 \zeta_3^2 \ln^2\frac{Q^2}{q_T^2} - 30 \zeta_7 \ln\frac{Q^2}{q_T^2} + 28 \zeta_3 \zeta_5 \biggr)
 + \cO(\GammaC^5)
\biggr]
\,.\end{align}
The result agrees with \eq{dsigma toy dis} up to the constant terms $- 10/3\,\zeta_3^2 \GammaC^3$ and $28 \zeta_3 \zeta_5 \GammaC^4$.
These nonlogarithmic terms are examples of different boundary terms induced by using the Fourier-space boundary conditions.
The important observation is that the logarithmic structure exactly matches in both attempts.

Lastly, we also compare this to the result from using the naive rapidity evolution of the soft function.
In this case, \eq{dsigma} is evaluated at $\mu = \mu_T \sim q_T$,
and the soft function is taken from \eqs{naive solution S}{naive V},
\begin{align}
\frac{\del\sigma}{\del Q^2 \del Y \del \qt}
&= \sigma_0\, H(Q,\mu_H) \exp\biggl[ \int_{\mu_H}^{\mu_T}\! \frac{\del\mu'}{\mu'}\, \gMuH(Q,\mu') \biggr]
\int\!\del^2\ka\, \del^2\kb\, \del^2\ki{s} \, \delta(\qt - \ka - \kb - \ki{s})
\nn \\ & \quad
\times B_a(\omega_a, \ka,\mu_T,\nu_a)\, B_b(\omega_b, \kb,\mu_T,\nu_b)
\nn \\ & \quad
\times \int\!\del^2\kt V_S(\ks-\kt,\mu_T,\nu_B,\nu_S) S(\ks,\mu_T,\nu_S)
\,.\end{align}
Assuming the LO boundary conditions for the hard, beam, and soft functions, we obtain
\begin{align}
\frac{\del\sigma}{\del Q^2 \del Y \del \qt}
&= \sigma_0\, f_a(\omega_a,\mu_T)\, f_b(\omega_b,\mu_T)
\exp\biggl[ \int_{\mu_H}^{\mu_T}\! \frac{\del\mu'}{\mu'}\, \gMuH(Q,\mu') \biggr]
\nn \\ & \quad
\times \biggl[ \delta(\qt) + \sum_{n=1}^\infty \frac{1}{n!} \ln^n \frac{\nu_B}{\nu_S} (\gNu \otimes^n)(\qt,\mu_T) \biggr]
\,.\end{align}
Setting $\mu_T = \nu_S = q_T\dis$, we get
\begin{align} \label{eq:dsigma toy naive}
\frac{\del\sigma}{\del Q^2 \del Y \del \qt}
&= \sigma_0\, \frac{1}{2\pi q_T} \frac{\del}{\del q_T} \theta(q_T)
  f_a(\omega_a,q_T)\, f_b(\omega_b,q_T) \exp\biggl[ - \frac{\GammaC}{2} \ln^2\frac{Q^2}{q_T^2} \biggr]
\nn \\ & \quad
\times \biggl[
 1 + \frac{2}{3} \GammaC^3 \ln^3\frac{Q^2}{q_T^2} \zeta_3 + \cO(\GammaC^5)
\biggr]
\,.\end{align}
The obtained highest logarithmic term $\GammaC^3 \ln^3(Q^2/q_T^2)$ matches the one in \eqs{dsigma toy dis}{dsigma toy FT 2}, but all other terms are missing. The second line of \eq{dsigma toy naive}
is nothing but the familiar factor $e^{-2\gamma_E \omega_s} \Gamma(1-\omega_s) / \Gamma(1+\omega_s)$
from \eq{V naive sol}, which diverges at $\omega_s = 2 \GammaC \ln(Q/q_T) = 1$. The fact that this term also appears
in the correctly resummed results in \eqs{dsigma toy dis}{dsigma toy FT 2} is quite surprising as one might
expect it to get modified in order to alleviate the spurious divergence.
On the other hand, since the naive rapidity evolution kernel \eq{V naive sol}
does correctly \textit{shift} the rapidity logarithms, it must in fact also appear in \eqs{dsigma toy dis}{dsigma toy FT 2}, and so it seems that both the $\pt$-space and $\bt$-space resummation still contain this divergence.
However, we know that \eq{dsigma toy FT 1} is well behaved, and hence all the additional logarithmic terms in \eqs{dsigma toy dis}{dsigma toy FT 2} that are not present in \eq{dsigma toy naive} must conspire to cancel the divergence in the spectrum.
Hence, we find the peculiar feature that apparent-NNLL and higher terms cancel the divergence caused by the apparent-NLL terms in the strict LL spectrum.
A detailed numerical study of this effect would be an interesting extension of this work.

This simple exercise clearly shows that counting logarithms $\ln(Q/q_T)$ in the $q_T$ spectrum is an intrinsically
ill-defined notion. Instead, the resummation order should be defined strictly and unambiguously through the perturbative order of anomalous dimensions and pure boundary terms entering all RGEs. This is done in both our momentum-space and the
Fourier-space resummation, which both lead to well-behaved results.

\section{Comparison to the literature}
\label{sec:comparison}

In this section, we discuss the relation of our results to the Fourier-space resummation
in the original CSS formulation and various existing implementations (\sec{lit Fourier}),
as well as other approaches for a direct momentum-space resummation (\secs{lit qt resummation}{lit coh branching}).

\subsection{Resummation in Fourier space}
\label{sec:lit Fourier}

\subsubsection{Comparison to CSS formalism}
\label{sec:lit CSS}

The equivalence of the Fourier-space resummation in the original CSS formulation
and performing the RG evolution in Fourier space in the context of SCET has been
discussed several times before~\cite{Becher:2010tm, Neill:2015roa}
(see e.g.~\mycites{Becher:2006mr, Sterman:2013nya, Almeida:2014uva, Bonvini:2014qga}
for one-dimensional cases like thrust or threshold resummation).
It is not unexpected that the two formulations are equivalent, since
both are based on the same underlying factorization in the soft-collinear approximation.

Interestingly, in their first paper~\cite{Collins:1981uk} in the context of back-to-back jets in $e^+e^-$-collisions, Collins and Soper formulated an evolution equation in transverse momentum space that precisely corresponds
to the rapidity RGE in our framework; compare eq.~(6.4) in \mycite{Collins:1981uk} with \eq{RGE nuS},
where the rapidity scale $\nu$ essentially corresponds to the scale $\zeta$ in the CSS formulation.
However, due to the much simpler formulation in Fourier space, as first noted by Parisi and Petronzio \cite{Parisi:1979se},
they carried out their analysis in Fourier space. Studying in detail the properties of the inverse transform,
with particular focus on the effect of the (nonperturbative) large $b$-region~\cite{Collins:1981va},
then paved the road to resumming the $q_T$-spectrum in hadronic collisions~\cite{Collins:1984kg}.

The CSS formula of \mycite{Collins:1984kg} can be easily related to the SCET formalism
by Fourier transforming the beam and soft functions and their RGEs given in \sec{TMD factorization}.
For example, for the soft function we have
\begin{align}
\mu \frac{\del \tilde S(\bt, \mu, \nu)}{\del\mu} &= \gMuS(\mu, \nu) \tilde S( \bt, \mu, \nu)
\,, \nn \\
\nu \frac{\del \tilde S(\bt, \mu, \nu)}{\del\nu} &= \tilde\gamma_\nu(\bt,\mu) \tilde S( \bt, \mu, \nu)
\,,\qquad
 \mu\frac{\del \tgNu(\bt,\mu)}{\del\mu} = -4 \GammaC[\as(\mu)]
\,,\end{align}
and most importantly the rapidity RGE becomes multiplicative.
Since logarithms in impact parameter space always depend on $\mu b_T/b_0$ or $\nu b_T/b_0$, all logarithms are minimized with boundary scale choice $\mu_0 = \nu_0 = b_0/b_T$. With this choice, the above RGEs are solved by
\begin{align}
\tilde S(\bt,\mu,\nu)
&= \exp \biggl[ \int_{b_0/b_T}^\mu \!\frac{\del\mu'}{\mu'}\, \gMuS(\mu',\nu)\biggr]
\exp\biggl[ \tgNu (\bt, b_0/b_T) \ln\frac{\nu b_T}{b_0} \biggr] \tilde S(\bt,b_0/b_T,b_0/b_T)
\,, \nn\\
\tgNu(\bt,\mu)
&= - \int_{b_0/b_T}^\mu\! \frac{\del\mu'}{\mu'}\, 4 \GammaC[\as(\mu')] + \tgNu[\as(b_0/b_T)]
\,.\end{align}
Note that by first solving the $\nu$-RGE at the low scale and then the $\mu$-RGE, $\tgNu$ enters evaluated at its natural scale
$\mu_0=b_0/b_T$ for which it reduces to the pure fixed-order boundary $\tgNu[\as(b_0/b_T)]$.
In this way, the resummation of $\tgNu$ becomes trivial in the Fourier space evolution.

The beam functions are similarly resummed with boundary scales $\mu_0 = b_0/b_T, \nu_0 = \omega_i$,
\begin{align}
\tilde B_a(\omega,\bt,\mu,\nu)
 &= \exp \biggl[ \int_{b_0/b_T}^\mu \frac{\del\mu'}{\mu'} \gMuB(\mu',\nu;\omega) \biggr] \exp \biggl[ -\frac{1}{2} \tilde \gNu (\bt, b_0/b_T) \ln\frac{\nu}{\omega} \biggr]
\nn\\&\quad\times
 \tilde B_a(\omega,\bt,\mu=b_0/b_T,\nu=\omega)
\,,\end{align}
and can furthermore be matched onto PDFs using an operator product expansion,
\begin{equation}
\tilde B_a(\omega,\bt,\mu,\nu)
= \sum_{i} \int \frac{\del z}{z}\, \tilde \cI_{ai}(z,\bt,\mu,\nu)\, f_{i}\Bigl(\frac{\omega/E_{\rm cm}}{z},\mu\Bigr)
\,,
\end{equation}
where $E_{\rm cm}$ is hadronic center-of-mass energy, $\tilde \cI_{ai}$ are perturbative matching coefficients, and $f_i$ are standard PDFs. Altogether, the resummed cross section \eq{factorization thm} takes the simple form
\begin{align} \label{eq:CSS like}
\frac{\del\sigma}{\del Q \del Y \del \qt}
&= \sigma_0 \int\!\frac{\del^2\bt}{(2\pi)^2}\, e^{i \bt \cdot \qt} \sum_{i,j}
\int\! \frac{\del z_a}{z_a} \frac{\del z_b}{z_b}\, f_{i}\Bigl(\frac{\omega_a/E_{\rm cm}}{z_a}, \frac{b_0}{b_T}\Bigr) f_{j}\Bigl(\frac{\omega_b/E_{\rm cm}}{z_b}, \frac{b_0}{b_T} \Bigr)
\nn\\ & \quad \times
\exp\biggl[ - \int_{b_0/b_T}^Q \frac{\del\mu'}{\mu'} \gMuH(Q,\mu') + \ln\frac{Q b_T}{b_0} \tgNu[\as(b_0/b_T)] \biggr]
\nn\\ & \quad \times
H(Q,\mu=Q)\, \tilde \cI_{ai}(z_a, b_T)\, \tilde \cI_{bj}(z_b, b_T)\, \tilde S(b_T)
\,.\end{align}
The functions $\tilde \cI_{ai}(z, b_T) \equiv \tilde \cI_{ai}(z, b_T,\mu_0=b_0/b_T, \nu_0=\omega)$ and $\tilde S(b_T) \equiv \tilde S(\bt,\mu_0=\nu_0=b_0/b_T)$ are all free of logarithms.
We have explicitly split the Sudakov exponential into a piece originating from the $\mu$-RGEs and one from the rapidity RGE.
This result should be compared to the CSS formula~\cite{Collins:1984kg},
\begin{align} \label{eq:CSS}
\frac{\del\sigma}{\del Q \del Y \del \qt}
&= \sigma_0 \int\!\frac{\del^2\bt}{(2\pi)^2}\, e^{i \bt \cdot \qt} \sum_{i,j} \int\!
 \frac{\del z_a}{z_a} \frac{\del z_b}{z_b}\, f_{i}\Bigl(\frac{\omega_a/E_{\rm cm}}{z_a}, \frac{b_0}{b_T}\Bigr)
 f_{j}\Bigl(\frac{\omega_b/E_{\rm cm}}{z_b}, \frac{b_0}{b_T} \Bigr)
\\\nn & \quad \times
C_{ai}(z_a, b_T)\, C_{bj}(z_b, b_T)
\exp \biggl[ - \int_{(b_0/b_T)^2}^{Q^2}\! \frac{\del\bar\mu^2}{\bar\mu^2}\, \Bigl( \ln\frac{Q^2}{\bar\mu^2} A[\as(\bar\mu)]
+ 2 B[\as(\bar\mu)] \Bigr) \biggr]
\,.\end{align}
Comparing \eqs{CSS like}{CSS}, we see that the functions $C_{aj}$ correspond to the product of hard and soft function with the beam function matching kernels $\tilde \cI_{ai}$, showing that they can be further split into a process-dependent hard virtual
and a process-independent soft-collinear piece, as was noticed also in \mycites{Catani:2000vq, Catani:2013tia}.
The evolution kernels are related by
\begin{align}
 A(\as) &= \GammaC(\as) + \beta(\as) \frac{\del\tgNu(\as)}{\del\as}
\,,\quad
 2 B(\as) = \gMuH(\as) - \tgNu(\as)
\,.\end{align}
The important observation that $A(\as)$ receives a contribution from the rapidity anomalous dimension in addition to the cusp anomalous dimension was first made in \mycite{Becher:2010tm}.

To compare to our canonical resummation in momentum space, first note that setting $\nu_0 = b_0/b_T$ inside the Fourier integral is the analog of choosing the scale $\nu_0 = k_T\dis$ inside the convolutions building up the $\nu$-evolution kernel.
Similarly, choosing $\mu_0=b_0/b_T$ for both beam and soft function inside the Fourier integral corresponds to choosing $\mu_0 = p_T\dis$ when solving the $\mu$-RGE for $S(\pt)$ and $B(\pt)$,
which then accordingly enters the convolutions $B \otimes B \otimes S$.
In practice, the Fourier-space resummation is of course more easily implemented because the measurement constraint $\qt = \sum_i\ki{i}$ is translated into a common impact parameter $\bt$ for all convolved functions,
and hence the iteratively defined convolution exponential in \eq{soft nu evolution} turns into a standard exponential function in Fourier space.
As discussed on general grounds in \sec{comparison rge tools} and in detail for $\gNu$ in \sec{gNu comparison},
the formal difference between the different spaces arises from the fact that one uses different boundary conditions
which induce different subleading terms to all orders in $\as$.
Hence with a robust estimate of theoretical uncertainties at any given order, one would expect that both techniques yield results compatible within their uncertainties, but a direct comparison of both results would provide another interesting way to assess theory uncertainties, in particular of nonperturbative effects.

\subsubsection{Practical implementations}
\label{sec:lit impl}

In the following we briefly comment on several implementations (without claiming to be exhaustive),
which perform the resummation fully or partially in $b$ space.
As reference to compare to we take either the canonical $b$-space or the canonical momentum-space evolution we have derived, as both techniques reproduce \textit{all} logarithms in the fixed-order reexpansion in their respective space.
We only focus on the effects of deviating from these canonical scale choices and the strict resummation order.
A detailed discussion regarding the phenomenologically important aspects of matching to the full fixed-order and assessing theory uncertainties can be found e.g.\ in \mycite{Neill:2015roa}.

The original CSS formula is the basis of \mycites{Balazs:1997xd, Balazs:2000wv, Balazs:2006cc, Wang:2012xs} as well as
\mycites{Catani:2000vq, deFlorian:2000pr, deFlorian:2001zd, Catani:2010pd, Catani:2013tia, Bozzi:2010xn, Catani:2015vma, Bozzi:2007pn, deFlorian:2012mx, Cieri:2015rqa}.
For the latter, the canonical Fourier-space logarithms are replaced by
\begin{equation}
 \ln(b_T^2 \mu^2 / b_0^2) \to \ln(1 + b_T^2 \mu^2 / b_0^2)
\,.\end{equation}
The benefit of shifting the argument of the logarithm is that it
suppresses the region $p_T \sim 1/b_T \gg Q$. This should effectively suppress the
contributions from energetic emissions, and it would be interesting to study its effect
on the small $q_T$ region compared to the strict canonical resummation.
Furthermore it ensures that integrating over the $\qt$-spectrum restores the inclusive cross section.
The Landau pole intrinsic to the calculation is treated using the minimal prescription \cite{Laenen:2000de,Kulesza:2002rh} by deforming the integration contour around the Landau pole, which is valid as long as $q_T$ is sufficiently perturbative.

The first SCET-based calculation\footnote{Earlier attempts of $\qt$-resummation in SCET~\cite{Gao:2005iu, Idilbi:2005er, Mantry:2009qz} missed the effects of rapidity divergences.}
was carried out in \mycites{Becher:2010tm, Becher:2011xn, Becher:2012yn} in the so-called collinear anomaly framework using an analytic regulator \cite{Chiu:2007yn,Beneke:2003pa,Becher:2011dz}.
The resummation of rapidity logarithms there corresponds to performing the $\nu$-evolution in Fourier space with fixed canonical scale choice $\nu_0 = b_0/b_T$.
However, the low scale $\mu_0$ is set to $q_T$, and the rapidity anomalous dimension $\gNu$ (corresponding to the anomaly coefficient $F$ in their framework) is not resummed but expanded to a certain fixed order (referred to as $\epsilon$-expansion).
This looses some of the strict resummation accuracy, and in light of our discussion, one might expect that energetic emissions are incorrectly treated, which could affect the region of very small $q_T$.
The naive divergence in the cross section (see \sec{TMD illustration}) is further avoided by choosing $\mu_0 = q_T + q_*$. The offset $q_* \approx 2~\GeV$ for Drell-Yan and $q_* \approx 8~\GeV$ for Higgs production is chosen large enough to explicitly avoid the divergence. However, it does so by essentially turning off the resummation below $q_T \lesssim q_*$, which then also drops the sensitivity to nonperturbative effects in the rapidity evolution kernel.

In \mycites{GarciaEchevarria:2011rb, Echevarria:2012js, Echevarria:2014rua}, a factorization theorem has been derived using the $\delta$-regulator \cite{Chiu:2009yx, GarciaEchevarria:2011rb}, and has been applied e.g.\ in \mycites{D'Alesio:2014vja, Echevarria:2015uaa}.
Here, the rapidity evolution is also performed in Fourier space, while the $\mu$-RGEs are solved in momentum space, but in contrast to the above the rapidity anomalous dimension (in their notation the $D$ function) is fully resummed.
The low scale $\mu_0$ is chosen as $\mu_0 = q_T + Q_0$, where $Q_0 \approx 2~\GeV$ for both Drell-Yan and Higgs production
serves as a cutoff for the nonperturbative region.
While at very small $q_T \lesssim Q_0$ deviations from the canonical resummation are expected,
these can be effectively absorbed into the nonperturbative contributions that become relevant in this region.
Indeed \mycite{D'Alesio:2014vja} takes great care to assess nonperturbative effects in the Drell-Yan spectrum.

Lastly, the factorization theorem of \mycites{Chiu:2011qc,Chiu:2012ir} has been applied to Higgs production in \mycite{Neill:2015roa}.
They employ a canonical resummation fully in $b$-space, similar to the CSS approach,
but instead of shifting the arguments of the $b$-space logarithms, the resummation is turned off with
profile scales in $b$-space whose form is based on the final value of $q_T$.
The nonperturbative large $b_T$-region is avoided by explicitly cutting off the Fourier integration at $b_T \le 2~\GeV^{-1}$,
while verifying that changing the cut in $1.5 - 3\,\GeV^{-1}$ only produces a negligible variation.
Because of the generic suppression of nonperturbative effects by $\LQCD^2/q_T^2$, this can be expected to hold
as long as $q_T$ is sufficiently perturbative.

\subsection{Early approaches for direct \texorpdfstring{$q_T$}{qT}-space resummation}
\label{sec:lit qt resummation}

There have been several attempts in the past to carry out the resummation directly in momentum space.
They typically attempt to explicitly count logarithms $\ln(Q/q_T)$ in the $q_T$ spectrum.
As we discussed before, this is dangerous as it can easily lead one
to discard apparent subleading contributions that are seemingly unimportant but are actually
relevant.
The original DDT approach~\cite{Dokshitzer:1978hw, Dokshitzer:1978yd} introduced the LL cross section in the form
of \eq{dsigma DDT},
\begin{equation} \label{eq:DDT}
\frac{\del\sigma^\text{(DDT)}}{\del Q^2 \del Y \del \qt}
= \frac{\sigma_0}{\pi} \frac{\del}{\del q_T^2} f_a(\omega_a,q_T)\, f_b(\omega_b,q_T)\, e^{\bS(Q,q_T)}
\,.\end{equation}
Its connection to our full LL solution was already discussed in \sec{implementation}.

Ref.~\cite{Frixione:1998dw} tried to extend \eq{DDT} to NLL, counting logarithms $L=\ln(Q/q_T)$
in the Sudakov exponent $\bS$. In this counting, \mycite{Frixione:1998dw} finds the cross section
\begin{align} \label{eq:sigma_FNR}
\frac{\del\sigma^\text{(DDT)}}{\del Q^2 \del Y \del \qt}
&\propto Q^2 \frac{\del}{\del q_T^2} \biggl[ e^{\bS} \frac{\Gamma(1+h/2)}{\Gamma(1-h/2)} \biggr]
\,,\qquad
h = 4 \GammaC[\as(q_T)] \ln\frac{q_T}{Q}
\,,\end{align}
which diverges for $h = -2$.
This can be directly related to our results, where the log-counting in the exponent corresponds to using the naive solution of the rapidity RGE in \eq{naive V} and keeping the rapidity anomalous dimension $\gNu$ at a fixed order rather than fully resuming it.
As we saw in \sec{TMD illustration}, the result for the rapidity evolution kernel in this approximation
[see \eq{V naive sol}] contains exactly the same spurious divergence at $\omega_s \equiv -h/2 = 1$ as \eq{sigma_FNR}.
As we have argued, this is caused by the incorrect treatment of energetic emissions, which become increasingly
important for small $q_T$. This agrees with \mycite{Frixione:1998dw}, where it is remarked that energy conservation
constraints are not correctly implemented in the resummation formula.
This is remedied by basing the logarithmic order counting on the anomalous dimensions,
and hence is not a flaw of the factorization theorem itself.

The second DDT-based approach, \mycite{Ellis:1997ii}, counts logarithms $L = \ln(Q/q_T)$ directly in the cross section, i.e.\ only counting $\as L^2 \sim 1$.
Interestingly, this actually happens to lead to a numerically well-defined prediction of the cross section.
To see this, consider again the naive rapidity evolution kernel \eq{naive V},
\begin{align}
V(\pt,\mu, \nu, \nu_0) &= \delta(\pt)  + \sum_{n=1}^\infty \frac{1}{n!} \ln^n\!\frac{\nu}{\nu_0} (\gNu \otimes^n)(\pt,\mu)
\,.\end{align}
Counting logarithms in the cross section is equivalent to truncating the $n$-fold convolution $\gNu \otimes^n$ at the desired accuracy. For example, at LL$_\sigma$ where $\gNu(\pt,\mu) = 2 \GammaC[\as(\mu)] \PlusLog{0}(\pt,\mu)$, one would only keep the first term of the sum
\begin{align}  \label{eq:gNu^n appr}
(\gNu \otimes^n)(\kt,\mu)
 &= (2\GammaC)^n n \biggl[ \PlusLog{n-1}(\kt,\mu) + 4 \zeta_3 \binom{n-1}{n-4} \PlusLog{n-4}(\kt,\mu) + \cdots \biggr]
\,,\end{align}
but treat the first subleading term $\PlusLog{n-4}$ as a N$^3$LL$_\sigma$ correction.
The evolution kernel would then be given by
\begin{align}
V(\pt,\mu, \nu, \nu_0) = \delta(\pt) + \omega_s \PlusPol{\omega_s}(\pt,\mu)
\,,\end{align}
where $\omega_s = 2 \GammaC[\as(\mu)] \ln(\nu/\nu_0)$.
Comparing to \eqs{V naive sol}{sigma_FNR}, where all terms in \eq{gNu^n appr} are kept,
only keeping the first term in \eq{gNu^n appr} completely removes the divergence in the kernel.
However, this cross-section counting is of course only applicable in an intermediate $q_T$ range and
only includes a small subset of logarithms compared to the full resummation.

A different approach was adopted in \mycites{Kulesza:1999gm,Kulesza:1999sg,Kulesza:2001jc},
which attempt to obtain an explicit momentum-space expression of the inverse Fourier transform of the $b$-space result.
The resummed $b$-space result $\del\tilde\sigma/\del b_T^2$ is then expanded in terms of logarithms $L_b = \ln(b_T^2 \mu^2 / b_0^2)$, whose inverse Fourier transform is known for arbitrary powers $L_b^n$,
see appendix \ref{app:plus distributions ips}. This allows one to construct a series in momentum space that approximates the Fourier transform to an in principle arbitrary precision.
However, the resummation itself is effectively still performed canonically in $b$ space.

\subsection{Coherent branching formalism}
\label{sec:lit coh branching}

Very recently, a different approach using the coherent branching formalism of
refs.~\cite{Banfi:2004yd,Banfi:2014sua} has been proposed in ref.~\cite{Monni:2016ktx},
which is not based on a factorization theorem or solving the $q_T$ evolution equations.

While a detailed comparison to the NNLL result given in \mycite{Monni:2016ktx} would be very interesting,
we leave it for future work and in the following concentrate on comparing to their NLL result, which
is already instructive.
The starting point of their derivation is the cumulative cross section obtained by summing
over any number of independent emissions
\begin{align} \label{eq:cumulant_MRE}
\Sigma(q_T)
= \int_0^{q_T}\! \del k_T\, \frac{\del \sigma(k_T)}{\del k_T}
&= \sigma_0 \int_0^\infty\! \braket{\del k_1}\, R'(k_1)\, e^{-R(\epsilon k_1)}
\nn\\ & \quad \times
\sum_{n=0}^\infty \frac{1}{n!} \prod_{i=2}^{n+1} \int_{\epsilon k_1}^{k_1}\! \braket{\del k_i}\, R'(k_i)\,
 \theta\biggl(|\qt| - \biggl\lvert\sum_j \vec k_j\biggr\rvert\biggr)
\,,\end{align}
where the phase-space measure is $\braket{\del k_T} = \frac{\del k_T}{k_T} \frac{\del\phi}{2\pi}$.
The differential spectrum follows to be
\begin{align} \label{eq:spectrum_MRE}
\frac{\del \sigma}{\del\qt}
 = \sigma_0 \int\!\del^2\ki{1}\, \frac{R'(k_1)}{2\pi k_1^2} e^{-R(\epsilon k_1)}
 \sum_{n=0}^\infty \frac{1}{n!} \prod_{i=2}^{n+1} \int_{\epsilon k_1 < |\ki{i}| < k_1} \hspace{-3ex}\del^2\ki{i}\, \frac{R'(k_i)}{2\pi k_i^2}\,
 \delta\biggl(\qt - \sum_j \vec k_j \biggr)
\,,\end{align}
where we use a notation resembling our convolution notation.
Here, the hardest emission $\ki{1}$ has been singled out.
The parameter $\epsilon \ll 1$ reflects that emissions with $k_i < \epsilon k_1$ are unresolved.
Correspondingly, the exponential $e^{-R(\epsilon k_1)}$ encodes the Sudakov suppression
of having no emission between scales $\epsilon k_1$ and $Q$.
The radiator $R$ is given at NLL by
\begin{equation} \label{eq:radiator}
 R(k_T)
 = \int_{k_T}^Q \frac{\del\mu'}{\mu'} \left( 4 \GammaC[\as(\mu')] \ln\frac{Q}{\mu'} -  \frac{\as(\mu')}{\pi} \beta_0 \right)
 = \int_{k_T}^Q \frac{\del\mu'}{\mu'} \gMuH(Q,\mu')
\,,\end{equation}
where we converted from the Catani-Marchesini-Webber scheme \cite{Catani:1990rr} used in ref.~\cite{Monni:2016ktx} to the $\overline{\mathrm{MS}}$-scheme.
Its derivative $R'$ evaluates to\footnote{%
The results in \eqs{cumulant_MRE}{spectrum_MRE} are technically only NLL accurate up to the fact that
the PDFs in $\sigma_0$ are evaluated at fixed $\mu_F$ rather than $k_1$, which however is
irrelevant for the present discussion.
We have dropped the constant $\beta_0 \as / \pi$ here, which would be canceled if the PDFs
were evaluated at $k_1$.}
\begin{equation}
 R'(k_T) = - k_T \frac{\del R(k_T)}{\del k_T} =  4 \GammaC[\as(k_T)] \ln\frac{Q}{k_T}
\,.\end{equation} \label{eq:radiator_derivative}

The structure of \eq{spectrum_MRE} is closely related to our results,
as it essentially contains an infinite number of convolutions,
with the exception that the hardest emission is explicitly singled out.
Furthermore the unresolved regions $0 < k_i < \epsilon k_1$ that are cut out of the $k_i$ integrals ($i\geq2$)
are already captured in the Sudakov exponent.
Letting $\epsilon\to 0$ is then equivalent to the cancellation of IR singularities
encoded in the plus distributions in our framework.
Each factor $R'$ corresponds to a $\nu$-anomalous dimension times its rapidity logarithm,
\begin{equation}
 \frac{R'(k_i)}{2\pi k_i^2}
 = \frac{4 \GammaC[\as(k_i)] \ln(Q/k_i)}{2\pi k_i^2}
 = \int_{k_i}^Q\! \frac{\del\nu'}{\nu'}\, \frac{4 \GammaC[\as(k_i)]}{2 \pi k_i^2}
 = \int_{k_i}^Q\! \frac{\del\nu'}{\nu'}\,  \gNu(\ki{i},\mu)
\,,\end{equation}
where we assumed $k_i > 0$ to drop the plus prescription in $\gNu$.
In summary, \eq{spectrum_MRE} agrees very well with our momentum-space resummed cross section
that follows from the factorization theorem,
up to the different treatment of the cancellation of IR divergences.

Differences arise when the resummed spectrum is further expanded to obtain NLL accuracy,
defined by counting logarithms in the cumulative cross section. Expanding the individual emission momenta
around $k_i \sim q_T$ and counting logarithms $\ln(Q/q_T)$, they also reproduce the spurious
divergence in the cross section. This is precisely equivalent to using the naive rapidity resummation,
which does not treat energetic emissions correctly, as discussed in \secs{TMD convolutions}{TMD illustration}.

To circumvent this, ref.~\cite{Monni:2016ktx} instead expands all radiators appearing
in \eq{cumulant_MRE} around the hardest emission $k_1$.
At NLL, the necessary expansions are
\begin{align}
 R(\epsilon k_1) = R(k_1) + R'(k_1) \ln\frac{1}{\epsilon} + \cdots
\,,\quad
 R'(k_i) = R'(k_1) + \cdots
\,.\end{align}
A simplification is that all radiators $R(k_i)$ and $R'(k_i)$
are now evaluated at $\as(k_1)$ rather than $\as(k_i)$, thereby
removing the nonperturbative effects that would otherwise be present in the rapidity evolution.
This procedure hence fundamentally resums logarithms of $\ln(Q/k_1)$ to NLL,
where the resummation accuracy is defined by explicitly counting logarithms in
the cumulative cross section (using exponent counting).
Ref.~\cite{Monni:2016ktx} then argues that the formal accuracy in terms of counting logarithms
$\ln(Q/q_T)$ will be the same, and only differ by subleading terms from the naive
result.

Compared to our exact solution, this procedure effectively corresponds to approximating the
rapidity logarithms $\ln(\nu/k_i)$ in the exact rapidity evolution kernel by logarithms
$\ln(\nu/k_1)$, which allows to pull them out of the convolutions.
Since this avoids the spurious singularity, one might expect that this approximation
is safer than the naive one of taking $\ln(\nu/k_i)\sim \ln(\nu/q_T)$. On the other hand,
these differences are all of apparent subleading nature, and it would be interesting to study
in more detail to what extent this approach reproduces the subleading terms in the $q_T$
spectrum that are included in the strict LL and NLL evolution in either momentum or Fourier space.

\section{Conclusion}
\label{sec:conclusion}

We have investigated solving differential equations for arbitrary distributions.
These arise naturally in differential spectra of observables resolving additional soft or collinear QCD radiation,
where the cancellation of IR singularities is encoded through plus distributions,
and the all-order logarithmic structure is fully encoded in (renormalization group) evolution equations.
Solving these equations allows the resummation of the logarithmic distributions to all orders,
provided that the boundary term in the solution is free of logarithms,
such that it can be reliably calculated in fixed-order perturbation theory.

We have introduced a technique for distributional scale setting $\mu = k\dis$ that allows
one to treat logarithmic distributions like ordinary logarithms. In particular, it
eliminates any logarithms contained in the boundary condition of plus distributions such as $[ \mu/k ]_+^\mu$.
It can be straightforwardly applied to solve distributional differential equations
for both one-dimensional and two-dimensional distributions,
where it ensures that the appearing boundary term is free of any logarithmic distributions.
This allows one to perform the RG evolution and resummation directly in distribution (momentum) space.
It thus enables the implementation of profile scales, the transition and matching to the full fixed-order distribution,
and the estimation of perturbative uncertainties through scale variations directly in distribution space.

The technique has been applied to obtain the resummation of the
transverse momentum ($q_T$) spectrum for the first time by solving the associated
evolution equations in momentum space.
We showed that a well-known spurious singularity in the spectrum arises from wrong scale setting,
i.e.~wrong boundary terms in the RG evolution, causing an incorrect treatment of energetic emissions,
and which is cured by a proper distributional scale setting.
This yields a well-defined resummation of the $q_T$ spectrum,
whose resummation accuracy is strictly defined by the perturbative expansion of the associated
anomalous dimensions (and boundary terms) without requiring to count explicit powers of logarithms.
We indeed find that trying to specify the logarithmic accuracy by explicitly counting logarithms $\ln(Q/q_T)$
in the spectrum can be ill defined, and partly is the reason for the spurious divergences encountered in previous attempts
to perform resummation in momentum space.

Previous attempts at a momentum-space resummation were partially motivated by trying to avoid nonperturbative effects in the
$q_T$ spectrum, which are unavoidable in the Fourier space resummation due to integrating $\as(1/b_T)$
over its Landau pole. We find that analogous nonperturbative effects also appear in the rapidity evolution kernel
in the strict momentum-space resummation, because real emissions with momentum $\ki{i}$ naturally scale with $\as(k_i)$, which is
necessary to suppress energetic emissions. We discussed how the nonperturbative contributions to the
rapidity anomalous dimension can be isolated in momentum space, which closely reproduces the common treatment in
Fourier space.
We also showed that nonperturbative effects arising from integrating over the $\ki{i}\to 0$ region inside
convolutions are generically suppressed as $\LQCD^2 / q_T^2$,
such that they do not spoil the predictivity of the resummation for perturbative $q_T$.

The correct momentum-space rapidity evolution involves an intricate iterative convolution structure.
Its numerical implementation is nontrivial, which we plan to address in future work.
The rapidity evolution has so far been performed in Fourier $b_T$ space,
where logarithms $\ln(Q b_T)$ rather than $\ln(Q/q_T)$ are resummed.
While both approaches are formally equivalent, our general analysis shows that the
boundary conditions employed in the evolution intrinsically differ to all orders.
In the future, it would be interesting to compare them numerically,
as they probe different subleading terms to all orders, and this could also provide
new insight into nonperturbative effects.

\acknowledgments
The authors thank Piotr Pietrulewicz for helpful comments on the manuscript and Stefan Liebler for useful discussions.
The authors thank the Erwin Schr\"{o}dinger Institute program ``Challenges and Concepts for Field Theory and Applications in the Era of the LHC'' and the NIKHEF for hospitality while portions of this work were completed.
This work was supported by the DFG Emmy-Noether Grant No. TA 867/1-1 and the PIER Helmholtz Graduate school.

\appendix
\section{Notation and conventions}
\label{app:notation}

Most of this paper is devoted to transverse momentum dependent distributions, but \sec{distr scale setting} also discusses one-dimensional distributions.
To distinguish the two cases, the momentum-space arguments are typically called $k$ for the one-dimensional case and $\pt,\kt,\qt$ for the two-dimensional case. The subscript $_T$ is always added to make the distinction clear.
For magnitudes of vectors, we drop the explicit vector sign and simply write $|\pt|^2 = p_T^2$ etc.

\subsection{Fourier transformations}

By default we work in distribution space.
The conjugate functions in Fourier space are always denoted with a tilde.
The Fourier conjugate variable in the one-dimensional case is typically called $y$, while for the two-dimensional case it is called $\bt$. Our conventions for the Fourier transformation for the one-dimensional case are
\begin{align}
f(k) &= \int\! \frac{\del y}{2\pi}\, e^{+i x y} \tilde f(y)
\,, \\
\tilde f(y) &= \int\! \del k\, e^{-i k y} f(k)
\,,\end{align}
and for the two-dimensional case
\begin{align}
f(\pt) &= \int\!\frac{\del^2\bt}{(2\pi)^2}\, e^{+i \pt \cdot \bt} \tilde f(\bt)
\,, \\
\tilde f(\bt) &= \int\! \del^2\pt\, e^{-i \pt \cdot \bt} f(\pt)
\,.\end{align}
For azimuthally symmetric functions, $f(\pt) \equiv f(p_T)$, the latter simplify to
\begin{align}
f(p_T) &= \frac{1}{2\pi}\int_0^\infty\! \del b_T\, b_T J_0(b_T p_T)\, \tilde f(b_T)
\,, \\
\tilde f(b_T) &=  2 \pi \int_0^\infty\! \del p_T\, p_T J_0(b_T p_T)  f(p_T)
\,,\end{align}
where $J_0(x)$ is the 0th-order Bessel function.

\subsection{Convolutions}

One-dimensional convolutions are defined as
\begin{equation}  \label{eq:conv 1d}
(f \otimes g)(k, \ldots) \equiv \int\! \del k'\, f(k-k',\ldots)\, g(k',\ldots)
\,,\end{equation}
where the dots stand for possible additional arguments of the functions.
Multiple convolutions are abbreviated as
\begin{equation} \label{eq:multiple convolutions 1d}
(f \otimes^n)(k)
\equiv \int\! \del k_1 \dots \del k_n\, f(k_1) \dots f(k_n)\, \delta(k - k_1 - \cdots - k_n)
\,.\end{equation}

Two-dimensional convolutions are defined as
\begin{align}  \label{eq:conv}
(f \otimes g)(\pt, \ldots)
&= \int\!\del^2\ki{1}\del^2\ki{2}\, f(\ki{1},\ldots)\, g(\ki{2},\ldots)\, \delta(\pt - \ki{1} - \ki{2})
\nn \\
&= \int\!\del^2\kt\, f(\pt - \kt, \ldots)\, g(\kt, \ldots)
\,,\end{align}
where the dots stand again for possible additional arguments of the functions.
Multiple convolutions are abbreviated as
\begin{equation}  \label{eq:multiple convolutions}
(f \otimes^n)(\pt)
\equiv \int\!\del^2\ki{1} \dotsb \del^2\ki{n}\, f(\ki{1}) \dots f(\ki{n})\, \delta(\pt - \ki{1} - \cdots - \ki{n})
\,.\end{equation}

\subsection{Fixed-order perturbative expansions}

We make frequent use of fixed-order expansions in $\as$. The expansion coefficients
of beta function and cusp anomalous dimension are defined as
\begin{align} \label{eq:Gammabetacoeffs}
\frac{\del\as}{\del\ln\mu} = \beta(\as) = -2 \as \sum_{n=0}^\infty \beta_n \Bigl(\frac{\as}{4\pi}\Bigr)^{n+1}
\,, \qquad
\GammaC(\as) = \sum_{n=0}^\infty \Gamma_n \Bigl(\frac{\as}{4\pi}\Bigr)^{n+1}
\,.\end{align}
Similarly, the constant noncusp pieces of all anomalous dimensions are expanded as
\begin{equation}
\gamma(\as) = \sum_{n=0}^\infty \gamma_n \Bigl(\frac{\as}{4\pi}\Bigr)^{n+1}
\,.\end{equation}
The soft function, and analogously other functions, are expanded as
\begin{equation}
S(\pt,\mu,\nu) = \sum_{n=0}^\infty S^{(n)}(\pt,\mu,\nu) \biggl[\frac{\as(\mu)}{4\pi} \biggr]^n
\,.\end{equation}

\section{One-dimensional plus distributions}
\label{app:plus distributions 1d}

For completeness we collect and extend the definitions and formulas for one-dimensional plus distributions from refs.~\cite{Ligeti:2008ac, Jain:2008gb}.

\subsection{Definition}

For a function $g(x)$ that has support for $x \geq 0$ and diverges less than $1/x^2$ for $x \to 0$,
the defining properties of its plus distributions with boundary condition $x_0 > 0$ are
\begin{align}  \label{eq:plus func definition 1d}
\Bigl[\theta(x) g(x) \Bigr]^{x_0}_+ &= \theta(x)\, g(x) \qquad \text{for $x \neq 0$}
\,,\\
 \int^{x_0}\! \del x'\, \Bigl[\theta(x') g(x')\Bigr]^{x_0}_+ &= 0
\,.\end{align}
The lower limit of integration is kept implicit and formally has to include the singularity at $x=0$.
An explicit definition is given by~\cite{Ligeti:2008ac}
\begin{align}
\Bigl[ g(x) \Bigr]_+^{x_0}
&\equiv \Bigl[ \theta(x) g(x) \Bigr]_+^{x_0}
\nn \\
&= \lim_{\epsilon \to 0} \frac{\del}{\del x}\! \biggl[ \theta(x-\epsilon) \int_{x_0}^x \del x'\, g(x') \biggr]
\nn \\
&= \lim_{\epsilon \to 0} \biggl[ \theta(x-\epsilon) g(x) + \delta(x - \epsilon) \int_{x_0}^x \del x'\, g(x') \biggr]
\,.\end{align}
Equivalently, we have
\begin{align} \label{eq:thetaderivative}
\lim_{\epsilon\to0} \frac{\del}{\del{x}} \bigl[ \theta(x-\epsilon)\, G(x) \bigr]
&= \lim_{\epsilon\to0} \frac{\del}{\del{x}} \bigl[ \theta(x -\epsilon) \bigl(G(x) - G(x_0) + G(x_0) \bigr) \bigr]
\nn \\
&= \biggl[\theta(x) \frac{\del G(x)}{\del x}\biggr]_+^{x_0} + \delta(x)\, G(x_0)
\,.\end{align}

From the above definitions, it follows that the boundary condition can be shifted using
\begin{equation} \label{eq:plus func rescaling 1d}
\Bigl[ \theta(x) g(x) \Bigr]_+^{x_0}
= \Bigl[\theta(x) g(x) \Bigr]_+^{x_1} + \delta(x) \int_{x_0}^{x_1}\! \del x' \, g(x')
\,.\end{equation}
The derivative with respect to the boundary value is thus given by
\begin{equation} \label{eq:plus func derivative 1d}
\frac{\del}{\del{x_0}} \Bigl[ \theta(x) g(x) \Bigr]^{x_0}_+ = - g({x_0})\, \delta(x)
\,.\end{equation}
More generally, if $g$ itself depends on $x_0$, we have
\begin{equation} \label{eq:plus func derivative 1d 2}
\frac{\del}{\del{x_0}} \Bigl[ \theta(x) g(x,x_0) \Bigr]^{x_0}_+
= \biggl[ \frac{\del}{\del x_0} g(x,x_0) \biggr]_+^{x_0} - g(x_0,x_0)\, \delta(x)
\,.\end{equation}

Following ref.~\cite{Ligeti:2008ac}, we denote the standard plus distributions with boundary condition $x_0 = 1$ as
\begin{equation} \label{eq:plusdef_standard}
\PlusLog{n}(x) = \biggl[ \frac{\theta(x) \ln^n x}{x}\biggr]_+^1
\,, \qquad
\PlusPol{a}(x) = \biggl[ \frac{\theta(x)}{x^{1-a}} \biggr]_+^1
\end{equation}
In addition, for dimensionful arguments, we define
\begin{equation}
\PlusLog{n}(k,\mu)
\equiv \frac{1}{\mu}\cL_n \Bigl(\frac{k}{\mu}\Bigr)
\equiv \biggl[ \frac{\theta(k)}{k}\ln^n\frac{k}{\mu}\biggr]_+^\mu
\,, \qquad
\PlusPol{a}(k, \mu)
\equiv \frac{1}{\mu}\cL^a \Bigl(\frac{k}{\mu}\Bigr)
\equiv \biggl[ \frac{\theta(k)}{k} \Bigl(\frac{k}{\mu}\Bigr)^a \biggr]_+^\mu
\end{equation}
They are related by
\begin{equation}  \label{eq:plus func relation 1d}
\cL_n(x) = \frac{\del^n}{\del a^n} \PlusPol{a}(x) \bigg\rvert_{a=0}
\,,\qquad
\PlusLog{n}(k, \mu) = \frac{\del^n}{\del a^n} \PlusPol{a}(k,\mu) \bigg\rvert_{a=0}
\,.\end{equation}

\subsection{Fourier transformation}
\label{app:Fourier trafo 1d}

\begin{table}[t!]
\centering
\begin{tabular}{c|l}
\hline\hline
$\PlusLog{n}(k, \mu) $ & $\text{FT}[\PlusLog{n}(k,\mu)]$ \\ \hline
$\PlusLog{0}(k, \mu)$
& $ - L_y $
\\ \hline
$\PlusLog{1}(k, \mu)$
& $+\frac{1}{2} L_y^2 + \frac{\pi^2}{12}$
 \\ \hline
$\PlusLog{2}(k, \mu)$
& $-\frac{1}{3} L_y^3 - \frac{\pi^2}{6} L_y - \frac{2}{3} \zeta_3$
\\ \hline
$\PlusLog{3}(k, \mu)$
& $+\frac{1}{4} L_y^4 + \frac{\pi^2}{4} L_y^2 + 2 \zeta_3 L_y + \frac{3}{80}\pi^4 $
\\ \hline
$\PlusLog{4}(k, \mu)$
& $-\frac{1}{5} L_y^5 - \frac{\pi^2}{3} L_y^3 - 4 \zeta_3 L_y^2 - \frac{3}{20}\pi^4 L_y - \bigl( \frac{2}{3}\pi^2\zeta_3 + \frac{24}{5} \zeta_5 \bigr)$
\\ \hline
$\PlusLog{5}(k, \mu)$
& $+\frac{1}{6} L_y^6 + \frac{5}{12} \pi^2 L_y^4 + \frac{20}{3}\zeta_3 L_y^3 + \frac{3}{8} \pi^4 L_y^2 + \bigl(\frac{10}{3}\pi^2 + 24 \zeta_5\bigr) L_y + \bigl(\frac{61}{1008}\pi^6 + \frac{20}{3} \zeta_3^2\bigr)$
\\ \hline\hline
\end{tabular}
\caption{Fourier transform of $\PlusLog{n}(k, \mu)$ for $n \le 5$. Results are expressed in terms of $L_y = \ln(i y \mu e^{\gamma_E})$. See  \eq{Ln to y}.}
\label{tbl:Ln to y space}
\end{table}

\begin{table}[t!]
\centering
\begin{tabular}{c|l}
\hline\hline
$L_y^n$ & $\text{FT}^{-1}[L_y^n]$ \\ \hline
1
& $\delta(k)$
\\ \hline
$L_y$
& $-\PlusLog{0}(k,\mu) $
 \\ \hline
$L_y^2$
& $2 \PlusLog{1}(k,\mu) - \frac{\pi^2}{6} \delta(k) $
 \\ \hline
$L_y^3$
& $-3 \PlusLog{2}(k,\mu) + \frac{\pi^2}{2} \PlusLog{0}(k,\mu) - 2\zeta_3\delta(k)$
 \\ \hline
$L_y^4$
& $4\PlusLog{3}(k,\mu) -2 \pi^2 \PlusLog{1}(k,\mu) + 8 \zeta_3 \PlusLog{0}(k,\mu) + \frac{\pi^4}{60}\delta(k)$
 \\ \hline
$L_y^5$
& $-5 \PlusLog{4}(k,\mu) + 5\pi^2 \PlusLog{2}(k,\mu) - 40 \zeta_3 \PlusLog{1}(k,\mu) - \frac{\pi^4}{12} \PlusLog{0}(k,\mu)
+ \bigl(\frac{10}{3}\pi^2\zeta_3 - 24 \zeta_5\bigr) \delta(k)$
 \\ \hline
$L_y^6$
& $6 \PlusLog{5}(k,\mu) - 10 \pi^2\PlusLog{3}(k,\mu) + 120\zeta_3 \PlusLog{2}(k,\mu) + \frac{\pi^4}{2} \PlusLog{1}(k,\mu)
+ (144 \zeta_5 - 20\pi^2 \zeta_3) \PlusLog{0}(k,\mu) $
\\ & $+ \bigl(40\zeta_3^2 - \frac{5}{168}\pi^6\bigr)\delta(k)$
 \\ \hline\hline
\end{tabular}
\caption{Fourier transform of $L_y^n = \ln^n(i y \mu e^{\gamma_E})$ for $n \le 6$. See \eq{Ln to k}.}
\label{tbl:Ln to k space}
\end{table}

\begin{table}[t!]
\centering
\begin{tabular}{c|c|c}
\hline\hline
 $\dR[1]{n}$ & Exact value & Numerical value  \\ \hline
 $\dR[1]{0}$ & $1$ &  1 \\ \hline
 $\dR[1]{1}$ & $0$ &  0 \\ \hline
 $\dR[1]{2}$ & $\frac{\pi^2}{6}$ & $\approx 1.64  = 0.822\times 2!$ \\ \hline
 $\dR[1]{3}$ & $-2 \zeta_3$     & $\approx -2.40  = -0.401\times 3!$ \\ \hline
 $\dR[1]{4}$ & $\frac{3 \pi^2}{20}$ &$\approx 14.6 = 0.609\times 4!$ \\ \hline
 $\dR[1]{5}$ & $-\frac{10\pi^2}{3} \zeta_3 - 24 \zeta_5$    & $\approx -64.4 = -0.537 \times 5!$  \\ \hline
 $\dR[1]{6}$ & $\frac{61 \pi^6}{168} + 40\zeta_3^2$  & $\approx 406.9 = 0.565 \times 6!$  \\ \hline
 $\dR[1]{7}$ & $-\frac{21 \pi^4}{2} \zeta_3 - 84\pi^2 \zeta_5 - 720 \zeta_7$  & $\approx -2815 = -0.559\times 7!$  \\ \hline\hline
\end{tabular}
\caption{The first values of $\dR[1]{n}$, defined in \eq{dR1}.}
\label{tbl:dR1}
\end{table}

\begin{table}[t!]
\centering
\begin{tabular}{c|c|c}
\hline\hline
 $\dRc[1]{n}$ & Exact value & Numerical value  \\ \hline
 $\dRc[1]{0}$ & $1$ &  1 \\ \hline
 $\dRc[1]{1}$ & $0$ &  0 \\ \hline
 $\dRc[1]{2}$ & $-\frac{\pi^2}{6}$ & $\approx -1.64$ \\ \hline
 $\dRc[1]{3}$ & $-2\zeta_3$ & $\approx -2.40$ \\ \hline
 $\dRc[1]{4}$ & $\frac{\pi^4}{60}$ & $\approx 1.62$ \\ \hline
 $\dRc[1]{5}$ & $\frac{10\pi^2}{3}\zeta_3 - 24\zeta_5$ & $\approx 14.7$ \\ \hline
 $\dRc[1]{6}$ & $-\frac{5\pi^6}{168} + 40\zeta_3^2$ & $\approx 29.2$ \\ \hline
 $\dRc[1]{7}$ & $-\frac{7\pi^4}{6}\zeta_3 + 84\pi^2\zeta_5 - 720\zeta_7$ & $\approx -2.96$ \\ \hline\hline
\end{tabular}
\caption{The first values of $\dRc[1]{n}$, defined in \eq{dR1c}.}
\label{tbl:dR1c}
\end{table}

The Fourier transformation of a plus function $\PlusPol{a}(k,\mu)$ with respect to $y = y - i0$ is given by
\begin{align}  \label{eq:La to y}
 \int\! \del k\, e^{-i k y} \PlusPol{a}(k,\mu)
 &= \frac{1}{a} \bigl[ (i \mu y)^{-a} \Gamma(1+a) - 1 \bigr]
  = \frac{1}{a} \Bigl[ e^{-a L_y} R_1(a) - 1 \Bigr]
\,,\end{align}
which holds for $a > -1$ through analytic continuation.
In the second step we introduced the abbreviations
\begin{align}
 L_y &= \ln(i y \mu e^{\gamma_E})
\,,\qquad
 R_1(a) = e^{\gamma_E a} \Gamma(1+a)
\,.\end{align}
An explicit form of the Fourier transform for plus functions $\PlusLog{n}(k,\mu)$ follows from \eq{plus func relation 1d},
\begin{align} \label{eq:Ln to y}
 \int\del k\, e^{-i k y} \PlusLog{n}(k,\mu)
 &= \frac{\del^n}{\del a^n}\bigg|_{a=0} \frac{1}{a} \Bigl[ e^{-a L_y} R_1(a) - 1 \Bigr] \nn \\
 &= \frac{1}{n+1} \sum_{k=0}^{n+1} (-1)^k \binom{n+1}{k} L_y^k\, \dR[1]{n+1-k}
\,,\end{align}
and the inverse is given by \cite{Jain:2008gb}
\begin{align} \label{eq:Ln to k}
 \int \frac{\del y}{2\pi} e^{i k y} L_y^n
 &= \frac{\del^n}{\del a^n} \left\{ \frac{-a}{R_1(-a)} \PlusPol{-a}(k) +  \frac{\delta(k)}{R_1(-a)} \right\}_{a=0} \nn \\
 &= \sum_{k=0}^{n-1} (-1)^{k+1} n \binom{n-1}{k} \dRc[1]{n-k-1} \PlusLog{k}(k,\mu) +  \dRc[1]{n}\delta(k)
\,,\end{align}
where the quantities $\dR[1]{n}, \dRc[1]{n}$ occurring in the coefficients are defined as
\begin{align}
\label{eq:dR1}
 \dR[1]{n} &= \frac{\del^n}{\del a^n}\Big\rvert_{a=0} e^{\gamma_E a} \Gamma(1+a)
\,, \\
\label{eq:dR1c}
 \dRc[1]{n} &= \frac{\del^n}{\del a^n}\Big\rvert_{a=0} \frac{e^{\gamma_E a}}{\Gamma(1 - a)}
\,.\end{align}
They fulfill the sum rule
\begin{equation}
 \sum_{k=0}^n \binom{n}{k} (-1)^k \dR[1]{n-k} \dRc[1]{k} = 0
\,.\end{equation}
The values can be conveniently obtained from the expansions
\begin{align}
 \sum_{n=0}^\infty \frac{\dR[1]{n} x^n}{n!} &= \exp\biggl[ + \sum_{n=1}^\infty \frac{\zeta(n+1)}{n+1} (-x)^{n+1} \biggr]
\,, \\
 \sum_{n=0}^\infty \frac{\dRc[1]{n} x^n}{n!} &= \exp\biggl[ - \sum_{n=1}^\infty \frac{\zeta(n+1)}{n+1} x^{n+1} \biggr]
\,.\end{align}
Numerically we find the asymptotic behavior
\begin{align}
 \dR[1]{n} &\approx 0.561 \times (-1)^n n! \qquad \text{for } n \gg 1
\,.\end{align}
The factorial behavior of $\dR[1]{n}$ reflects that the Taylor series of $R_1(a) = e^{\gamma_E a} \Gamma(1+a)$ only converges for $|a| < 1$.
For $\dR[1]{n}$ there is no such simple asymptotic behavior, but due to the infinite radius of convergence of $\tilde R_1(a) = e^{\gamma_E a} / \Gamma(1-a)$, it is expected to grow much less severely.
This is confirmed by the first values shown in \ref{tbl:dR1c}.

For illustration,
table \ref{tbl:Ln to y space} shows the Fourier transforms of $\PlusLog{n}(k,\mu)$ for $n \le 5$, table \ref{tbl:Ln to k space} the Fourier transforms of $L_y^n$ for $n \le 6$.
The first few $\dR[1]{n}$ and $\dRc[1]{n}$ are given in tables \ref{tbl:dR1} and \ref{tbl:dR1c}.

\subsection{Convolutions}

The convolution of two plus distributions is given by
\begin{align}
\int\!\del x'\, \cL^a(x - x')\, \cL^b(x')
&= \biggl[\cL^{a+b}(x) + \frac{\delta(x)}{a+b} \biggr] V_1(a,b) \nn\\&\quad
  + \Bigl(\frac{1}{a}+\frac{1}{b}\Bigr) \cL^{a+b}(x)
  - \frac{1}{b}\, \cL^a(x) - \frac{1}{a}\, \cL^b(x)
\,,\end{align}
where $V_1(a,b)$ is defined by
\begin{equation}
V_1(a,b) = \frac{\Gamma(a)\,\Gamma(b)}{\Gamma(a+b)} - \frac{1}{a} - \frac{1}{b}
\,,\end{equation}
which satisfies $V_1(0,0) = 0$.  Taking derivatives with respect to $a$ and $b$ we
can get the corresponding formulas for convolutions of the form $\cL_n\otimes \cL^a$
and $\cL_m \otimes \cL_n$. The explicit results can be found in Appendix B of \mycite{Ligeti:2008ac}.

An important special case are multiple convolutions of $\cL_n$, which is given by
\begin{align} 
(\PlusLog{0}\, \otimes^n)(k, \mu)
&= (-1)^n\sum_{i=0}^{n-1} (-1)^{i+1} n \binom{n-1}{i} \dRc[1]{n-i-1} \PlusLog{i}(k,\mu) + (-1)^n\dRc[1]{n} \delta(k)
\,.
\end{align}

\section{Two-dimensional plus distributions}
\label{app:plus distributions pt}

\subsection{Definition}

Plus distributions in momentum space are uniquely defined by the two conditions
\begin{align}  \label{eq:plus func definition}
\Bigl[g(\pt) \Bigr]^\mu_+ &= g(\pt)
\qquad\text{for } |\pt| > 0
\, \\
\int_{|\pt| \le \mu}\! \del^2\pt\, \Bigl[g(\pt)\Bigr]^\mu_+ &= 0
\,,\end{align}
where $g(\pt)$ diverges at most as $1/p_T^2$ for $p_T \to 0$.
The boundary condition can be shifted using
\begin{equation} \label{eq:plus func rescaling 2d}
\Bigl[g(\pt)\Bigr]^\mu_+
= \Bigl[g(\pt)\Bigr]^{\xi}_+ \mp \delta(\pt)  \int_{\xi < |\kt| < \mu} \hspace{-3ex}\del^2\kt\, g(\kt)
\,,\end{equation}
where the $-$ sign holds for $\xi < \mu$ and the $+$ sign for $\xi > \mu$.
For azimuthally symmetric functions, $g(\pt) \equiv g(|\pt|)$, this simplifies to
\begin{equation} \label{eq:plus func rescaling}
\Bigl[g(|\pt|)\Bigr]^\mu_+ = \Bigl[g(|\pt|)\Bigr]^{\xi}_+ - 2\pi \delta(\pt) \int_{\xi}^\mu\! \del k_T\,k_T\,g(k_T)
\,.\end{equation}
It is important to keep in mind that it is nevertheless defined as a two-dimensional distribution, even though the distribution effectively contains a scalar function.

It follows that the derivative with respect to the boundary condition is given by
\begin{equation} \label{eq:plus func derivative}
\mu \frac{\del}{\del\mu} \Bigl[g(|\pt|) \Bigr]^\mu_+ = - 2 \pi \mu^2 g(\mu)\, \delta(\pt)
\,.\end{equation}
More generally, if $g$ itself depends on $\mu$, then
\begin{equation}
\label{eq:plus func derivative 2}
\mu \frac{\del}{\del\mu} \Bigl[g(|\pt|, \mu) \Bigr]^\mu_+
= - 2 \pi \mu^2 g(\mu,\mu)\, \delta(\pt)  + \biggl[ \mu \frac{\del g(|\pt|,\mu)}{\del\mu} \biggr]_+^\mu
\,.\end{equation}

For a scalar input function $f(p_T)$, the azimuthally symmetric two-dimensional plus distribution
can be defined by the derivative
\begin{align} \label{eq:cumulant inverse}
\frac{1}{2\pi p_T} \frac{\del}{\del p_T} \theta(p_T) &= \delta(\pt)
\,,\qquad
\int_{|\kt| \le p_T}\! \del^2\kt\, \delta(\kt) = \theta(p_T)
\,, \nn \\
\frac{1}{2\pi p_T} \frac{\del}{\del p_T} \bigl[ \theta(p_T) f(p_T) \bigr]
&= \biggl[ \frac{1}{2 \pi p_T} \frac{\del f(p_T)}{\del p_T} \biggr]_+^\xi + \delta(\pt) f(\xi)
\,.\end{align}
Note that the appearance of the $\theta$-function is crucial to render the derivative well defined in form of a plus distribution.

We define the frequently occurring logarithmic plus distributions as
\begin{align} \label{eq:plus func pol}
\PlusPol{a}(\pt, \mu)
&\equiv \frac{1}{\pi \mu^2} \biggl[\biggl(\frac{\pt^2}{\mu^2}\biggr)^{a-1} \biggr]_+^\mu
\equiv \frac{1}{\pi \mu^2} \PlusPol{a}\biggl(\frac{\pt^2}{\mu^2}\biggr)
 \,, \\
\label{eq:plus func log}
\PlusLog{n}(\pt, \mu)
&\equiv \frac{1}{\pi \mu^2} \biggl[ \frac{\mu^2}{\pt^2} \ln^n\frac{\pt^2}{\mu^2} \biggr]_+^\mu
\equiv \frac{1}{\pi \mu^2} \PlusLog{n}\biggl(\frac{\pt^2}{\mu^2}\biggr)
\,,\end{align}
where the $\PlusPol{a}(x)$ and $\PlusLog{n}(x)$ on the right-hand side are the standard ones for
dimensionless arguments with boundary condition $x_0 = 1$ as defined in \eq{plusdef_standard}.
They are related by
\begin{equation} \label{eq:plus func relation}
\PlusLog{n}(\pt, \mu) = \frac{\del^n}{\del a^n} \mathcal{L}^a(\pt, \mu) \bigg\rvert_{a=0}
\,.\end{equation}
Their cumulant and inverse cumulants are
\begin{align}   \label{eq:cumulant to plus func}
\int_{|\kt| \le p_T}\! \del^2\kt \, \PlusLog{n}(\kt,\mu) &=  \frac{\theta(p_T)}{n+1} \ln^{n+1} \frac{p_T^2}{\mu^2}
\,, \nn \\
\frac{1}{2\pi p_T} \frac{\del}{\del p_T} \biggl[\theta(p_T) \ln^{n}\frac{p_T^2}{\mu^2} \biggr]
&= n\, \PlusLog{n-1}(\pt,\mu)
\qquad\text{for } n\ge1
\,.\end{align}

Finally note that our definition of $\cL_n(\pt, \mu)$ is related to that in ref.~\cite{Chiu:2012ir} by
\begin{equation}
 \PlusLog{n}(\pt, \mu) = 2 (-1)^n \cL_n^0(\mu,\pt;\mu)
\,,\end{equation}
where $\cL_n^0(\mu,\pt;\mu)$ is defined in eq.\ (F.1) in ref.~\cite{Chiu:2012ir}.

\subsection{Fourier transformation}
\label{app:plus distributions ips}

The Fourier transform of $\PlusPol{a}(\pt,\mu)$ is given by
\begin{equation} \label{eq:La to b}
\int\! \del^2\pt\, e^{-i \pt \cdot \bt} \PlusPol{a}(\pt,\mu)
= \frac{1}{a} \biggl[ \biggl( \frac{b_T^2 \mu^2}{4} \biggr)^{-a} \frac{\Gamma(1+a)}{\Gamma(1-a)} - 1 \biggr]
\,.\end{equation}
It is convenient to express the Fourier transform of $\PlusLog{n}(\pt,\mu)$ as polynomials of
\begin{equation}
L_b \equiv \ln\frac{b_T^2 \mu^2}{b_0^2}
\,, \qquad
b_0 = 2 e^{-\gamma_E} \approx 1.12291\dots
\,.\end{equation}
The Fourier transform of $\PlusLog{n}(\pt,\mu)$ then follows from \eq{La to b} using \eq{plus func relation},
\begin{align} \label{eq:Ln to b}
\int\!\del^2\pt\, e^{-i \pt \cdot \bt} \PlusLog{n}(\pt,\mu)
&= \frac{\del^n}{\del a^n} \frac{1}{a} \biggl[ \biggl( \frac{b_T^2 \mu^2}{4} \biggr)^{-a} \frac{\Gamma(1+a)}{\Gamma(1-a)} - 1 \biggr] \biggr|_{a=0}
\nn \\
&= \frac{1}{n+1} \sum_{k=0}^{n+1} (-1)^k \binom{n+1}{k} \dR{n+1-k} L_b^k
\,.\end{align}
The inverse transformation is given by
\begin{align} \label{eq:Ln to pt}
\int\! \frac{\del^2\bt}{(2\pi)^2}\, e^{i \pt \cdot \bt} L_b^n
&= \frac{\del^n}{\del a^n} \bigl[ - a R_2(a) \PlusPol{-a}(\pt,\mu) + R_2(a) \delta(\pt) \bigr]  \biggr|_{a=0}
\nn \\
&= \sum_{k=0}^{n-1} (-1)^{k+1} n \binom{n-1}{k} \dR{n-k-1} \PlusLog{k}(\pt,\mu) +  \dR{n} \delta(\pt)
\,.\end{align}
The constants $\dR{n}$ occurring in the coefficients are defined as
\begin{equation} \label{eq:R(n)}
\dR{n} = \frac{\del^n}{\del a^n}\, e^{2\gamma_E a} \frac{\Gamma(1+a)}{\Gamma(1-a)}
\bigg\rvert_{a=0}
\,.\end{equation}
They fulfill the useful property
\begin{equation}
\sum_{k=0}^n (-1)^k \binom{n}{k} \dR{k} \dR{n-k} = 0
\,.\end{equation}
By integrating \eq{Ln to pt} over $|\pt| \le \mu$, it also follows that
\begin{equation} \label{eq:dR2 integral}
\dR{n} = \int_0^\infty \del x\, J_1(x) \ln^n\frac{x^2}{b_0^2}
\,.\end{equation}
They can also be obtained from the relation \cite{Kulesza:1999gm}
\begin{equation}
\sum_{n=0}^\infty \frac{\dR{n}}{n!} x^n
= \exp\biggl[ -2 \sum_{n = 1}^\infty \frac{\zeta(2n+1)}{2n+1} x^{2n+1} \biggr]
\,.\end{equation}
Approximating the zeta-function with its asymptotic value $\zeta(n) \approx 1$, it is easy to see from this equation that $\dR[2]{n}$ scales as $n!$ for large $n$. Numerically we find
\begin{equation} \label{eq:R2n_largen}
\dR[2]{n} \approx 0.315 \times (-1)^n n!
\qquad\text{for } n \gg 1
\,.\end{equation}

For illustration, we list in table~\ref{tbl:Ln pt to b space} the Fourier transforms of $\PlusLog{n}(\pt,\mu)$ for $n \le 5$, and in table~\ref{tbl:Ln b to pt space} the inverse Fourier transforms of $L_b^n$ for $n \le 6$.
The first few $\dR{n}$ are given in table \ref{tbl:Rn}.

\begin{table}[t!]
\centering
\begin{tabular}{c|l}
\hline\hline
$\PlusLog{n}(\pt, \mu) $ & $\text{FT}[\PlusLog{n}(\pt,\mu)]$ \\ \hline
$\PlusLog{0}(\pt,\mu)$
& $-L_b$
\\ \hline
$\PlusLog{1}(\pt,\mu)$
& $+\frac{1}{2} L_b^2$
 \\ \hline
$\PlusLog{2}(\pt,\mu)$
& $-\frac{1}{3} L_b^3 - \frac{4}{3}\zeta_3$
\\ \hline
$\PlusLog{3}(\pt,\mu)$
& $+\frac{1}{4} L_b^4 + 4 \zeta_3 L_b$
\\ \hline
$\PlusLog{4}(\pt,\mu)$
& $-\frac{1}{5} L_b^5 - 8 \zeta_3 L_b^2 - \frac{48}{5}\zeta_5$
\\ \hline
$\PlusLog{5}(\pt,\mu)$
& $+\frac{1}{6} L_b^6 + \frac{40}{3}\zeta_3 L_b^3 + 48 \zeta_5 L_b +\frac{80}{3} \zeta_3^2$
\\ \hline\hline
\end{tabular}
\caption{Fourier transform of $\PlusLog{n}(\pt, \mu)$ to $\bt$-space for $n \le 5$. Results are expressed in terms of $L_b = \ln(b_T^2 \mu^2 e^{2\gamma_E}/4)$. See  \eq{Ln to b}.}
\label{tbl:Ln pt to b space}
\end{table}

\begin{table}[t!]
\centering
\begin{tabular}{c|l}
\hline\hline
$L_b^n$ & $\text{FT}^{-1}[L_b^n]$ \\ \hline
$1$
& $\delta(\pt)$
\\ \hline
$L_b$
& $-\PlusLog{0}(\pt, \mu)$
 \\ \hline
$L_b^2$
& $+2 \PlusLog{1}(\pt, \mu)$
 \\ \hline
$L_b^3$
& $-3 \PlusLog{2}(\pt, \mu) - 4 \zeta_3 \delta(\pt)$
 \\ \hline
$L_b^4$
& $+4 \PlusLog{3}(\pt, \mu) + 16 \zeta_3 \PlusLog{0}(\pt, \mu)$
 \\ \hline
$L_b^5$
& $ -5 \PlusLog{4}(\pt, \mu) - 80 \zeta_3 \PlusLog{1}(\pt, \mu) - 48 \zeta_5 \delta(\pt)$
 \\ \hline
$L_b^6$
& $+6 \PlusLog{5}(\pt, \mu) + 240 \zeta_3 \PlusLog{2}(\pt, \mu) + 288 \zeta_5 \PlusLog{0}(\pt, \mu) + 160 \zeta_3^2 \delta(\pt)$
 \\ \hline\hline
\end{tabular}
\caption{Fourier transform of $L_b^n = \ln^n(b_T^2 \mu^2 e^{2\gamma_E}/4)$ to $\pt$-space for $n \le 6$. See \eq{Ln to pt}.}
\label{tbl:Ln b to pt space}
\end{table}

\begin{table}[t!]
\centering
\begin{tabular}{c|c|c}
\hline\hline
 $\dR{n}$ & Exact value & Numerical value \\ \hline
 $\dR{0}$ & $1$ &  $1$ \\ \hline
 $\dR{1}$ & $0$ & $0$ \\ \hline
 $\dR{2}$ & $0$ & $0$ \\ \hline
 $\dR{3}$ & $-4 \zeta_3$     & $\approx -4.81 =- 0.801 \times 3!$ \\ \hline
 $\dR{4}$ & $0$ & $0$ \\ \hline
 $\dR{5}$ & $-48 \zeta_5$    & $\approx -49.8 = -0.415 \times 5!$ \\ \hline
 $\dR{6}$ & $160 \zeta_3^2$  & $\approx 231.2 = 0.321 \times 6!\,\,\,$ \\ \hline
 $\dR{7}$ & $-1440 \zeta_7$  & $\approx -1452 = - 0.288 \times 7!$  \\ \hline\hline
\end{tabular}
\caption{The first values of $\dR[2]{n}$, defined in \eq{R(n)}.}
\label{tbl:Rn}
\end{table}

\subsection{Convolutions}
\label{app:convolutions}

Convolutions of plus functions can be conveniently calculated by transforming to impact parameter space with \eqs{La to b}{Ln to b}, where convolutions become simple products, and then Fourier transforming back using \eq{Ln to pt}.

The convolution of two $\PlusPol{a}$ is given by~\cite{Chiu:2012ir}
\begin{align} \label{eq:La conv Lb}
 (\PlusPol{a} \otimes \PlusPol{b})(\pt, \mu) = V_2(a,b) \mathcal{L}^{a+b}(\pt, \mu) - \frac{\mathcal{L}^a(\pt, \mu)}{b} - \frac{\mathcal{L}^b(\pt, \mu)}{a} + \biggl(\frac{V_2(a,b)}{a + b} - \frac{1}{a b} \biggr) \delta(\pt)
\,,
\end{align}
where
\begin{align}
 V_2(a,b)
 &= \frac{\Gamma(1-a-b)}{\Gamma(1-a)\Gamma(1-b)} \frac{\Gamma(a)\Gamma(b)}{\Gamma(a+b)} \,.
\end{align}
Convolutions of type $\PlusLog{n}\otimes\PlusPol{a}$ and $\PlusLog{n}\otimes\PlusLog{m}$ can be obtained by applying \eq{plus func relation} to both sides of the above relation by carefully taking the derivative with respect to $a$ and $b$.
An important special case are multiple convolutions of $\PlusLog{0}(\pt,\mu)$, which are found to be
\begin{align}  \label{eq:Plus0^n}
(\PlusLog{0}\, \otimes^n)(\pt, \mu)
&= \int\!\frac{\del^2\bt}{(2\pi)^2}\, e^{i \pt \cdot \bt} (-L_b)^n
\\ \nn
&= (-1)^n\sum_{k=0}^{n-1} (-1)^{k+1} n \binom{n-1}{k} \dR{n-k-1} \PlusLog{k}(\pt,\mu) +  (-1)^n \dR{n} \delta(\pt) \,.
\end{align}

\subsection{Integral relations}
\label{app:cumulants}

Here we collect a few important integrals with distributional scale setting according to \eq{integral scale setting},
\begin{equation}
\int_{p_T\dis}^\mu\! \frac{\del\mu'}{\mu'}\, f(\pt,\mu')
\equiv \frac{1}{2\pi p_T} \frac{\del}{\del p_T} \int_{k_T \le p_T}\! \del^2\kt \int_{p_T}^\mu\! \frac{\del\mu'}{\mu'} f(\kt,\mu')
\,,\end{equation}
which are useful to resum the soft function iteratively using \eqs{soft iterative solution}{soft iterative solution 2}
\begin{align}
 \int_{p_T\dis}^\mu \frac{\del\mu'}{\mu'} \delta(\pt) \ln^n\frac{\mu'^2}{\mu^2} &= - \frac{1}{2} \cL_n(\pt,\mu)
\,, \\
 \int_{p_T\dis}^\mu \frac{\del\mu'}{\mu'} \cL_m(\pt,\mu') \ln^n\frac{\mu'^2}{\mu^2} &= - \frac{1}{2} \frac{m!\,n!}{(n+m+1)!} \cL_{n+m+1}(\pt,\mu)
\,, \\
 \int_{p_T\dis}^\mu \frac{\del\mu'}{\mu'} \delta(\pt) \ln^n\frac{\mu'}{\nu = p_T\dis} &=  (-1/2)^{n+1} \cL_n(\pt,\mu)
\,,  \\
 \int_{p_T\dis}^\mu \frac{\del\mu'}{\mu'} \cL_m(\pt,\mu') \ln^n\frac{\mu'}{\nu = p_T\dis} &=  \frac{(-1/2)^{n+1}}{m+1} \cL_{n+m+1}(\pt,\mu)
\,.\end{align}
Setting $\nu = p_T\dis$ inside the last two integrals only makes sense within the distributional prescription.

Furthermore, the following two-dimensional generalization of \eq{deq_prop1} is also useful,
\begin{align}
 \delta(\pt) \ln^{n+1} \frac{\mu_0^2}{\mu^2} \Biggr|_{\mu_0 = p_T\dis} &= (n+1) \cL_{n}(\pt,\mu)
\,, \nn\\
 (m+1) \cL_m(\pt,\mu) \ln^n \frac{\mu_0^2}{\mu^2} \Biggr|_{\mu_0 = p_T\dis} &=  (n+m+1) \cL_{n+m}(\pt,\mu)
\,, \nn\\
 \cL_m(\pt,\mu_0) \ln^n \frac{\mu_0^2}{\mu^2} \Biggr|_{\mu_0 = p_T\dis} &= 0
\,,\end{align}
which holds for $n, m \ge 0$.

\section{Fixed-order expansion of the soft function}
\label{app:soft func}

For completeness we give the fixed order expansion of the soft function $S(\pt,\mu,\nu)$
to three loops, derived using distributional scale setting in \sec{soft func},
and show the relation to the soft function predicted through scale setting in Fourier space.
Similar expressions for the rapidity anomalous dimension $\gNu$ are given in \secs{gNu iterative}{gNu comparison}.
To simplify the formulas, we make use of the known one-loop result
\begin{equation}
\gNuCNC{0} = 0
\,,\qquad
\gMuSCNC{0} = 0
\,.\end{equation}
The soft function is expanded as
\begin{equation}
S(\pt,\mu,\nu) = \sum_{n=0}^\infty S^{(n)}(\pt,\mu,\nu) \ASmu^{n}
\,.\end{equation}
The coefficients through $\cO(\as^3)$ are expressed in terms of $L_\nu \equiv \ln(\mu/\nu)$ and $\PlusLog{n} \equiv \PlusLog{n}(\pt,\mu)$,
\begin{align}
S^{(0)} &= \delta(\pt)
\,, \\
S^{(1)}
&= - \Gamma_0\, \PlusLog{1}
  - 2\Gamma_0\, L_\nu\, \PlusLog{0}
 + S_1 \delta(\pt)
\,, \\
S^{(2)}
&= \frac{\Gamma_0^2}{2}\, \PlusLog{3}
 + \PlusLog{2} (3\Gamma_0^2\, L_\nu + \beta_0\Gamma_0 )
+ \PlusLog{1} ( 4\Gamma_0^2\, L_\nu^2 + 2\beta_0\Gamma_0\, L_\nu - \Gamma_1 - \Gamma_0 S_1 )
\nn \\ & \quad
+ \PlusLog{0} \Bigl[ L_\nu ( -2\Gamma_1 - 2\Gamma_0 S_1) - \frac{1}{2}(\gMuSCNC{1} + \gNuCNC{1}) - \beta_0 S_1
+ 2 \Gamma_0^2 \zeta_3 \Bigr]
\nn \\ & \quad
 + \delta(\pt)   \bigl[ L_\nu(-\gNuCNC{1} + 4\Gamma_0^2 \zeta_3) + S_2 \bigr]
\,, \\
S^{(3)}
&= -\frac{\Gamma_0^3}{8}\, \PlusLog{5}
 + \PlusLog{4} \Bigl( - \frac{5}{4}\Gamma_0^3 L_\nu -\frac{5}{6}\beta_0\Gamma_0^2 \Bigr)
\nn \\ & \quad
 + \PlusLog{3} \biggl( -4\Gamma_0^3\, L_\nu^2 - \frac{14}{3} \beta_0 \Gamma_0^2\, L_\nu - \beta_0^2\Gamma_0 + \Gamma_0 \Gamma_1 + \frac{1}{2}\Gamma_0^2 S_1 \biggr)
\nn \\ & \quad
 + \PlusLog{2} \Bigl[
  -4 \Gamma_0^3 L_\nu^3
  -6 \beta_0 \Gamma_0^2 L_\nu^2
  + L_\nu   \bigl( -2 \beta_0^2 \Gamma_0 + 6\Gamma_0\Gamma_1 + 3\Gamma_0^2 S_1\bigr)
\nn\\ & \qquad\quad
  + \beta_1\Gamma_0 + 2\beta_0\Gamma_1 + \frac{3}{4}\Gamma_0(\gMuSCNC{1} + \gNuCNC{1}) + \frac{5}{2} \beta_0\Gamma_0 S_1 - 5\Gamma_0^3\zeta_3
 \Bigr]
\nn \\ & \quad
 + \PlusLog{1} \Bigl[
    L_\nu^2 \bigl( 8 \Gamma_0 \Gamma_1 + 4 \Gamma_0^2 S_1 \bigr)
\nn\\ & \qquad\quad
  + L_\nu \bigl( 2\beta_1\Gamma_0 + 4\beta_0\Gamma_1 + 2\Gamma_0 \gMuSCNC{1}+ 3 \Gamma_0 \gNuCNC{1} + 6\beta_0 \Gamma_0 S_1 - 20\Gamma_0^3\zeta_3 \bigr)
\nn\\ & \qquad\quad
   -\Gamma_2 + \beta_0 \gMuSCNC{1} + 2\beta_0\gNuCNC{1} + 2\beta_0^2 S_1 - \Gamma_1 S_1 - \Gamma_0 S_2 - 12\beta_0 \Gamma_0^2 \zeta_3
\Bigr]
\nn \\ & \quad
 + \PlusLog{0} \Bigl[
    L_\nu^2 ( 2\Gamma_0 \gNuCNC{1} - 16\Gamma_0^3 \zeta_3 )
  + L_\nu ( -2\Gamma_2 + 2\beta_0\gNuCNC{1} - 2\Gamma_1 S_1 - 2\Gamma_0 S_2 - 16\beta_0 \Gamma_0^2 \zeta_3 )
\nn\\ & \qquad\quad
  - \frac{1}{2} (\gMuSCNC{2} + \gNuCNC{2}) - \frac{1}{2}(2\beta_1 + \gMuSCNC{1} + \gNuCNC{1}) S_1 - 2\beta_0 S_2 + 4 \Gamma_0 \Gamma_1 \zeta_3
   + 2\Gamma_0^2 S_1 \zeta_3 - 6\Gamma_0^3 \zeta_5
 \Bigr]
\nn \\ & \quad
+ \delta(\pt) \Bigl[
  -\frac{16}{3}\Gamma_0^3 \zeta_3 L_\nu^3  - 8 \beta_0 \Gamma_0^2 \zeta_3 L_\nu^2 \nn\\&\hspace{10ex}
  + L_\nu \bigl( -\gNuCNC{2} - \gNuCNC{1}S_1 + 8\Gamma_0\Gamma_1\zeta_3 + 4\Gamma_0^2 S_1 \zeta_3 - 12\Gamma_0^3 \zeta_5 \bigr)
  + S_3
 \Bigr]
\,.\end{align}
The results through $\cO(\as^2)$ agree with the explicit calculation using the $\eta$-regulator in \mycite{Luebbert:2016itl}.
The $S_n$ denote the constants as derived using distributional setting, see \sec{soft func}.
They are related to $\tilde S_n$, the correspond constants with scale setting in Fourier space, by
\begin{align}
 S_1 &= \tilde S_1
\,, \\
 S_2 &= \tilde S_2 + \frac{4}{3}\beta_0\Gamma_0\zeta_3
\,, \\
 S_3 &= \tilde S_3 + \frac{4}{3}\beta_1\Gamma_0\zeta_3 + \frac{8}{3}\beta_0\Gamma_1\zeta_3 + \Gamma_0(\gMuSCNC{1}+\gNuCNC{1})\zeta_3 +\frac{10}{3}\beta_0\Gamma_0S_1\zeta_3
 - \frac{10}{3}\Gamma_0^3\zeta_3^2 - 8\beta_0\Gamma_0^2\zeta_5
\,.\end{align}


\phantomsection
\addcontentsline{toc}{section}{References}

\bibliographystyle{jhep}
\bibliography{literature}

\end{document}